\documentclass[11pt]{article}
\usepackage[letterpaper,margin=1in]{geometry}

\usepackage{multirow}
\usepackage{graphicx}
\usepackage{subfigure}
\usepackage{color}
\usepackage{import}
\usepackage{xcolor}
\usepackage{amsmath}
\graphicspath{{figures/}}
\usepackage{listings}
\usepackage{comment}
\usepackage{color,soul}
\usepackage{longtable}
\usepackage{multirow}
\usepackage{pdflscape}
\usepackage{lscape}
\usepackage{changepage}
\usepackage{textcomp}

\usepackage{pgfplots}
\pgfplotsset{compat=1.14}
\usepackage{tikz}
\usepackage{framed} 
\usepackage{textcomp}
\usepackage{float}
\usepackage{url}
\usepackage{hyperref}

\title{Thread Homeostasis: Real-Time Anomalous Behavior Detection\\ for Safety-Critical Software}
\author{Mohamed Alsharnouby \& Anil Somayaji\footnote{msharnouby@gmail.com, soma@scs.carleton.ca}}
\date{September 30, 2019\\Version 0.5}

\begin{document}
\maketitle
\begin{abstract}
Safety-critical systems must always have predictable and reliable behavior, otherwise systems fail and lives are put at risk.  Even with the most rigorous testing it is impossible to test systems using all possible inputs.  Complex software systems will often fail when given novel sets of inputs; thus, safety-critical systems may behave in unintended, dangerous ways when subject to inputs combinations that were not seen in development.  Safety critical systems are normally designed to be fault tolerant so they do not fail when given unexpected inputs.

Anomaly detection has been proposed as a technique for improving the fault tolerance of safety-critical systems.  Past work, however, has been largely limited to behavioral parameter thresholds that miss many kinds of system deviations.  Here we propose a novel approach to anomaly detection in fault-tolerant safety critical systems using patterns of messages between threads.  This approach is based on techniques originally developed for detecting security violations on systems with UNIX-like system call APIs; here we show that they can be adapted to the constraints of safety critical microkernel-based hard real-time systems.  We present the design, implementation, and initial evaluation of tH (thread Homeostasis) implemented on a QNX-based self-driving car platform.
\end{abstract}

\section{Introduction}

A safety-critical system is one that either directly or indirectly ensures the safety of its operators and users. The correct function of these systems is critical and their failure to mitigate or control a hazard could result in financial loss, injury, death, loss of vital equipment or damage to the environment \cite{nasa_2004}. Safety-critical systems are pervasive in our everyday lives, from microwave ovens to vehicles, planes, nuclear power stations, medical and space systems. With our increasing dependence on such systems and their growing complexity, safety-critical software is becoming much harder to develop while ensuring its correctness and safety \cite{6504883}. 

Despite the highest level of standards used in developing safety critical systems, runtime errors and bugs are never completely eliminated. Deficiencies in the software development process usually get a fair share of the blame due several reasons including incorrect or incomplete requirements, misleading specifications, poor design, miscommunication between teams and insufficient quality assurance. No matter how careful a safety-critical system is designed and developed, mistakes are made and errors are introduced along the way.

Given this reality, it is vital for a safety-critical system to have some form of error detection and recovery capabilities to prevent latent software errors from becoming faults and causing hazards \cite{Global}. Such systems are known as fault-tolerant systems \cite{nasa_2004}\cite{hobbs_2017}.  Several safety standards such as ISO-26262 for automotive functional safety and DO-178C for airborne systems recommend the use of some form of anomaly detection during the safety-critical system's operational phase. Anomaly detectors would enable a safety-critical system to detect errors and engage its fail-safe modes or other error correction functions before errors become disastrous. 

Much of the fault-detection research work focuses on detecting anomalies from individual components that are elected to be most critical. Since most systems are composed of highly interconnected components, we believe that in order to effectively detect faults, the system should be monitored as a whole. Isolating specific processes or components might distort the view of what is actually happening. This could result in delaying an action that could have been taken at an earlier and more appropriate stage and preventing an error from becoming a fault. Such a technique could prove to be even more powerful on a message-passing microkernel where high interconnectivity between its microservices is a core part of its design. An anomaly detector that monitors the behavior of the entire system however, must have a negligible impact on the system's performance and its ability to fully achieve its functional requirements.
 
Using the safety-certified message-passing QNX microkernel, we implemented an adaptive anomaly detector using a technique inspired by a detector first proposed by Forrest \texttt{et al.} \cite{forrest1996sense}: short-sequences of system calls. However, since most of the system calls on a QNX microkernel are implemented using message passing, our anomaly detector builds behavioral profiles for system processes using short-sequences of system calls \textit{and} messages passed between a process and the kernel and between a processes and another. We leverage the high interconnectivity and the distributed micro-architectural nature of processes running on a message-passing microkernel to our advantage: the effects of a misbehaving process can be quickly seen throughout the entire system.

One key challenge for any adaptive anomaly detector is obtaining sufficient training data in order to build a model that will produce few false positives.  
We take advantage of the fact that safety-critical software is subjected to an excruciating level of testing and build behavioral profiles during the verification phase.  In production, profiles would be static and no learning would normally occur.

This paper makes the following contributions:
\begin{itemize}
    \item Aa novel online anomaly detection method using short sequences of messages emitted by individual processes' threads. This work is an extension of the previous work of building models using short-sequences of system calls emitted by processes \cite{somayaji2002operating} and detects behavioral anomalies on a process thread-level. The design, implementation details and validation of the proposed detection method along with the framework for collecting the required traces at runtime are described.
    \item A lean and efficient mechanism to extract the information we need from a safety-critical microkernel is implemented.
    \item Real-life self-driving car technology is used to evaluate the effectiveness of our methodology for its ability to detect true positives with very low false-positive rates. We demonstrate that our solution can be used in an embedded environment to monitor all the processes in the system efficiently, requiring in our tests an average of 1.2\% of the CPU resources and 4.5 MB of memory on a highly active self-driving car system.
\end{itemize}

The rest of this paper is organized as follows: Section \ref{background} outlines a high-level overview of the software-engineering processes involved in developing fault-tolerant systems along with the deficiencies found in the processes. This is followed by background material on anomaly-detection for safety-critical systems including the anomaly detector that this work is based on.   In Section \ref{rationale} we argue our point-of-view for adopting behavioral-based anomaly detection for safety-critical systems and explain the gap we found in the current research. We also introduce the design guidelines we adopted for developing our anomaly detector and the rationale behind them. 
The design details of our implementation are explained in Section \ref{implementation} followed by an evaluation on a real production system in Section \ref{evaluation_chapter}. Finally, Section \ref{discussion} has the discussion of the results, our conclusion and future work.

\section{Background}\label{background}

Simple errors in safety-critical software can have severe consequences.  In June 1996 an Ariane 5 rocket exploded due to an overflow of a floating point to signed integer conversion \cite{ariane2}.  In February 1991 a timing-based calculation error in a US Patriot missile-defense system resulted in an intercept failure that resulted in killing 25 US soldiers and injuring 97 \cite{4085640}. More recently in October 2018 a Boeing 737 MAX dove into the Java sea off the coast of Indonesia, killing all 189 people on board \cite{stewart_2018} due to a misbehaving flight computer.

Safety-critical systems are designed, implemented, and evaluated to try and prevent such tragedies.  Safety requirements affect the system's requirements by defining the hazards that must be avoided and ones that must be handled during the operation of the final product \cite{Lutz:2000:SES:336512.336556}. Safety analysis techniques such as functional hazard assessment, fault tree analysis and failure mode effects analysis are some of the most prominent and proven techniques in the field \cite{1053007}.

Fault-tolerant systems attempt to mitigate potential failures arising from  design, implementatation, and testing errors or omissions using a variety of techniques.  Software Fault Prediction (SFP) uses machine learning techniques to build models that can be used to predict potential and fault prone software modules in the early stages of the development life cycle \cite{fault_survey}.  N-Version programming runs multiple, independently-developed versions of the same software in parallel, with faulty modules being detected in real time via conflicting outputs \cite{nasa_2004, 56849}.  Lockstepping can mitigate errors by running software or hardware in tandem, synchronous operations \cite{5590258, 7575387}.  Fault-Containment Regions, areas on the system to isolate safety and non-safety critical components, prevent faults from migrating between the different critical regions \cite{nasa_2004, 1193942, 1193942}.  High-assurance backup systems mitigate failures through the use of a trimmed-down, highly trusted backup system that can take over when the primary module exhibits problems \cite{Bodson1994,nasa_2004,Seto1998TheSA,4840571}.

Anomaly detection has been used extensively in fault-tolerant safety critical systems \cite{Pawar2015ACS} and such work  spans multiple disciplines such as chemical, nuclear and aerospace engineering \cite{5282515}.  Anomalies can be classified in two categories; point anomalies and contextual anomalies \cite{hobbs_2017}.  Point anomalies are specific data points that are outliers with respect to the rest of the data. Contextual anomaly detectors use behavioral and contextual attributes in order to detect anomalous behavior.  Most anomaly detection research in safety-critical systems is directed towards detecting point anomalies \cite{Chandola:2009:ADS:1541880.1541882}.

Anomaly detectors can either be online or offline. Online anomaly detectors build a model they believe best represents the system's behavior from various predetermined sources during the system's execution. This model is continuously updated with the runtime data as the system is executing, using efficient online algorithms \cite{pmlr-v9-kloft10a}. Once the model is ready for use, deviations from this model are marked as anomalies and can be used to raise alarms as the system runs. On the other hand, offline detectors require all the data generated during the system's execution to be presently available in order to build their models and detect anomalies. Offline detectors such as SiPTA \cite{Zadeh:2014:SSP:2656045.2656071} have the advantage of executing on machines that are more powerful than the typical low to moderate power embedded systems.

Salem \textit{et al.} \cite{7557872} presented an offline behavioral anomaly detector to analyse discrete process event trace data generated by an operating system  kernel. Their work relies on the fact that embedded systems typically have recurrent behavior. They propose a technique, inter-arrival curves, which is based on arrival-curves to extract high-level features and build a training model. The model uses the upper and lower bounds to the number of events that can occur within a specified time period to represent system behavior. During the training phase, event curves are generated per trace and then aggregated to provide input into a two-stage curve classifier that detects and measures the deviation from the training model. During testing time (offline), the inter-arrival curves from the trace data are generated and the classifier is used to detect whether the traces are anomalous.

Offline detectors are not suitable for all use cases; they suffer from some issues that might make them less ideal than their online counterpart. They require an extensive amount of data from previously executed system runs which might not be feasible in some use cases, especially those where the operating environment of the system changes frequently. Some offline detectors run post-mortem (after the system has finished execution), or remotely, on a different machine. This is another major limiting factor as it prohibits them from raising alarms as bugs occur in real time \cite{10.1007/11693017_23}. Raising alarms as bugs occur in real time is necessary for detecting anomalies before they become major failures. Offline systems usually do not have the flexibility of updating their models on-the-fly, at runtime, if new system behavior needs to be incorporated in its model of normal behavior.

Much of the literature on anomaly detection for safety-critical systems revolves around error detection in data collected from hardware components such as sensors and other key system components. This data is usually collected non-invasively, using components external to the core system or via embedded event tracing facilities. A model is built and different methods are then used to analyse the model and detect outliers or events that do not conform to the general pattern in the model. Gosh \textit{et al.} \cite{7786853} built a finite state automaton-based model to represent the behavior of Programmable Logic Controllers (PLC) used in manufacturing control. They use simple events, such as the boolean status of signals. The model, which is a simple set of state transition rules, is built using captured data from what they believe to be a fault-free manufacturing system. At runtime, the model is used to detect two types of behavioral anomalies: errors occurring in state transitions and errors that occur in the duration of those transitions. Chen \textit{et al.} \cite{8079160} developed a hardware module that detects anomalies in unmanned aerial vehicles at runtime. Their anomaly detector is based on the common supervised learning algorithm, least-squares support vector machine (LS-SVM). They use LS-SVM to predict and classify flight data generated from key sensors and components into anomalous and normal groups at runtime. Liu \texttt{et al.} pre-process continuous and discrete flight data then convert them into discrete sequences \cite{8622818}. Each sequence is further divided into smaller subsequences upon which variable length n-grams are obtained. The Unique n-grams are identified and used as the feature space for the training data-set. Each subsequence is an input feature vector to a one-class Support Vector Machine (SVM). The learnt classifier is then used to detect anomalous flight data in realtime.

Bovenzi \textit{et al.} propose using online statistical analysis at an OS level to detect anomalous user-level processes. Their solution does not have a training or profiling phase to understand how the system is expected to behave. They use indicators such as system call errors, disk I/O timeouts, process scheduling delays and time to acquire semaphores \cite{6847216}. The indicators are collected at regular intervals and the time series is analysed for anomalies based on the computation of upper and lower thresholds. Suspicious changes in the features of their model are considered to be a result of a fault activation. They base their analysis on injected faults at runtime and show that their algorithm can detect the breach of their calculated upper and lower threshold ranges (as well as global threshold). They then assume that this ``anomaly" is a result of an activation of a fault and is not part of the normal system behavior.

Several operating system reliability techniques are based on detecting system hangs or delays, be it the operating system or applications running atop. These methods are often concerned with the liveness property of a system versus its correctness. A hang condition can be either a result of infinite loops, also known as active hangs, or a result of permanent or extended wait conditions, known as passive hangs \cite{Carrozza:2008:OSS:2227461.2227474}. Irrera \textit{et al.} \cite{5703221} argue that critical systems are too complex and have highly non-deterministic behavior that renders online fault detection for safety-critical systems too difficult of a task. This non-determinism is the result of multiple factors such as multi-threading and the use of shared resources. This can cause a safety-critical application to hang, and hence their development of an application hang detector. They model normal behavior through various operating system parameters, namely system calls, OS signals, scheduling timeout, waiting times on semaphores, holding times for critical sections, processes and threads exit codes, and I/O throughput. For each of the monitored events, they associate pre-configured upper and lower threshold values which they determine from a training phase. At runtime, if an event deviates above or below the threshold within a given specified temporal window, an alarm is raised.  Wang \textit{et al.} \cite{4378410} developed a framework for runtime system hang detection and application checkpointing. One of the functions of their kernel module is to detect hangs in the operating system and its applications in order to provide a low-latency error detection and recovery mechanism. Their work relies on using various CPU debugging facilities and counters to non-invasively \footnote{In their opinion. Our opinion differs.} instrument the operating system. An example of one such instrumentation is keeping track of a process context switching frequency; a breakpoint is set to trap on the kernel's scheduling entry point and their profiling work is performed in their custom exception handler when the generated software exception is raised. Some instrumented values are profiled in order to provide more accurate thresholds before assuming a hang condition. Their related works section provides an excellent list of similar solutions in both academia and industry. 

While these reliability techniques have merit for certain use cases, for example in systems were high-availability is required, we believe that they cannot be the sole fault-tolerance technique in a system. Without doubt, system hangs represent a critical system failure, but they are merely one type of failure that could possibly occur. We also argue that reaching a state where the system has already hung, is a late stage if not a final one in the fault-detection life cycle. Such solutions are not suitable for all the complex operating environments that systems could be subjected to. For those, as we have argued earlier, a holistic view of the system's behavior must be taken into account. We would rather have a system revert to its design or fail-safe state if it is performing the incorrect function or misbehaving than have a lively system that never hangs but executes erroneously. Furthermore, in some cases, the hang detection algorithms require a pre-determined threshold that is tuned during the training phase such as the work by Carrozza \textit{et al.} \cite{Carrozza:2008:OSS:2227461.2227474}. This is not very well suited for systems with highly dynamic environments and might require adaptive threshold techniques like the one presented by Bovenzi \textit{et al.} \cite{10.1007/978-3-642-24270-0_10}.

Program control flow is considered to be a powerful indicator of normal behavior that can be leveraged to detect behavioral anomalies and software bugs. Argus \cite{10.1007/11693017_23}, an online bug detection tool, collects samples of pre-determined events that best represent the normal control-flow behavior of programs at runtime. Using the samples, an extended finite state automaton is used as a model of the programs' normal control flow. Each state in the ext-FSA, represents an event and the transitions between them are augmented with the distribution of their transition frequency. During detection time, a window of a fixed size is slid over the generated control-flow events and the contents of the window are then checked against the model raising an alarm if a certain threshold is crossed. In Lorenzoli \texttt{et al.}'s work, a more detailed model is automatically generated in an attempt to fully capture the behavior of software using an ext-FSA \cite{lorenzoli}. The ext-FSA uses both the constraints on the data as well as the interaction between components via method invocations. The relationship between them is also incorporated in the model.

Huang, H. \textit{et al.} created a runtime error detector for avionics software using the common machine learning clustering technique, K-means \cite{Huang2016}. The detector collects data from carefully hand-selected program variables such as aircraft altitude and vertical velocity during testing. A set of time-series signals is built from the collected events. Features and clusters are then extracted from the time-series signals. Using logistic regression, a fault-probability model is generated and an alarm is raised if the probability that certain data is erroneous exceeds a predetermined threshold. Their binary classifier was tested using a normally behaving autopilot simulator and another with two injected faults: an integer overflow and a unit conversion error.

Even though it is an offline diagnostic tool, Chopstix \cite{Bhatia:2008:LHM:1855741.1855749} attracted our attention. Chopstix is an offline diagnostic tool developed primarily for production systems. With a very low overhead, Chopstix continuously collects system-level events that it believes best describe the behavior of the entire system. Events such as CPU utilization, L2 Cache misses, page allocation and scheduling delays are logged at a very high frequency using a probabilistic data structure along with detailed contextual information. A human operator can then use Chopstix to analyse misbehavior. Chopstix uses a predefined set of rules in order to correlate anomalies and uncover the root cause before they become major system faults. Chopstix has a few commonalities with our proposed anomaly detector: i) it uses low-level operating system events for detecting behavioral anomalies; ii) it stores the events in a compact efficient manner with a very low CPU overhead (less than 1\%).

Langer \textit{et al.} propose two methods for anomaly-based error detection. In \cite{Langer_1}, the Angluin learner is used to infer a minimal deterministic finite state automation (DFA) that represents the normal behavior of a distributed system. They claim that the system states and control flow are directly correlated to its communication behavior and use network trace data to build a DFA. They show that in theory and by using synthesized trace data, detecting anomalies using this technique works very well. However, they could not conclude the same using real-world data since it was more complex that expected. The DFA is built using data generated during the testing phase. This closely aligns with our philosophy. Based on the exhaustive testing any safety-critical system is expected to be subjected to, they assume that the normal behavior of a system during its operational phase should not be different than any of its verification test cases. Thus, generating the model of normal behavior during the final product testing should be enough. Langer \textit{et al.} made an attempt to train an artificial neural network (ANN) on the discrete events generated by a distributed system's network traffic \cite{10.1007/978-3-642-04617-9_52}. The ANN is then used to forecast message sequences, comparing its prediction to the values generated at runtime. Deviations in the predicted versus actual values allows them to detect behavioral anomalies in the internal state of the software generating this traffic. Their choice of using an ANN comes from the fact that other mathematical models cannot fully cover the behavior of complex systems that exhibit discontinuous behavior or that the model would be infeasible to calculate. Again, this work has great similarity to ours; the messages being sent on a network are directly correlated to the internal state of the software sending or receiving them. A deviation from the normal traffic is used to indicate an anomaly in the system's behavior. Considering that our work is based on a message-passing system, our anomaly detector also builds a behavioral model out of the messages being sent and received by processes, be it kernel calls or others. It is worth noting that Langer \textit{et al.} have also proposed using their anomaly detector to detect the lack of testing during the product verification phase in order to detect missed test cases \cite{Langer}. To the best of our knowledge, this is the only literature other than ours that makes such a proposal.

Fault screeners are simple algorithms that detect errors in data based on unary program invariant \cite{908957} checking \cite{10.1007/978-3-642-14819-4_5}. Systems that are exposed to radiation, such as space systems, have a higher probability of hardware failure due to single-event upsets. Racunas \texttt{et al.} propose a fault screener as the processor fault-tolerance algorithm to handle such events \cite{4147658}. The static instructions generated by a process and their expected valid values are used to learn program behavior. They show that consistent upper and lower bounds can be calculated for the valid value space and show that these bounds are violated in the presence of a fault or abnormal program behavior. Abreu \textit{et al.} \cite{10.1007/978-3-642-14819-4_5} investigated and compared the use of several fault screeners: Bloom filters, bitmask and range screeners.

One of the techniques that had a common occurrence in the literature was the use of Markov models. Baah \textit{et al.} built an offline anomaly detector that can detect faults at runtime using a fully observable Markov model \cite{Baah:2006:OAD:1188895.1188911}. The model is built during a training phase and its parameters are estimated using the Baum-Welch algorithm. During training, the program to be monitored is instrumented to collect predicate information and the results collected during a test suite execution. The predicates are then translated into states which represent some of the program semantics and capture its behavior. The generated states are then used as an input to a clustering algorithm and the unique clusters are used to represent states in the Markov model. During detection time, if the probabilities of state transitions exceed a pre-computed threshold, an alarm is raised.
Bowring \textit{et al.} use the sequence of method calls and program branches as features for a Markov model representing program behavior \cite{Bowring:2004:ALA:1007512.1007539}. In order to create an aggregate representation of a program's executions, they devised a technique to cluster Markov models of a program's executions and create Markov model-based classifiers that model a collection of the program's executions. The classifiers are enhanced using an incremental active learning technique and the classifier is applied to a set of unlabeled program execution data. The outputted labeled data is then used as an input to the classifier in the next stage or increment, thus retraining the classifier.

Lu \textit{et al.} built a behavioral model from various system timing measures such as instruction and data cache mistimings, interrupt service routines, system and function calls and various other software execution timings \cite{Lu:2019:DAD:3319359.3279949}. They used three separate classification-based anomaly detection methods: i) a range-based classifier that utilizes the upper and lower bounds of executions times, ii) a distance-based classifier that creates a three dimensional sphere from timing data and uses the Euclidean distance to the sphere's center compared with the sphere's boundary as a measure of anomaly, iii) a one-class SVM that defines a normal class for timing data. They implemented their detector on a hardware module that taps into the processor's trace port.

Yoon \textit{et al.} tackle the safety and security of industrial plant control systems \cite{6531076}. Using the Gaussian Kernel Density estimation, they created a statistical learning-based intrusion detection mechanism based on software execution times. They utilize the multicore nature of realtime embedded systems and use one of the cores to monitor the other. A hardware Timing Trace Model (TTM) is used to monitor a specific application for deviations in runtime execution signatures. In case of an anomaly, the monitoring core takes control of the system to ensure an uninterrupted error and fault free execution (fault-tolerance). The software being monitored is modified to add a special trace instruction at various locations around code blocks. When the TTM receives a specific trace instruction from the monitored core it reads some of its processor state, the timestamp and the program counters are well as the process identifier of the executing task. The non-parametric probability density estimation function of execution times is estimated using the Kernel Density Estimation method with a Gaussian kernel function.

Non-intrusively, Moreno \textit{et al.} introduce a novel runtime monitoring technique based on power consumption analysis of a CPU for anomaly detection \cite{Moreno2018}. They employ signal processing techniques for signal pattern recognition for classifying system behavior and detecting program anomalies. The core principal is that a CPU generates a unique power consumption profile when executing different software loads. This power profile can be detected using external monitoring devices. Anomalous behavior or perturbations in the running software causes a deviation from this power consumption profile that can be easily detected at runtime. This work is one of the few that closely matches our idea of behavioral anomaly detection. No specific program features or variables were selected to build an understanding of normal behavior, rather, the resulting outcomes and effects a running system has from an outside observer's perspective: an external device monitoring power consumption in their case and a trusted entity (the kernel) monitoring emitted messages and kernel calls.

\section{Process Homeostasis}
\label{ph}
Process Homeostasis, or pH, is the name of a real-time host-based anomaly detection and response prototype developed by Somyaji A. as part of his PhD dissertation \cite{somayaji2002operating}. pH employs a simple, yet powerful heuristic to detect diversions in a program's normal behavior utilizing short-sequences of system calls that was first introduced by Forrest \texttt{et al.} as a profiling technique \cite{forrest1996sense}. The prototype developed for this thesis is based on the same core principal.
\subsubsection{Process Homeostasis and Immunological Basis}

Process homeostasis refers to the biological organism's ability to maintain a stable internal environment suitable for the organism's ideal functioning \cite{somayaji2002operating}. The destruction of foreign pathogens, the constant monitoring and correction of temperature, acidity and other chemical imbalances are all part of an organism's homeostatic nature \cite{somayaji2002operating}.

The natural immune system plays a vital role in protecting animals from harmful pathogens. The immune system is capable of generating detectors such as T cells, B cells and antibodies by which it distinguishes the body's own cells from foreign ones \cite{forrest1997computer}. Once an antigen has been recognized, the detector cells bind to it, thus signaling the start of a destruction process via general purpose scavenger cells. To recognize a pathogen, the immune system's detectors bind to short-sequences of proteins or peptides found in the pathogen. The binding regions on the detector cells are generated through a pseudo-random genetic process. The detector cells that could bind to the body's own cells (autoimmune disease) are eliminated    \cite{forrest1997computer}.

pH attempts to mimic how the biological immune system works by giving the system a sense of self or the ability to distinguish between the system's self and non-self. Such separation enables pH to maintain a homeostatic and stable system. Once pH defines the system under normal operating conditions, it is capable of detecting any foreign activity or anything that is non-self. This powerful biological technique enables the immune system to defend against foreign bodies that it hasn't seen before or hasn't previously encountered their signature. The definition of self cannot be too specific, or too narrow, that the immune system would mistakenly identify variants of itself as foreign. It also cannot be too general that it would allow infections to go unnoticed. 

 Forrest \textit{et al.} \cite{forrest1994self} deduced important properties from the natural immune system that a security system must have in order to be effective. First, just like the immune system, detection must be distributed over multiple sites; detection is not localized to one part of the body. Similarly, for software ecosystems, detection must be spread across the entire system as well. Different running copies of software on the same local system, as well as software running on interconnected systems must have multiple detection mechanisms. Furthermore, these detection mechanisms must be unique to each running instance of software. This diversity allows a more powerful localized detection mechanism and prevents a single vulnerability from compromising the entire system or a network of systems. Second, the immune system utilizes multiple detectors each with their own input in the decision making process, thus detection is probabilistic. This greatly decreases the misclassification of legitimate events (false-positives) in the system as a whole, on the account of reducing the detection rate of malicious activity in local sites. Lastly, the probabilistic nature of the immune system's detection enables it to recognize foreign objects that it has not encountered previously. This would allow a security system to detect zero-day vulnerability.

Forrest \textit{et al.} \cite{forrest1996sense} showed that short-range, temporarily-ordered sequence of system calls (modeled after peptides \cite{forrest1997computer}) can be used as a reliable fingerprint of processes' normal behavior. They showed that the fingerprint uniquely identifies system processes, has low variance and most importantly, sequences generated by an ill-behaving process can be easily and efficiently identified at runtime when comparing to a baseline sequence. The short-range sequence of system calls are those invoked during the normal operation of a process and are only a subset of all the possible permutations of system calls found in the program's source code. A profile of normal behavior must withstand variations under normal operating conditions and hence be resilient to false-positives. At the same time the profile should be easily disturbed by real intrusions in order to have a high detection rate \cite{hofmeyr1998intrusion}. 

pH defines a system's self by building a profile of normal behavior for each process in the system. Naturally and by definition, a system's self is the manifestation of its normal behavior. The normal profile is built using the type and relative order of system calls issued by a process. Using system calls allows pH to observe process behavior without knowing its internals or implementation details. pH treats the running instance of program code as a black box emitting observable data and employs only one type of detector and effector \cite{hofmeyr1998intrusion}, an abnormal system call detector and an effector that delays anomalous ones \cite{somayaji2002operating}.

It is important to differentiate between a process' normal behavior, legal behavior and all possible behavior. Normal behavior is all behavior seen when a process is operating under its normal, intended operating conditions, \textit{i.e.}, the software's primary function. Legal behavior is all normal behavior in addition to exceptions that might potentially occur during normal operation, for example, low disk space or a missing file. All possible behavior encompasses legal and normal behaviors in addition to all possible branches and illegal paths that the software could potentially take. This is due to imperfections in code and/or hardware and includes exploited vulnerabilities such as buffer overflows and hardware faults \cite{somayaji2002operating}. Given this definition, pH observes normal program behavior under the assumption that normal program behavior is almost always safe program behavior \cite{somayaji2002operating}.

 One evident problem with pH is the difficulty of defining self, given the dynamic nature of systems, without having to modify that definition later on. Consider a system where  profiles of normal behavior have been established; what then, if a new program was installed? Without the ability to modify the definition of the normal behavior, the newly installed program and the changes in behavior it creates in the system will be wrongfully flagged as foreign or anomalous. pH handles this problem by allowing the system's profiles to be modified at runtime.
 
\subsubsection{Overview}
pH is implemented as a patch to the Linux kernel and has three main parts: a utility to trace system calls invoked by all processes on the system, a utility to analyze the traced system calls, and a utility to react to anomalies by slowing down the invocation of the offending call. pH sits between user space processes and the kernel right at the system call entry point in the kernel. System calls are a critical component to software running on any operating system. System calls are a predefined set of function calls used by processes to interact with the kernel in order to perform functions that the process itself does not have the privilege or ability to perform, such as allocating system memory, opening a file or communicating with hardware. System calls are implemented through a pre-defined software interrupt that traps into the kernel and is serviced by a privileged (Ring 0) routine.
    
\begin{figure}
	\centering
	\includegraphics[width=\textwidth]{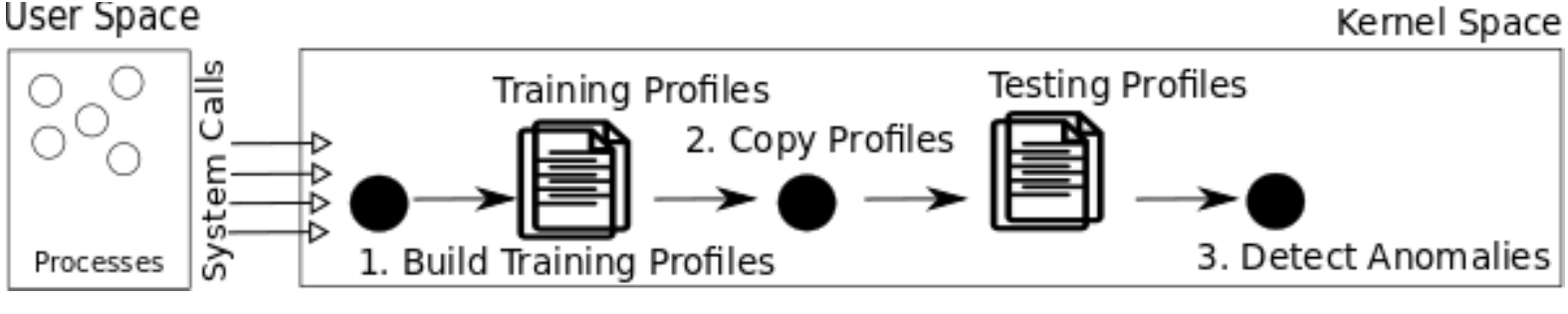}
	\caption{Training and testing overview.}
	\label{fig:fig1}
\end{figure}

Monitoring a process at the system call level does not detect or prevent corruptions within the process itself, for example, a buffer overflow attempting to change the result of a calculation within the process' own address-space. However, if a vulnerability, or an error in the process were to cause any kind of system-wide malicious activity, such as spawning a root shell or sending a command to another driver or another critical system-level component, it would certainly need to interface with the kernel by invoking system calls. 

pH starts in training or learning mode. It builds a profile of system call sequences collected under normal operating conditions. A profile exists per monitored process. pH monitors all processes on the system. Once pH decides that it has seen enough system calls and that it has adequate information that enables it to start detecting anomalous system calls sequences, it locks the profile down, creates a copy and switches into detection or testing mode. In testing mode, if an anomaly is detected, pH delays the offending system call by putting the invoking process to sleep for a calculated amount of time. Both the original and the copied profiles are stored on disk and loaded into kernel memory on process execution. Figure \ref{fig:fig1} shows an overview of the process.

\subsubsection{Training Phase}
During pH's training phase, pH builds a compact structure that represents a generalization of all the system calls observed per monitored process by using a simple heuristic called the lookahead pairs method. pH's designer has set a few constraints on the method to be used. Primarily, the method should be able to:

\begin{itemize}
	\item Converge as fast as possible at a fixed state during training phase.
	\item Efficiently determine the membership of patterns during the detection phase.
	\item Capture different and unique system call types with an acceptable limit on the size of generated data.
	\item Permit fast and incremental updates.
	\item Detect anomalous events that occur with low-frequency.
	\item Run high speed with a low and acceptable performance overhead.
\end{itemize}

\begin{figure}
	\centering
	\includegraphics[width=\textwidth]{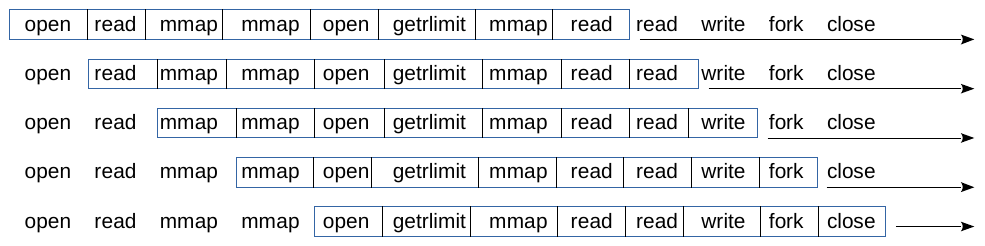}
	\caption{A sliding window over the system call sequence.}
	\label{window}
\end{figure}

As shown in figure \ref{window}, a window of a pre-determined size is slid across the system call sequence. For the calls found within the same window, pH records which system calls come before the one currently being invoked and at which position. This is implemented as a circular buffer of size (window length + 1) that gets updated every time a system call is invoked.

\begin{figure}
	\centering
	\small
	\includegraphics[width=\textwidth]{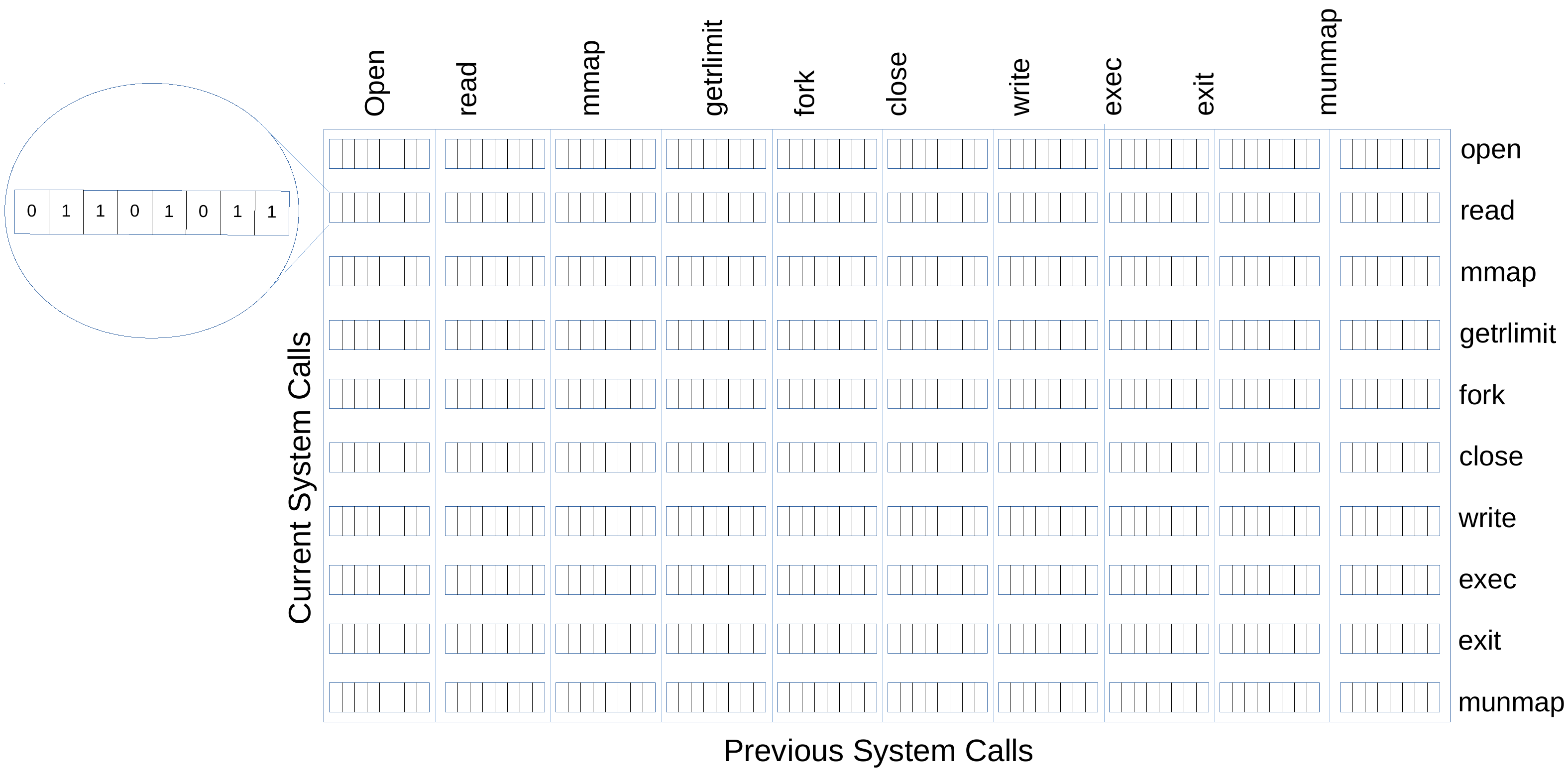}
	\caption{An NxN table of windows of size 8.}
	\label{fig:ts}
\end{figure}

A two-dimensional N-by-N table (shown in \ref{fig:ts}), where N is the number of discrete system calls in the system, is used to store the relative positions of the calls. Each cell in the table contains a one dimensional array representing the sliding window of a fixed size. The rows of the table represent the current call being invoked and the columns represent the previous calls. Their intersection yields the one dimensional array that indicates the relative positions between the current and all previous system calls that have occurred in the past. Specifically, ``window-size" previous system calls.

To give an example of how the table is updated, Figure \ref{fig:fts} shows a sample sequence (open, read, mmap, mmap, open, getrlimit, mmap, read, read, write, fork, close) with a window size of eight. The first open call is inserted into the circular buffer and causes no update since it is the only element in the buffer. Next, read is inserted. This causes pH to iterate through the circular buffer starting from the element that precedes the current one, i.e open. Next, the row indexing the current system call or ``read" is selected along with the column indexing the previous call or ``open". The intersection yields an array of size eight (the window size) that needs to be updated with open's relative position to the current read call within the window. Since open is the first system call preceding read, the first position in the window is marked.

The third mmap call causes the mmap row and the read column to be updated at position one. As well, it causes the mmap row and the open column to be updated at position two (since open is two positions away from mmap within the window). This same operation is repeated as more system calls are invoked. Skipping ahead to the ninth invocation of read, the read row is updated at columns read, mmap, getrlimit, open, mmap, mmap, read and open at positions 1,2,3,4,5,6,7 and 8 respectively.

To that extent, for any given window with size L and a current system call S, there are L - 1 pairs: \{ (S , S-1), (S, S-2), ... (S, S-L)\} can be formed for any window given. The set of all unique pairs form the compact model of behavior under normal operating conditions. The model serves as an approximation or a generalization of how the program is expected to behave.

\subsubsection{Termination of the Training Phase}
Once pH is convinced that it has seen enough system calls that represent the process' normal behavior or is instructed to do so by the system administrator, a duplicate of the training data is created. This copy is used at runtime during the detection phase to classify system call invocations whether they are benign or anomalous. The copy of the training profile will be referred to as the testing profile from here on. pH then turns on a flag stating that this particular training profile belonging to a program is complete, thus signaling the start of the detection phase for this program. The testing profile is never directly updated by pH. Both the training and testing data are stored in files on a permanent storage medium to persist across system reboots. Every program on the system has (unless explicitly ignored) a pair of training and testing files.

In order for pH to autonomously terminate the training phase for an executable, a couple of conditions must be met. pH must make this crucial decision with utmost care. A premature termination of training can result in a high percentage of false-positives, since pH would not have captured enough sequences to represent the program's normal behavior. To determine stabilization in training data, pH uses a simple heuristic that allows it to monitor the occurrence of new system call sequences both in terms of system calls and in terms of unit time.

\begin{figure}
	\centering
	\small
	\includegraphics[width=\textwidth]{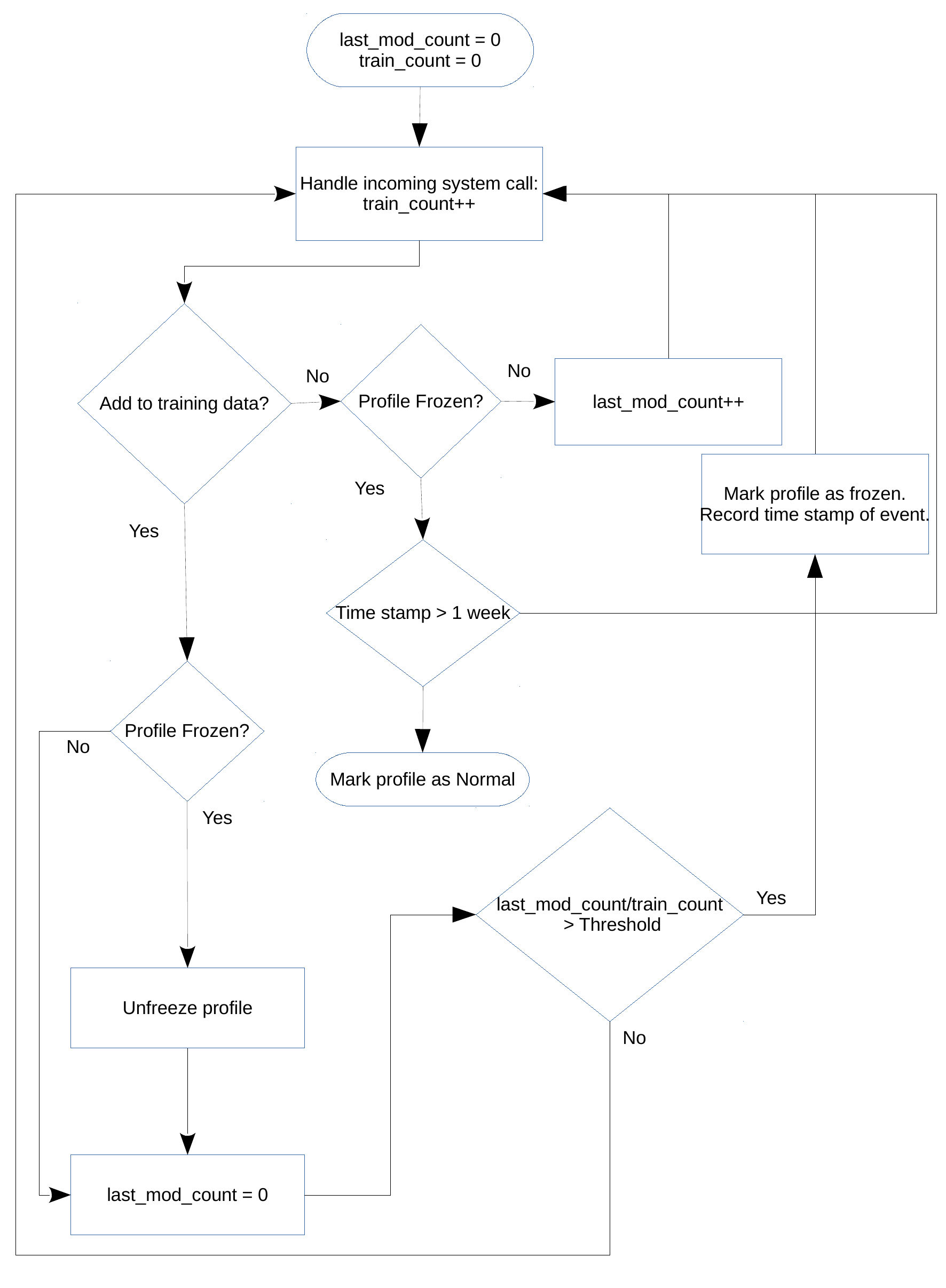}
	\caption[Profile Normalization Flow Chart]{last\_mod\_count keeps count of the number of system calls ignored. train\_count is the total number of system calls ever seen for that process.}
	\label{fig:lmc}
\end{figure}

\begin{figure}
	\centering
	\small
	\includegraphics[width=\textwidth]{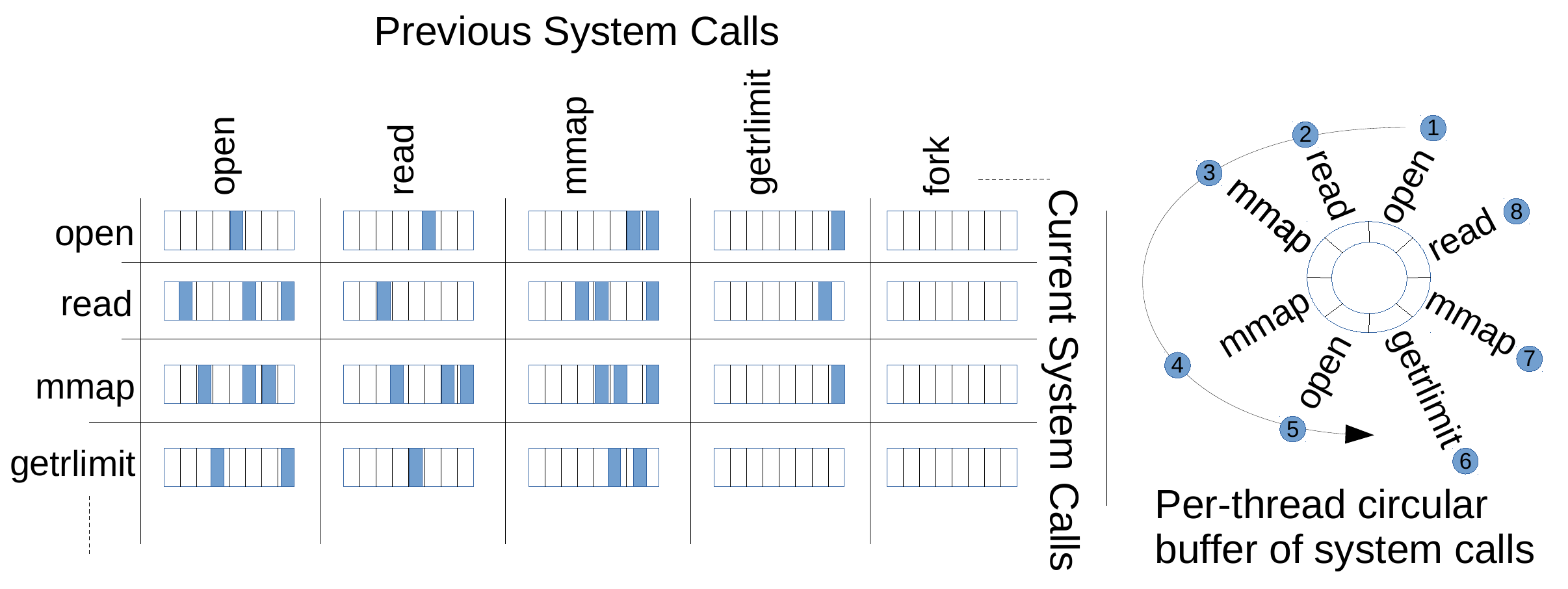}
	\caption{Building a training profile from a system call sequence: open, read, mmap, mmap, open, getrlimit, mmap, read, read, write, fork, close and window size 8.}
	\label{fig:fts}
\end{figure}

Figure \ref{fig:lmc} shows the flow chart of the heuristic. Two main measures are required for this decision:
\begin{enumerate}
	\item A ratio of the number of consecutive calls ignored (last\_mod\_count in Figure \ref{fig:lmc}) to the total number of calls made (train\_count in figure \ref{fig:lmc}). In other words, how many calls occurring in sequence have been ignored and not added to the training data because they already exist in relation to the total number of sequences ever seen. The count of ignored calls since the profile was last updated would also indicate that the profile has not been updated for count calls and hence frozen. If this ratio crosses a certain threshold, pH marks the training profile as 'frozen'. The threshold is a configurable runtime parameter; by default, pH sets this value to 4.
	\item The amount of time the profile has been frozen. Every time a profile is marked as frozen, the time stamp of the event is recorded. Once a preset amount of time has passed (a week by default) the profile is marked as 'normal'
\end{enumerate}

\subsubsection{Detection Phase} 

On program startup, pH will attempt to locate its testing profile. If the file has already been loaded in memory due to the same program being started before, pH will use the in-memory testing data. Otherwise, pH will locate the file on disk and load it in memory. Since program behavior is being profiled, it is important to note the distinction between a process and a program. A process is the runtime manifestation of some program code; multiple instances of the same program code will point to the same training data. 

Similar to the training phase, incoming system calls are added to a circular buffer. On each insertion, the compact testing data table (previously the training table) is checked for mismatches. For a system call \textit{n} inserted, the table at row \textit{n} is checked for every call preceding it in the circular buffer. Thus, for every call \textit{n} preceded by a call \textit{m} at position \textit{p}, the table is checked at row \textit{n} and column \textit{m}. The row and column intersection contains the relative position array, which in turn is checked at position \textit{p} for a true or false value. A call that has never been seen during training within that particular window (or circular buffer) will result in \textit{window size - 1} mismatches.

\subsubsection{Reaction Phase}
pH reacts to anomalies by delaying the anomalous system call's invocation, this includes the processes' ability to load new programs with the execve system call. The length of the delay is calculated in proportion to the number of recently illegitimate calls and is meant to give the system administrator the chance to evaluate whether this is the correct behavior and to delay the potential attacker. The idea behind the proportional delays is that true security violations will generate more illegitimate calls in succession, or very close to one another, thus, the more clustered the anomalies are, the larger the delay. pH maintains a locality frame count or LFC to keep track of such clusters.

\subsubsection{Tolerization and Sensitization}\label{tolerize}
 In order to help reduce the false positives in the field, pH implements two additional processes. If pH detects more than \textit{anomaly\_limit} anomalies at runtime pH decides that normalizing the process was incorrect or perhaps pre-mature. pH accepts the behavior as normal or tolerizes it, cancels the processes' monitoring, and restarts its training.
 
 On the other hand, in order to prevent pH from automatically learning anomalous behavior through tolerization, a threshold \textit{tolerize\_limit} for the locality frame count (LFC) is set in place. Since LFC is an indication of clustered anomalies, the likelihood that the anomalous sequence will bring a true positive is high. If this threshold is ever exceeded, the profile's training dataset is deleted and all previously learned sequences are deleted.

\section{System-Wide Behavioural-based Anomaly Detection: Rationale and Design Principals}\label{rationale}

In this chapter we present our own opinion and argue the need for adopting behavioral-based anomaly detection for fault-tolerant systems. After having reviewed the literature and presented the state of the art in the field, we could clearly see the gap in the research and product development that our solution fills. We also present the critical design guidelines that we adopt for our anomaly detector and the rationale behind them.

\subsection{The Need for Behavioral Anomaly Detection}
Detecting behavioral anomalies can play a vital role in safety-critical software, in particular, detecting deviations from what is believed to be normal behavior. Conventional methods of error detection, such as assertion-based methods, are not enough for detecting errors at runtime. The highly variable nature of the processes' operational environments produce new relationships between system entities and program variables \cite{Huang2016}. These new conditions might not have been accounted for during development and verification, leaving bugs that do not violate the rules of the program undetected \cite{10.1007/11693017_23}. Thus the need for fault detection mechanisms that could learn the intended behavior of software instead of what has actually been implemented. 

Faults do not have clear and concise signatures, due to the complex relationships between components, some of which are off-the-shelf components. Failure signatures cannot be predicted. Asserting specific conditions at runtime is important, however, runtime asserts alone are an incomplete method of being fault-tolerant. Software faults can be detected either explicitly, via detecting a specific pattern that is known to be produced by a particular fault or implicitly, by detecting an anomaly in a component that is indirectly linked \cite{980003}. Even though a safety-critical process could appear to be behaving normally, an abnormally high CPU usage or a misbehaving sensor driver can be an early indicator of a fault that is about to occur. Some faults cannot be detected through explicit patterns and assertions; they can only be detected through recognizing anomalous behaviour that does not match normal expectations.
\par 

As we have stated before, validating the correctness and completeness of the software specifications and requirements unto which the entire system is based is no trivial task. The software specifications are often wrong or incomplete \cite{Langer}, rendering some of the verification efforts useless. To this date, based on our literature and market product reviews, we are not aware of a system or process that achieves this task as well as correlates these requirements to source code and test specifications ensuring the correctness of the full development life cycle. This increases the dependence on the observed product behavior and common sense. In many cases, important corner cases and system stress tests are missed. A behavioral anomaly detector can make sure that a safety critical system does not fail as it experiences situations it was not designed to handle. An anomaly detector can provide an early warning as it detects abnormal behavior, allowing the system to handle the potential fault and fail safely.

No matter how much effort is being put into testing and verification, it will always be insufficient to catch all latent software errors before the system is deployed and used in the field. Researchers have categorized bugs into two major categories: deterministic and non-deterministic. Deterministic bugs can be easily detected and are a result of a particular combination of input to the program \cite{6405499}, also known as Bohrbugs. Non-deterministic bugs seem chaotic and result from the presence of a particular set of inputs along with other external factors \cite{6405499}, also known as Mandelbugs. These external factors are dependent on the system's operational environment and state and are highly variable. Thus, an unknown combination of events in a given environment can result in triggering a hidden Mandelbug. This makes Mandelbugs very difficult to occur during the verification phase. Meanwhile, it is not possible to detect errors that do not result in system breakdowns \cite{Langer}. This showcases the need for a runtime mechanism that can learn how software should behave; making it feasible to detect errors, or anomalies leading to errors, that have not been seen during the controlled verification tests. Test coverage is inherently insufficient and cannot be relied on to cover all possible system states that are produced when integrating different system modules and testing them in all their operational environments. Even though best efforts have to be made, covering every single possible operational environment does not seem feasible. This is further exacerbated by the presence of the elusive Mandelbug.

\subsubsection{The Difficulty of Feature Selection} \label{feature}

Most anomaly detectors we have seen use specific software features or variables as basis of profiling software (e.g. in  \cite{6847216} \cite{4378410} \cite{10.1007/11693017_23}\cite{Huang2016}\cite{Bhatia:2008:LHM:1855741.1855749}\cite{Baah:2006:OAD:1188895.1188911}\cite{Lu:2019:DAD:3319359.3279949}). This might not be the most accurate technique; while these carefully chosen features might provide some form of behavioral anomaly detection, they do not provide a holistic and complete view into how system components truly behave. These select features are nothing but a small window into the outcome of some of the actions a process takes. What if the choice of features was incomplete? How can one determine those features when there is an unbounded number of system states that result from the complex interactions of the different components in the system. Sometimes the features selected are incorrect, such as error logs \cite{4781208} that have no formal definition, or the presence of concrete requirements, and are left up to the developer's prerogative \cite{5544279}. Sometimes the features are not resilient to the dynamic nature of systems and the dynamic work loads they can be subjected to, such as process scheduling times \cite{6847216}, CPU utilization \cite{Bhatia:2008:LHM:1855741.1855749} or other general system timings \cite{Lu:2019:DAD:3319359.3279949}. These might represent point anomalies in a specific context, but are not necessarily anomalous in a different context. Using specific features is analogous to a security guard keeping a list of features of employees that are allowed to enter a restricted area, say eye and hair color, height and weight. This security guard can be easily fooled by anyone mimicking the features on his list. Furthermore, an employee changing their hair color or gaining weight would be considered an anomaly and disallowed into the building. The security guard did not have enough foresight to consider the complex interactions that occur in people's lives that could result in weight gain, making this a difficult task to achieve. A more intelligent security guard would be able to profile people based on their behavior, the way they talk, walk and conduct themselves. 

Feature selection is a difficult task and the effectiveness of an anomaly detector is heavily dependent on this selection; an inferior set of features can result in an anomaly detector learning an ineffective model of normality \cite{Kloft:2008:AFS:1456377.1456395}. In addition to the inaccuracy of using a select subset of features, determining this subset is in itself a non-trivial problem that fault prediction algorithms face, given the vast amount of features that a complex system has \cite{5703221}. Various research have gone to length in order to accurately identify the most valuable features for fault prediction such as the one presented by Irrera \textit{et al.} \cite{5703221} and the work presented by Kloft \textit{et al.} \cite{Kloft:2008:AFS:1456377.1456395} in which they propose using machine learning techniques to aid in feature selection. 

\subsection{Thread Homeostasis: The Design Principals}
This section describes the general and high level design principals behind our system-wide thread behavior anomaly detector using short sequences of messages and system calls. The details of the implementation are described in chapter \ref{implementation}.

\subsubsection{Sequences of System Calls as a Behavioral Profile}
From an outside observer's point of view, be it the kernel or a system process, we view the behavior of a process as the way it interacts with other system components. A process can bootstrap itself, open files, create shared memory objects and semaphores, use pipes, send messages and invoke the kernel among other things. Thinking back to the security guard analogy, this is how the process walks and talks. In order for a process to complete any of these functions, it has to interact with the kernel via system calls. For the design of our anomaly detector, we build upon previous seminal work in the field by Forrest \textit{et al.}, that a process' normal behavior can be defined via the short sequence of system calls it emits \cite{forrest1996sense}. Our anomaly detector's design follows pH's, the behavior anomaly detector designed and developed by Somayaji \text{et al.} \cite{somayaji2002operating} and described in detail in chapter \ref{ph}. pH uses the system call types along with their relative order to build a behavioral model for every process. 

Using pH's heuristic, the lookahead-pairs method (described in section \ref{ph}), a generalization of the process behavior is modeled in a compact profile. This model allows us to profile the process's natural and general behavioral without being too specific as to the exact sequence of system calls it should be emitting. The system calls emitted by a processes are bound to vary from one run to another, a model that is not representative of such variance in behavior will be rendered useless with too many false-positives. We monitor the process in its natural operating environment until we are convinced we have seen enough of all of its legitimate natural behavioral. Any deviance from this model is regarded as un-natural and thus anomalous behavior. Thus, we build an approximation of the process' normal behavior while maintaining the sensitivity towards novel sequence and anomalies as described in pH's design guidelines \cite{somayaji2002operating}.

One could argue that sequences of systems calls, as a selected profiling feature, could suffer from the same feature selection issues we described in section \ref{feature}. However we believe that building a compact model of system calls that processes emit allows us to determine its behaviour at a very granular and yet more generic level. Consider an error detector that decides to track the number of open file descriptors that belong to a process; instead of keeping count, we monitor the sequence of system calls a process has been invoking. An anomalous number of open file descriptors will surely be reflected in the erroneous calls to \textit{fopen()} found in the sequence, without having to specifically identify what we're looking for beforehand. This removes the burden of having to select variables or extract important features to build a model of what the normal operating characteristics are believed to be. The number of open file descriptors is not in itself the behavior of a process but rather is the result of an action, the call to \textit{fopen()}.

\par

\subsubsection{System-wide Profiling Using Sequences of Messages}
Given that our anomaly detector is specifically designed to run on a message-passing microkernel, modeling sequences of system calls is not enough. Many critical system services, such as the filesystem and process manager are implemented as user-level system services. These services can only be communicated with via messages. In addition, many system calls are actually implemented using the microkernel's message passing utilities. For this reason, our anomaly detector builds a compact profile of both system calls and messages that a process sends. This allows us to capture the full behavior of a process with regards to its interaction with microkernel as well as any other process in the system it communicated with. Profiling all inter-process messaging allows us to create a system-wide behavioral profile of all the system processes' interactions and enables us to detect anomalies in the process network.

\subsubsection{Thread-based Profiling}
Since the behavior of a process is the result of the aggregate behavior of its running threads, our anomaly detector builds bahvioral profiles for every thread running on the system. We utilize the concept of a process only to group its thread's profiles together. Profiling individual thread behaviors allows us to eliminate unwanted changes in behavior that occur due to factors outside of the thread's control, such as thread scheduling or thread pools. This is further expanded on in Chapter \ref{implementation}. We explicitly define the normal behavior of a system in terms of the short sequences of system calls and messages emitted by each thread in the system and hence the name Thread Homeostasis (tH).

\subsubsection{The System As a Network of Message-Passing Threads}
Sufficiently complex systems are composed of highly-coupled components. An abnormal behavior in a minor component that is not directly connected to the core operation of the system can potentially bring the entire system down. While isolating specific components in a safety-critical system for monitoring might have its merits, monitoring the entire system for faults and anomalies is essential for effective anomaly detection. One does not know where and how a fault can be introduced in the system. A perturbation of anomalies in one part of the system can be a strong early warning sign that a fault is about to happen in a critical system component. This would allow the fault-tolerant system to react in a timely manner. A hang or a system crash are far too late in the fault detection stages. We view the system as a network of highly communicating, interconnected threads and attempt to learn the entire network's behavior as well as detect anomalies over the network as a whole.

\subsubsection{A Lightweight Anomaly Detector}
An anomaly detector cannot change system properties in such a way that the core functional requirements of a system are violated. Sensitive functional requirements such as stringent timing deadlines that must be met might be violated given a detector that presents a heavy load and overburdens the system. As such, tH must have a low memory and CPU utilization overhead that fits the criteria of running on embedded systems in order to learn and detect anomalies in an online, real-time manner. The detector should be as minimally intrusive as possible as not to alter the underlying system's behavior. We view system threads and processes as black boxes and observe their behavior from an external entity using a reliable source of information, the kernel. Modifications to the running software or the underlying operating system might alter the state of the system and require further verification, testing, and perhaps even re-certification, which might decrease the acceptance of adopting such technology. Having to modify the software to support anomaly detection means that third-party, off-the-shelf components cannot be monitored as-is and must be either modified or excluded from being monitored. 

\subsubsection{tH as a User-level System Process}
An error in the anomaly detector itself might compromise the safety of the entire system if not properly isolated from other critical functions. Another advantage of using a microkernel-based operating system for our implementation is that the anomaly detector will be running as a regular user process, just like all the other system services. Its failure to operate correctly is completely isolated from other safety-critical functions.

\par

\subsubsection{Building Behavioral Profiles During Verification}
Safety-critical systems are subjected to an excruciating level of testing and verification. Many safety-critical systems require certification before being allowed to operate in the field. Certification bodies such as ISO 26262 for automotive functional safety usually have very high standards when it comes to fault-tolerance testing and verification of the correctness of these systems \cite{SINHA20111349} \cite{icsoft-ea13}. As a result, they are subjected to an excruciating level of verification. We are of the opinion that safety-critical software should experience a large percentage of the modes of operation it will be subjected to when deployed. We make the assumption that if a misbehavior is not seen during testing it should not be seen during deployment. That is not to say that the operational environment of the system must exactly match the test environment, but it must be as close as humanly possible. We utilize this to our advantage and use the verification phase to learn the expected behavior of software before it becomes operational. More importantly, since we know that verification would most likely be incomplete, we cannot claim that we are learning how the system should behave from the verification phase, as this would be incomplete learning. Rather, we ought to claim that \textbf{we are learning how the system should not be behaving} in the field.

If an anomaly is observed in the field, this might be an indicator that i) an actual anomaly is present or ii) an incomplete test coverage was performed, or iii) a false-positive was detected. However, an anomaly occurring during the verification phase can be an indicator that i) some normal behavior was never experienced during the testing stage, indicating to a shortcoming in the test coverage, or ii) unnecessary functions exist in the product that do not conform to the system's specification. These are serious enough indicators of a problem somewhere in the development life cycle that warrants an investigation. Using an anomaly detector as part of the critical software development life cycle might prove to be a useful tool.

\subsubsection{Low False-Positive Rate}
The false alarm rate is defined as the rate of misclassified normal behavior. An ideal anomaly detection system would be one that has a 100\% detection rate and a 0\% false alarm rate. False alarms depend on how well the model captures the normal behavior of a program while ignoring the information that does not generalize well. The false alarm rate remains one of the top limiting factors of how usable and effective an anomaly detector is \cite{Axelsson:2000:BFD:357830.357849}. Even for very low false positive rates, the amount of data generated, thousands of kernel calls per second in our case, would translate into a considerable number of false-positives \cite{Axelsson:2000:BFD:357830.357849}. This renders the system completely useless or even dangerous to use, since users habituate and learn to ignore warning messages \cite{anderson2014users}, which sometimes are very serious true positives. Thus, an anomaly detector must strike the perfect balance between maintaining a tolerable rate of false positives, while having a higher rate of true positive detection.  In contrast to general purpose desktop machines, embedded safety-critical systems naturally have more concise and limited operations. They are mostly created to perform a few precise functions. This makes them a fertile ground for more successful behavioral profiling and modeling efforts. We hypothesize that safety-critical processes might show an acceptable level of behavioral determinism that makes it easier to profile their behavior. Deviations from a learnt behavioral model can be marked as anomalous with a higher degree of confidence than their general purpose desktop machine counterpart.

The next chapter details the technical design based on the guidelines presented in this chapter, followed by a field evaluation. Much could be achieved by learning how to not behave.

\section{Thread Homeostasis on a Message-Passing Microkernel}\label{implementation}
Based on pH's core concept of modeling the behavior of processes using short sequences of system calls, we propose a new online anomaly detector, Thread Homeostasis (tH). Our system is specifically tailored for use on a message-passing microkernel, namely, the QNX\texttrademark\ realtime safety-critical operating system. tH differs from pH in two key areas: i) as opposed to profiling user-level processes, it profiles and builds behavioral models for every thread in a user-level process in the system, and ii) in addition to using system calls for profiling behavior, tH extends this profiling mechanism to include unique identifiers of messages used in all inter-process communication, including the messages sent to system-level services. 

tH is an online and non-invasive anomaly detector. It relies on QNX's highly granular tracing facilities to capture data at runtime and update its models. This chapter describes the design, technical details and the rationale behind the newly proposed system and its operating environment.

\subsection{The Design approach}
Before we get into the details of tH's design, it is important to shed some light on how we came to our design decisions. We adopted an iterative development methodology. Initially, we started with an attempt to port pH directly onto QNX as is, without any change. However, due to the message-based system call implementation, this proved to be impossible. A new technique had to be created to account for the fact that not all system calls are implemented as traps into the microkernel but rather some are delivered to the microkernel as messages. This led to the extension of profiling to all message-based interprocess communication. We developed a mock client-server application to verify the correctness and performance of our detector. Stress testing demonstrated issues that led to further design changes to tH and to the QNX kernel. After our controlled testing showed positive results, we started field evaluations with real deployed applications. This led to a decision to build per-thread as opposed to per-process models. The following section describes tH's technical details and design.

\subsection{The QNX\texttrademark~Operating System}

QNX is a POSIX-compliant, real-time, message-passing, preemptible microkernel. QNX is primarily used in embedded safety-critical systems such as medical equipment and automotive systems. QNX is a true microkernel; applications, device drivers and other system critical services run in isolation in user-space. The microkernel implements a few core functions such as thread, timer, synchronization, message-passing, signal, and scheduling services. Other common operating system services such as process and memory management are not part of the microkernel and run in regular user-space. This allows for a greater isolation between critical services such as drivers and the kernel itself. In addition, it enables critical system components to be restarted upon failure without the need to restart the entire system. 

Message passing is QNX's main interprocess communication mechanism and is a core fundamental function upon which QNX relies on. Message passing allows a process to send a message to another receiving process synchronously. In order for a message to be received, the message receiving process, or server, creates what is known as a channel. Each channel has a unique identifier. A process wanting to send a message to another process must first connect to this unique channel ID along with the receiving process ID, thereby creating a connection. Each connection has a unique identifier and refers only to the receiving process, thread, and channel IDs. Messages are the primary communication mechanism between regular user processes and system-level processes such as the process manager.

\begin{figure}
	\centering
	\includegraphics[width=\textwidth]{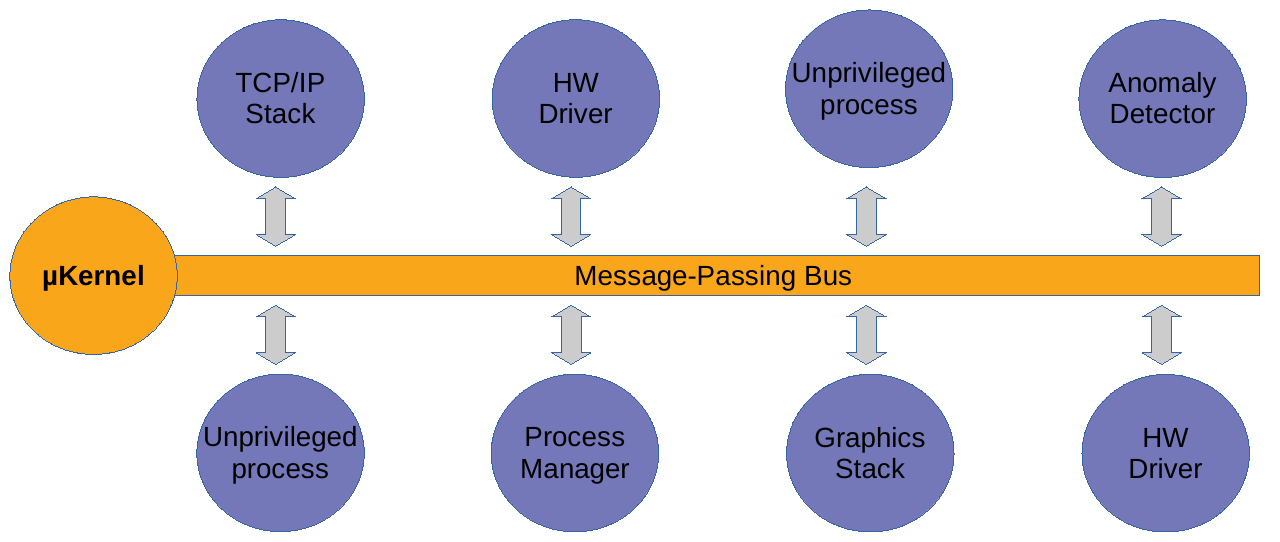}
	\caption{The QNX message-passing bus.}
	\label{fig:ipc}
\end{figure}

Thus, as shown in figure \ref{fig:ipc}, the QNX microkernel, along with a set of user-space privileged and non-privileged services connected to the message-passing bus, form a fully functional operating system. 
QNX only has 88 system calls \footnote{As of QNX version 7.0} that are implemented as an exception or a trap into the microkernel. The rest of the system calls are implemented as messages to the appropriate system service such as the process or the memory manager. As an example, in a traditional monolithic kernel such as Linux, a process wishing to create a new process calls the \textit{spawn()} system call, however, on QNX\texttrademark, calling \textit{spawn()} results in a message being sent to the process manager, using the \textit{MsgSend*()} API,  with a \textit{\_PROC\_SPAWN} message header type. The rest of the message payload would include all the required spawn parameters and flags. Since messages are synchronous, the process manager receives and performs the request while the sending process is blocked, waiting for its reply. The \textit{MsgSend()} API itself is a true system call implemented as a trap into the microkernel, but its only function is to deliver a payload on a specific channel ID to a specific process. Most of the system calls on the QNX operating system are implemented as messages and delivered via the message-passing infrastructure. For the rest of this thesis, we refer to system calls implemented via messages as \textbf{kernel calls} and calls implemented as a direct trap into the kernel as \textbf{system calls}.

QNX provides comprehensive instrumentation utilities. The instrumented version of the micokernel includes tracing facilities that provide a highly customizable and configurable trace data stream. This enables real-time monitoring of a QNX system at runtime. QNX states that there is a small 2\% performance overhead when using the instrumented versus the non-instrumented version of the microkernel. QNX provides APIs (the System Analysis Toolkit (SAT)) for programs wishing to use the logging facilities. Since the data rate of the trace logging stream is quite high, logging applications usually enable tracing for a few seconds before saving the massive amount of information to a file for later offline processing. As the system operates, the microkernel fills the trace data buffer. When a preset buffer threshold is reached, the kernel calls a handler function pre-configured via the trace logging application. The application would then consume trace data, freeing up the kernel trace buffers to free up room for more trace data. The kernel buffers themselves are pre-allocated and set up by the trace logging application and installed in the kernel via the trace logging setup kernel calls. The trace data generated by the microkernel is highly configurable; a trace logging application can use the tracing APIs to set filters and exclude or include specific types of data. The different types of data are divided into classes and events. For example, one can enable the \textit{THREAD} class, which generates thread-related trace events such as which thread is currently running on which CPU, or the \textit{KERNEL\_CALL} class which generates trace events when any kernel call is made by any process in the system. There are two generic tracing modes, a wide mode and a fast mode. Wide mode generates more in-depth logging information, such as the full argument list for kernel calls. Fast mode provides only a subset of the full information in a very compact form. Wide mode generates more data, while fast mode generates data faster. A \textit{The size of a MsgSendv()} trace data is 8 bytes in fast mode and 24 bytes in wide mode.

\begin{figure}
	\centering
	\includegraphics[width=\textwidth]{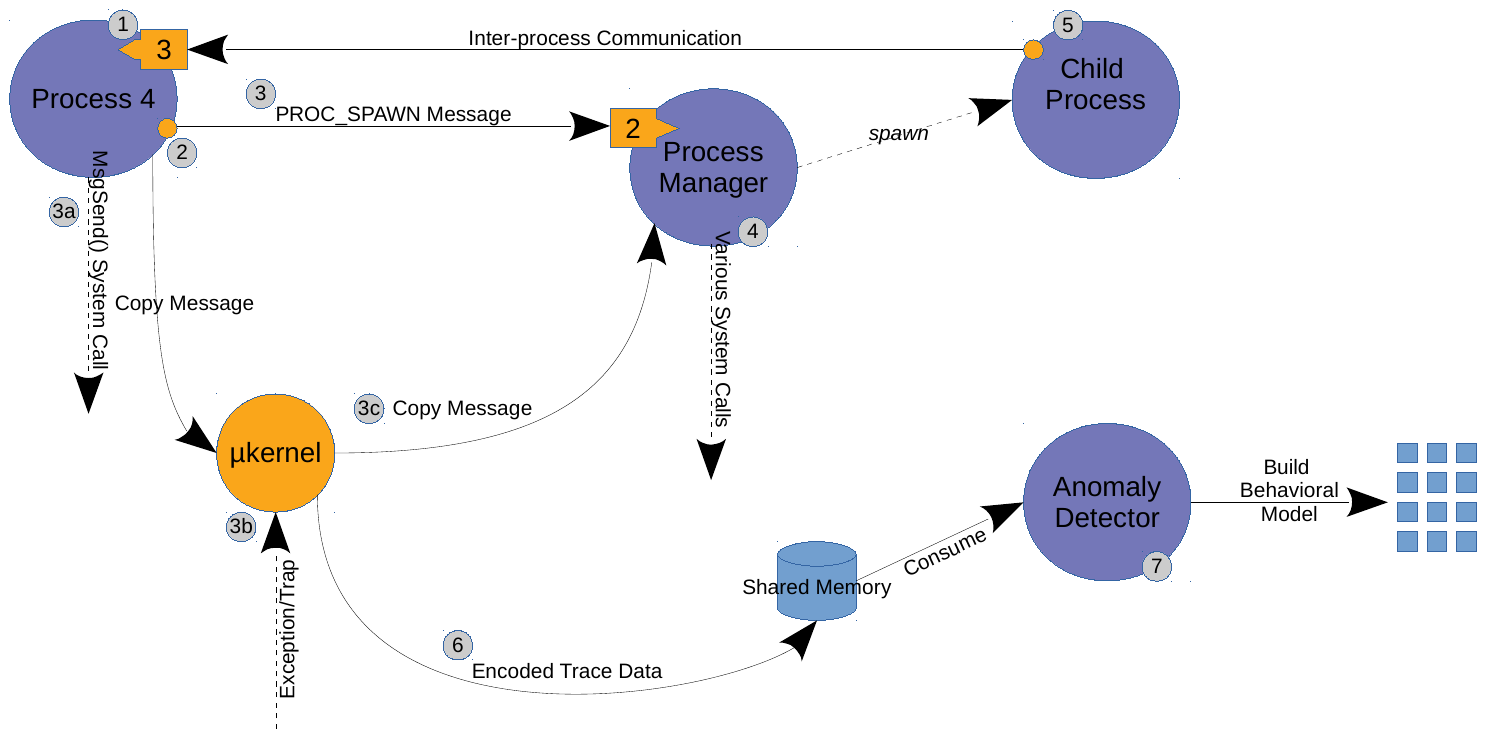}
	\caption{An overview of message-passing and trace data collection.}
	\label{fig:msg_overview}
\end{figure}

To piece the different components together, Figure \ref{fig:msg_overview} shows an overview of the operations involved when a process spawns a child. The numbered steps are described as follows:

\begin{itemize}
    \item \textbf{(1)} Process 4 is started. It will be acting as a server to receive communication via messages. It creates a channel via a system call and the microkernel assigns the channel ID 3 for Process 4.
    \item \textbf{(2)} Process 4 needs to spawn a child process. This is done via calling the kernel call spawn. \texttt{Spawn()} is implemented in libc as a message to the process manager. When process 4 was first initialized, a connection to the process manager communication channel was automatically created.
    \item \textbf{(3)} Process 4 (libc) calls \texttt{MsgSend()} with the appropriate payload. The destination is Process 1, the process manager and the payload include a message header (\texttt{PROC\_SPAWN}) and the message body which contains the rest of the spawn parameters.
    \item \textbf{(3a)} Since \texttt{MsgSend()} itself is a system call, it traps into the microkernel \textbf{(3b)} which copies the message from the source, Process 4, to the destination process manager \textbf{(3c)}.
    \item \textbf{(4)} The process manager creates a new process with the help of a few system calls into the microkernel and responds back to process 4 with the appropriate \texttt{spawn()} return value, unblocking it.
    \item \textbf{(5)} The child process happens to be a client of its parent, Process 4. It is now free to connect to process 4's known channel 3 in order to send requests. The same set of steps are repeated but this time, the destination process is not the process manager but rather Process 4.
    \item \textbf{(6)} Meanwhile, the microkernel tracing facilities are enabled and update a pre-setup shared memory area with the encoded trace data of the various operations described.
    \item \textbf{(7)} The anomaly detector, which includes a trace logging component, is sent an interrupt \footnote{a pseudo-interrupt to be accurate as it is not a result of a true interrupt generated by the interrupt controller} from the microkernel indicating that there is data available. The anomaly detector then consumes the data from the shared memory and updates its models.
\end{itemize}

Qnet, QNX's native networking protocol, enables message-passing across distributed QNX nodes on a network. This allows an entire network of trusted nodes to appear as one operating system running on distributed network of CPUs. One can send a message to spawn a process transparently on a remote node on a Qnet network. For this reason, any message being sent to a channel must include the Qnet node identifier. This identifier is 0 if the message is intended for the local node, otherwise the remote node ID is used.

QNX offers PPS, a persistent publish and subscribe utility. PPS enables writers to easily publish information and readers to subscribe to notifications from certain publishers. The published information is persistent across system reboots. tH uses PPS to publish different types of information that can be easily subscribed to from a command line. 

The QNX operating system has been chosen for our implementation for a few key technical reasons. First, QNX's microkernel architecture means that all services (such as drivers, filesystem, \textit{etc.}) are running outside the kernel as user-level processes. These processes mainly communicate via messages. The system is heavily dependent on message-passing so much so that most of the highly used system calls on UNIX-like systems are implemented via messages to system-level services that provide core OS functionality such as process and memory management. This means that regardless of the way applications are implemented, communication via message-passing is guaranteed to occur. Second, since drivers and other critical system processes are not part of a big monolithic kernel, they too can be monitored and profiled. This means that tH can profile and detect anomalies in critical operating system services. Lastly, the QNX kernel is a safety certified microkernel \footnote{ISO 26262 and IEC 61508 certified}; its unique position in the embedded safety-critical market makes QNX a suitable target for this type of anomaly-based fault detection work as we can test our solution with real safety-critical products and collect meaningful data results from real-life applications.

\subsection{Profiling Thread Behaviour}
Thread Homeostasis profiles threads and not executables. An executable has one profile that contains a single profile for every thread created by the process. This design decision was made after going through the first round of experimentation with profile-based models. In the first round of the detector's implementation, we collected trace data and built models for every process in the system. We ran experiments to determine the false-positive rates for the new detector and calculated the average normalization times. The details of the experiment are described in detail in the following chapter. The results were astonishingly satisfactory yet, not perfect. For some deterministic processes in the system, we expected false positive rates to be as low as 0\% and the time taken to learn the process behavior to be low. However, this was not the case. 

 A simple controlled test application was built to try and further analyze the problem. The test application included a server and client processes that communicate via messages. Both the amount of communicated messages and the variability of the message types were configurable. A series of test cases were built to verify the correctness of our implementation. All the stress tests passed excepts for one, the multi-threaded client test. It took almost twice as much time to learn the normal behavior of a client with 4 threads than it did with just one thread. Any variability introduced in the messages sent by any of the threads would greatly affect the results. To add to this, right after the client has normalized, anomalies start being detected without any change in behavior.
 
Even though, every thread in the process has a separate sequence buffer, the per-process model was being updated at non-deterministic times depending on which thread is currently executing. Essentially, thread messages were being interleaved. This caused the model to think it has observed a new sequence during training, when in fact, it did not. This resulted in a much higher normalization or learning time. Performance measures were being recorded and we have seen this effect in the amount of excessive profile thawing (frozen to unfrozen) as a result of the believed-to-be-new sequence. Also, given a multicore system, threads are running concurrently on different CPUs; the trace data is received and combined from all the different cores. There is no guarantee that they will be correctly ordered except if we read their timestamps. This has two issues: i) if two threads have the same timestamp, which one comes first? and ii) the message trace data comes in at a very high rate, stopping to parse and compare the time stamps and waiting long enough to make sure that no future event would be received with a timestamp earlier than the current one would result in a great performance hit. We have implemented such timestamp parsing mechanism and seen the performance penalty. The CPU usage rises from an average of 1.2\% to a whopping 15\%, an unacceptable performance hit.

 After further analysis, we determined that we had to build profiles for individual threads and not for the entire process. Hence the name \textbf{Thread Homeostasis}.
 
\subsection{A Thread Network Behavioral Model}
Given pH's promising experimental results and the nature of the  target implementation environment, we decided to further extend pH's modeling technique to include inter-process communication in addition to system calls. Since a system can be viewed as a network of interconnected processes and threads that communicate with the kernel as well as with one another, we hypothesised that including communication ID used for inter-process communication in the short sequences of calls would allow further insight into the behavior of a process and its threads. This is particularly true in environments that heavily rely on IPC to perform their functions. This new model would be a powerful indicator of the system's natural behavior. The process is no longer viewed as an isolated entity, but rather a node in a larger network. Given a per-thread behavioral model, it will be much harder for an anomalous user-space process to send messages to another process without going unnoticed. This should result in an anomaly at many different levels in the system:

\begin{enumerate}
	\item In the profile of the anomalous process, since it has never been seen sending such messages.
	\item In the profile of the receiving process since it never received such messages.
	\item Possibly in the profile of the receiving process as it sends a response to the anomalous process.
	\item Possibly in the profile of the anomalous process as it receives the response.
	\item Possibly in the profile of other processes: if as a result of the anomalous messages, the receiving process has performed an action that requires sending a message to yet another process or issued a system call.
\end{enumerate}

Thus, anomaly detection would be performed on the system as a whole and not on an isolated instance of a binary. A fault that causes a process to communicate with another in such a way that has never been seen during verification should be easily detected through our model. We believe that this holistic view of the system would allow errors to be observed at a much earlier stage before becoming faults. We will no longer refer to a sequence of system calls but rather the \textbf{sequences of messages} used for both process-to-process and process-to-kernel communication.

Typically, in multi-process scheduling-based operating systems, different processes communicate with one another. However, there is no mandate or rule that restricts system architects from building their entire safety-critical functionality using just one process along with a monolithic kernel that includes all the required drivers. By implementing tH on QNX, we guarantee a system in which a high level of inter-process connectivity and communication is built into its core design philosophy. 

\subsubsection{System call identification}
As discussed, system calls are either implemented via a trap into the kernel or via a message-send to the system-level services. In order to build the model of message-sequences, the first question we had to answer was: If we are building a two-dimensional table of look-ahead pairs of messages, how can we uniquely identify a message from a thread to another so that we use this identifier as an index in our table? Messages do not have a clear unique identifier that we can use. In a traditional monolithic kernel, system calls are traps into the kernel and are assigned unique numbers to identify them. Similarly, on QNX, trap-based system calls can be uniquely identified by their number. However, this is not the case for the message-based calls. A message payload consists of a message header or the command number followed by the command parameters. The same message header can be sent by any process to another or by a process to the kernel. As an example, \texttt{\_PROC\_SPAWN} is the message head that is sent to the process manager when calling \texttt{spawn()}; sending a message header with the same value as \texttt{\_PROC\_SPAWN} (numerical value 0x0010) to the serial driver would perform a completely different action. Message headers alone cannot uniquely identify a message. Rather, information about the source and destination must be included: the source and destination process and thread identifiers, pid and tid, the pid of the receiving process, the channel identifier unto which the message is received, the node identifier (node id) on a Qnet network and finally, the message identifier or the message head. These values combined uniquely identify a message.    
\par
\begin{figure}
	\centering
	\includegraphics[width=\textwidth]{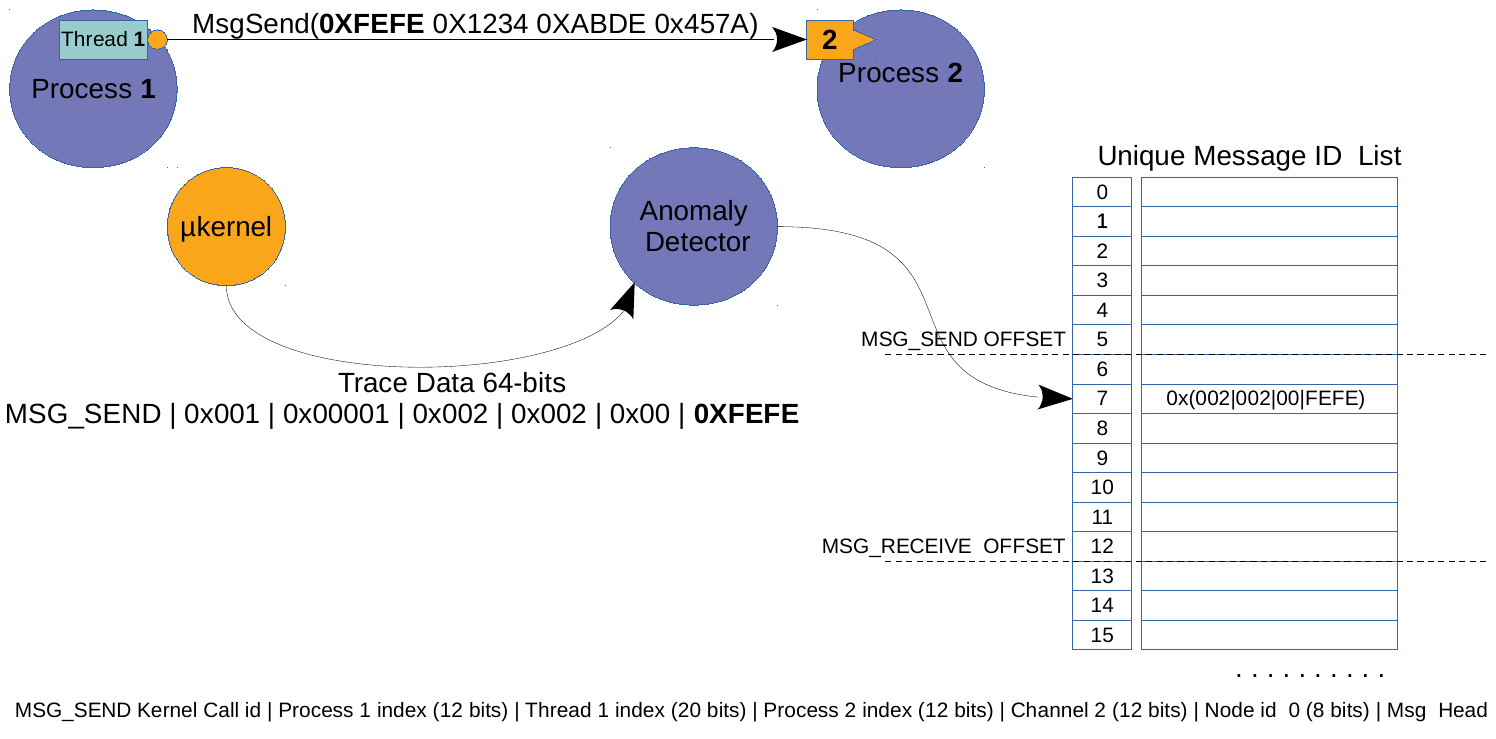}
	\caption{How a message is traced and stored in tH's data structures.}
	\label{fig:msg_hash}
\end{figure}
\begin{figure}
	\centering
	\includegraphics[width=\textwidth]{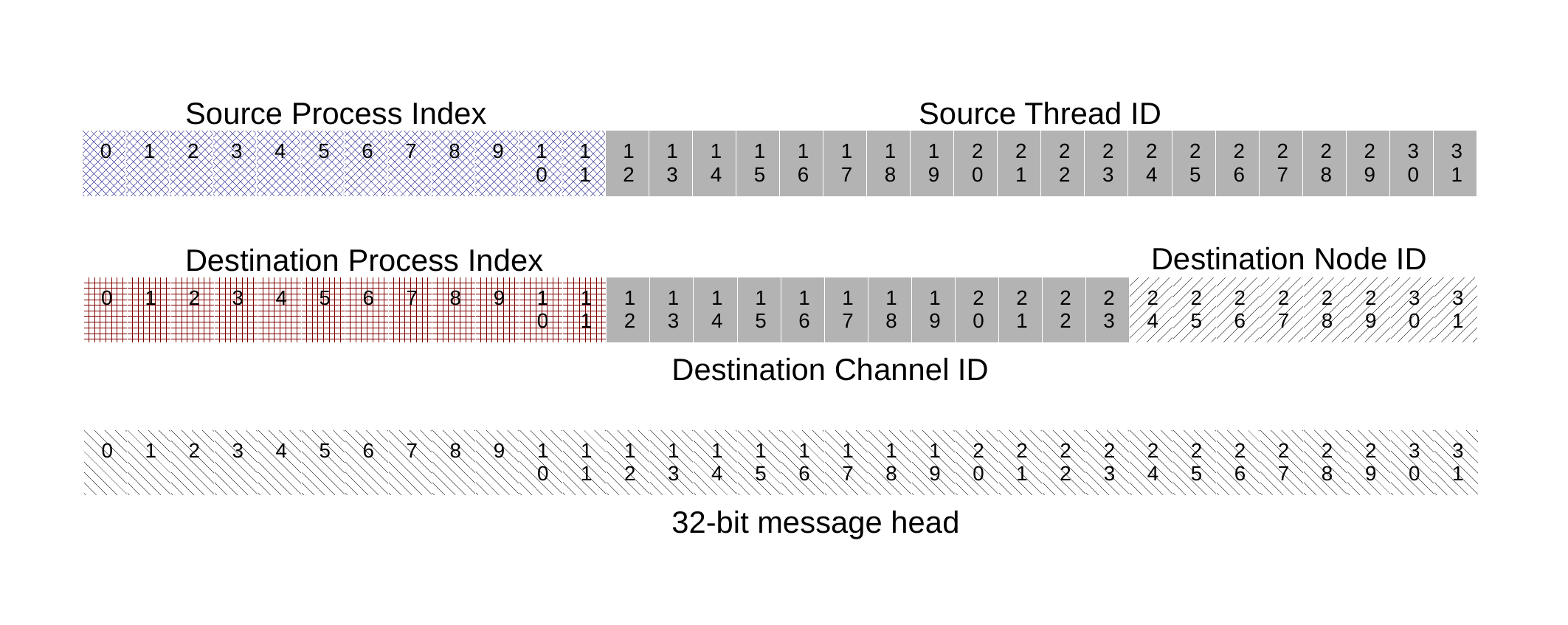}
	\caption{Kernel exit trace message bits.}
	\label{fig:bits1}
\end{figure}

We have modified the QNX microkernel to encode this data and send it to tH through the trace logging facilities every time a message is sent. Figure \ref{fig:bits1} shows the minimal 96-bits received by tH in addition to the kernel calls number. The source process index and thread ID are used by tH to identify the appropriate thread profile to update. The remaining 64-bits are composed of the 32-bit message head OR'd with the process index \footnote{The process index is the unique bottom 12-bits of a process and are all what's need to identify a process.} of the receiving process, the channel ID and the node ID to produce \texttt{\textbf{[PID TO (12-bits)| CHID (12-bits) | NID (8-bits) | MESSAGE HEAD (32-bits)]}}. Messages sent to a kernel service always have a destination process ID of 1, the process manager. This message identifying 64-bit value is then stored in an array. The array's index is then used to directly update the look-ahead pairs table. Thus, large 64-bit values are converted into zero-based array index values. In case of a trap-based system call, the microkernel only sends the 32-bit system call number along with the source process and thread IDs. Figure \ref{fig:msg_hash} shows an example data flow as a process sends a message to another until the unique message ID is stored in the message ID list.

\subsubsection{Interprocess Message Identification}
Messages sent from a process to another are uniquely identified using the same technique that the message-based kernel calls are identified. From the microkernel's point of view, there is no difference between sending a kernel-call message to the processes manager or sending a message to another process.

\subsection{Modifying the Instrumented Kernel}\label{mod}
Due to some performance challenges (discussed in section \ref{challenges}), the QNX microkernel was modified to tailor it to tH's specific needs. We modified the instrumented kernel to replace the current \texttt{KERNEL\_CALL\_EXIT} trace event \footnote{A trace event sent every time a kernel call concludes and exits the kernel} with our own. The new event contains the exact information that tH needs without requiring all the extra event types.  Recall that tH creates a single 64-bit value of the tuple (process ID, channel ID, node ID) of the receiving process OR'd with the message head (the first 32-bits of the message). Instead of emitting all this data from the kernel, we had the kernel create this 64-bit mask of the values we needed and include it in the trace so that tH could use it directly. The sending process ID was also replaced with the first 12 bits of the process ID; enough data for tH to directly index into its process list to uniquely identify a process. Similarly, for the thread and node IDs, only the bottom 12-bits were used. Thus, the process, thread, and node indexes (indexes versus IDs now) were OR'd together to generate one 32-bit field. The 32-bit field for the timestamp was not required (see section \ref{challenges}) and therefore was replaced with the process, thread, and node indexes. The final emitted information is shown in Figure \ref{fig:bits1}. This single \texttt{KERNEL\_CALL\_EXIT} trace event now contains the required information that tH needs in as little space as possible.

\subsection{Runtime Structure and Organization}
tH is divided into three main core components. The first is a core anomaly detection component that runs agnostic of the operating system, it maintains a generic view of the process and thread information, builds the models and detects anomalies. This core component is also responsible for saving and loading saving profiles to and from disk. The second component performs the QNX dependent functions, such as setting up trace logging with the kernel, parsing trace information and presenting it to the anomaly detector through a predefined API. The third component is the user interface component. For this, we used persistent-publish and subscribe (PPS) to present various pieces of information to the user.
\begin{figure}[ht]
	\centering
	\includegraphics[width=\textwidth]{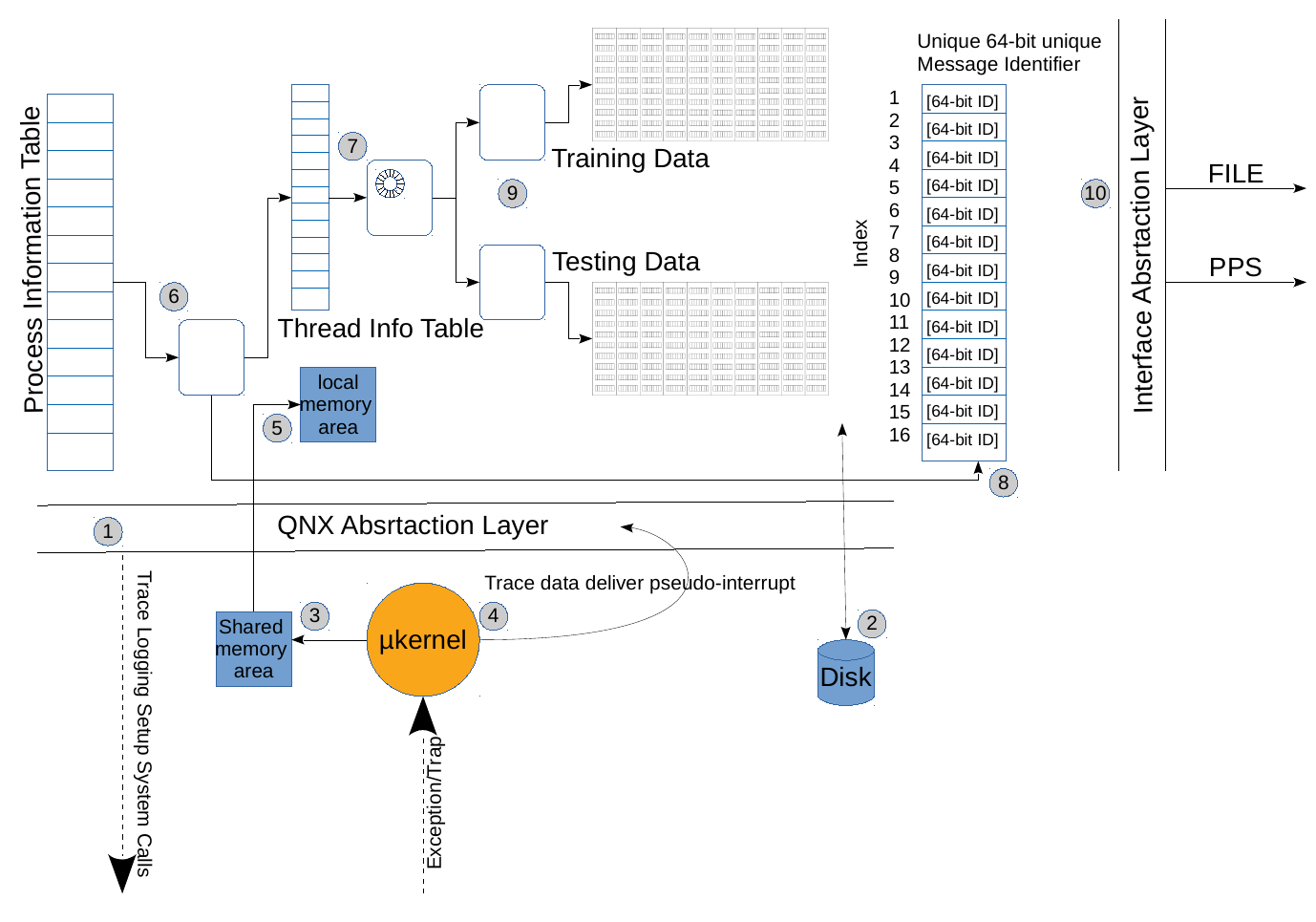}
	\caption{tH's components and data flow.}
	\label{fig:rs1}
\end{figure}

Figure \ref{fig:rs1} shows an enumerated outline of the main data structures used at runtime. The components and the data flowing into them are described as follows: 

\begin{enumerate}
    \item tH starts and sets up tracing with the microkernel. Using the trace logging API, tH installs a shared memory area that the kernel will be using as a buffer to store the trace data. tH also sets up the handler or pseudo-interrupt handler. This is used by the kernel to inform tH of the availability of the trace data. 
    \item tH reads the json configuration file and extracts runtime parameters. These include the size of the local memory area to copy the trace data into, the size of the sequence window, the time tH has to wait before a frozen profile is marked as normal and the location of the profile's folder on disk. If training or testing profiles exist on disk, tH reads and loads them into memory.
    \item Based on the pre-setup tracing filters (`kernel call exit' events) the microkernel writes the trace data into the pre-installed memory region.
    \item The kernel interrupts tH via the pseudo interrupt handler, notifying it that new data exists.
    \item tH copies the trace data into its local buffer, freeing up the shared memory area for further kernel traces.
    \item Using the source pid index in the trace data, the master process information table is accessed to retrieve the process structure. The trace data tells us that a specific thread, belonging to specific process has issued a kernel call. If this kernel call is a \texttt{MSG\_SEND} kernel call, it means that the process has sent a message and the trace data will contain the unique message identifier described above.
    \item The thread specific structure is retrieved along with the thread specific sequence window.
    \item The unique message identifier is stored in the message identifier list and its index is retrieved. If it already exists, its index is returned. The thread-specific sequence window is updated with the index.
    \item Depending on whether the thread is in the testing or training phase, the correct look-ahead pairs table is updated. Given that the unique 64-bit message ID is now translated to a zero-based index, this is a fast O(1) access.
    \item The PPS interface is updated accordingly. If the thread is still in training, the interface is updated to reflect statistical data, such as normal count, \texttt{last\_mod\_count} and profile status (frozen, thawed). If the thread has normalized and is in testing phase, the interface is updated to show if the trace contained any anomalies.
\end{enumerate}

Given the per-thread model, the different profile states, frozen, thawed and normal, happen at the thread level.

\subsection{Other Differences from pH} 
This section presents some other minor differences from the original pH implementation. These design decisions were made to better suit the embedded safety-critical nature of tH's use case.

\subsubsection{Dual Runtime Modes}\label{dualmodes}
To reduce false positives, pH has a unique algorithm to allow the addition of training data during testing (see Section \ref{tolerize}). If during testing, a certain sequence is seen frequently enough, the testing profile is unlocked, the new sequence added and training restarts for that process. We believe this is ideal for a general-purpose system, however, not for our use case. If tH is to be used on critical infrastructure, then it is safe to assume that the expectation would be to have the testing profiles generated during production then simply installed and used during deployment. From our experience in the field, critical systems that do not have deterministic behavior are not popular and allowing for tH to train in an operational system would greatly hamper its adoption. In addition, we make the assumption that safety-critical systems are deterministic in nature and that exhaustive verification and system tests will be performed for safety and certification purposes. This verification phase is when tH will be profiling and learning the system behavior. We do not expect any deviance in the learnt behavior. Any occurrence of anomalous behavior during testing time is only considered a failure of complete test coverage or actual anomalous behavior but not a false-positive.

Having said that, we allow the addition of anomalies to the profiles if the entire system is considered to be in the learning phase. Even if the thread profile has normalized, if the system is in learning mode, we add the ``anomaly" as a valid sequence to the profile. After having run the first round of experiments, we noticed that one of the processes, the sensor process, generated anomalies on the same training dataset. The sensor has multiple threads, and two of them raised anomaly alarms. The first found 36 anomalies in 1423946 calls and the other 14 anomalies in 4110899 calls. After further investigation, it became clear that another process, Parallax, was the main user of the sensor. Parallax had not yet normalized and thus, was exhibiting new behavior and interacting with the sensor in new ways. Meanwhile, the sensor had normalized and locked its profile thinking that these new interactions were anomalous. One might think that increasing the time-to-normal time might solve this problem, but it doesn't. The sensor needs to know that Parallax has normalized before it locks its profiles regardless of the time-to-normal time. There was an easy and straight forward solution to this problem: If we are in learning mode, any anomaly encountered would be regarded as normal and added as a legitimate sequence to the \textbf{testing} profile even if the thread profiles have normalized and locked themselves. Running tH in detection mode disables this feature and never adds new sequences. the entire tH process now has two runtime modes, learning mode and detection mode. Both Parallax and Sensor are described in detail in the next chapter.

\subsubsection{Monitoring Critical Processes only}
Even though this goes against our design philosophy, we profile and monitor the behavior of the system's thread network. tH can be configured to monitor specific processes only. This was useful to us when we ran the mock client/server application; we did not need to be bothered by any other system events and excluded all other processes in the system. This is also useful if a specific subset of processes that are deemed critical need to be monitored excluding other not-so-important ones.

\subsubsection{Configurable window size}
pH's original implementation used a sliding window size of 9. Luckily, this allowed the use of a single byte in the master sequencing table to represent an intersection of two message identifier entries in the table (current and previous). Each bit in the window indicated whether the previous system call has been seen before the current system call, during training. This is optimal in terms of storage space, since a byte is the least amount of space that could be used. In terms of run-time performance, the worst case scenario is shifting at most eight bits to flip the last bit.
 
tH allows the efficient use of a variable configurable size window. During initialization, a global small table is built that allows an O(1) lookup of the physical bit location within the n-sized window, given the current and previous message IDs. The physical bit location can then be used to flip the required bit in the master sequencing table. This configurable window size would allow us to easily experiment with different window sizes in the future. For now, the window size is 8. 

\subsubsection{System Call Delays}
As discussed earlier, pH reacts to anomalies by delaying the corresponding anomalous system calls. The concept behind a delay is to give a system administrator enough time to determine the validity of such a sequence, approving or rejecting the application behavior. Reacting to anomalies on a safety-critical system is a huge undertaking that can have serious consequences. We believe that reaction to anomalies must be done on a case-by-case basis by those who have intimate knowledge of the system. A generic reaction that fits all use cases in such sensitive environments is not suitable. As such, tH does not react to anomalies but rather has a functional interface to notify those who subscribe to alarms from anomalous traces. The reaction is left up to them.

\subsection{Implementation Challenges} \label{challenges}
\subsubsection{Customizing Trace Logging for tH}
Initially, we used the original stock instrumented kernel. We instructed the microkernel through the trace logging API to send us trace data every time a kernel call was entered in the kernel. This allowed us to trace both trap-based system calls and any message sent via the \texttt{MsgSend()} kernel call. Trace logging was enabled in wide mode, since we needed the extra information associated with the events. Unfortunately, the `kernel call entered' trace event was not enough; it lacked other information we needed. We were forced to ask the kernel for more types of trace events in order for us to have the required information. This drastically increased the size of the trace data and negatively affected tH's performance. Some of the missing data included: 

\begin{enumerate}
    \item The channel ID of the receiving process. This is required for the unique message ID creation. The `kernel call entered' trace event only included the connection ID used by the sending process. This meant that we had to ask the kernel to send us trace events every time a process connects to a channel. This way we could maintain internal lists of connection ID to channels for cross reference. Now that we are tracing channel connection events, we had to trace connection detach events as well so that we could remove the channel ID from our internal lists.
    \item The issuing thread ID. The `kernel call entered' trace event only included the ID of the CPU the kernel call was running on. In order to obtain this information, the `thread running' event was enabled. This event is issued every time a thread becomes runnable on a CPU. An internal list of CPU IDs and current threads running on them was created for cross reference.
\end{enumerate}

Table \ref{tab:trace_events} shows a detailed description of all the trace events collected from the kernel. The trace data generated by the instrumented kernel is extensive and highly granular; its primary purpose is to debug a system over a brief period of time. Even if one could provide enough storage space to store all the generated trace data, parsing this information was meant to be performed offline and not in a live, online, and continuous fashion. tH has a hard requirement to run full-time and monitor all the processes in the system with a minimum overhead. Given all the extra information we were requesting from the kernel, tH's performance degraded and became CPU intensive \footnote{The instrumented kernel's overhead did not increase but rather tH's parsing of data in realtime}. At times, CPU usage rose to 20\% on particular quad core machines running a fairly quiet system. We started seeing warning messages indicating that trace events were being dropped because tH could not keep up with parsing this sheer amount of trace data while updating multiple cross-reference lists and behavior models. Something had to be done; after all, what good is an anomaly detector running on an embedded safety-critical system that consumes all the CPU resources?

This led to the initial decision to modify the QNX kernel in order to tailor the \texttt{KERNEL\_CALL\_ENTER} \footnote{An tracelogging event logged every time a system call is called} trace event to tH's specific needs. Each event now included:

\begin{itemize}
    \item A time stamp of when the kernel call exit occurred. This is a 32-bit field and is required for the receiver of the trace data to sort out of order trace events as we will see later.
    \item The process ID of the process that issued the kernel call. This is a 32-bit field.
    \item The thread ID of the thread that issued the kernel call. This is a 32-bit field.
    \item The kernel call number, part of a 32-bit header.
    \item If this is a message send kernel call then the following is also included:
    \begin{itemize}
        \item The process ID (32-bits) of the message receiving thread.
        \item The channel ID (32-bits) of the message receiving thread.
        \item The node ID (32-bits) of the message receiving thread.
        \item The first 32-bits of the message header being sent.
    \end{itemize}
\end{itemize} 

This brings the total trace data size to 228 bits in addition to a constant header size of 32 bits. The instrumented QNX kernel generates two types of trace data events; simple and combine event \cite{qnxsat}. A simple event is a trace data event that is at most 96 bits in size. These events are used when an entire event's data can fit in 96 bits or less. A combine event however is an event type that has a variable length. Combine events are used to represent events that need to supply more than 96 bits of data. The kernel creates multiple combine events and adds them to a queue ready to be consumed. However, due to performance constraints, the kernel does not wait for all the combine events belonging to one trace event to be added to its internal trace queue holding up other kernel calls. The kernel services other kernel calls and adds their trace data to the queue before going back to enqueue the original combine events. Thus, the combine events of different traces are interleaved \footnote{See interleaved events in the System Analysis Toolkit Guide \cite{qnxsat}}. 

As a result, tH would then have to recognize that the trace data is a combined event (part of the trace data header) and use the time stamp to re-order events in order to get the correct view of the trace data. This re-ordering of combine events is time consuming and extremely CPU intensive. Looking at the required data, we realized that we could indeed compress our data into 96 bits (from an original 128 bits) and turn the multiple combine events emitted on every \texttt{KERNEL\_CALL\_ENTER} to a single simple event. This eliminated the need to have a  32-bit timestamp field as we are guaranteeing that the newly added kernel trace event will never generate a combine event. As a result, we were able to reuse the time stamp field and optimize the trace logging message sent to tH as described in section \ref{mod} to a single simple event, 96-bits in size. 40\% of the code in tH was eliminated due to this kernel change. tH's CPU usage dropped from 20\% to an average of 1.2\% CPU on a heavily utilized system.

\subsubsection{Kernel Call Restarts}

Having incorporated the above performance enhancement, we were ready to run a real-life test on one of the services used in production, the Sensor service described in the next chapter. The test ran for over three days without successfully normalizing and freezing the profile. We went back to our test harness and analysed the trace data generated for our proximity client that sends two different types of messages in a tight loop, message header number 1024 followed by 1025. The expectation was to see this sequence in the trace data:      

\begin{lstlisting}
MSG_SEND:1024,MSG_SEND:1025,MSG_SEND:1024,MSG_SEND:1025...
\end{lstlisting}

Except the following sequence is what was actually observed: 
\begin{lstlisting}
 MSG_SEND:1024, MSG_SEND:1024, MSG_SEND:1024, MSG_SEND:1025,
 MSG_SEND:1025, MSG_SEND:1024, MSG_SEND:1025, MSG_SEND:1025 ...
\end{lstlisting}
 
When tH receives the \texttt{MSG\_SEND} trace from the kernel, it indicates that the process in question has entered the kernel via the \texttt{MsgSend} kernel call and that the kernel is about to start processing its request but the process has not yet exited the kernel. The QNX kernel was further instrumented using the existing kernel tracing APIs to determine when a kernel call is done and exits the kernel back to the user process. Here is the result:                                                                                         
\begin{lstlisting}
 MSG_SEND:1024, MSG_SEND:1024, MSG_SEND:1024, MSG_SEND_EXIT: 1024,
 MSG_SEND:1025, MSG_SEND:1025, MSG_SEND_EXIT: 1025,
 MSG_SEND:1024, MSG_SEND_EXIT:1024, MSG_SEND:1025,
 MSG_SEND:1025, MSG_SEND_EXIT:1025 ...
\end{lstlisting}

 Using the mock client that sends two types of messages, the sequence shows that indeed, the QNX kernel was behaving correctly and as expected. We were seeing the same kernel calls in succession 4.3\% of the time (69043 out of 1572634 calls). However we have a major philosophical bug in our implementation of the anomaly detector: kernel-call restarts. 

Any thread wishing to have the microkernel perform an operation on its behalf issues a system call. Only one CPU is allowed to be in the kernel at a time. Usually, system calls are very short in duration to reduce the oveall latency of the system. However, a situation may arise that a low priority thread is in the kernel while a higher priority thread is waiting to get into the kernel. To reduce the latency for the more important system operations, the microkernel is fully preemptible. This means that the higher priority thread will kick the lower priority one out of the kernel and take its place. All the operations that the lower priority thread has completed in the kernel will be unwound. The next time this lower priority task runs, the preempted system call will be restarted from the beginning. Having said that, there might be a point in the system call's lifetime where the kernel would need to manipulate certain shared data structures. If this is the case, the kernel locks itself. This blocks preemption by a higher priority thread and allows the kernel to safely modify common variables without the risk of yielding and leaving them in an inconsistent state. If needed, the kernel could potentially disable interrupts to block an interrupt service routine (ISR) from interrupting it. The kernel is usually locked for a very small amount of time to minimize interrupt and higher priority thread latency.

Unfortunately, given the kernel call restarts, we had no hope in having deterministic behaviour that we could extract a normal pattern from. The kernel call restarts were happening 4.3\% of the time; their occurrence was unpredictable depending on timing, interrupts and scheduling priorities of other threads in the system. A change had to be made. 

Instead of asking the kernel to send us trace data whenever a kernel call starts, we flipped things around. We asked the kernel to send us trace data when the kernel call is exiting. This guarantees the elimination of the kernel call restart problem and provides determinism. This meant that whenever we receive a \texttt{KERNEL\_CALL\_ENTER} trace event, we would then have to wait for the \texttt{KERNEL\_CALL\_EXIT} in the trace data since we need the information from both the \texttt{KERNEL\_CALL\_ENTER} and the \texttt{KERNEL\_CALL\_EXIT} events. Unfortunately, asking the kernel to send us \texttt{KERNEL\_CALL\_EXIT} trace events in addition to all the \texttt{KERNEL\_CALL\_ENTER} trace events only doubled the size of the trace data \footnote{One \texttt{KERNEL\_CALL\_EXIT} event for every successful non-preempted \texttt{KERNEL\_CALL\_ENTER} event.}, requiring more CPU usage. In addition, more complex tH code had to be added to synchronize between the kernel call starts and the kernel call exit trace events so that the former could be eliminated appropriately. This change was more error prone.

Eventually, the decision was made to add a new \texttt{KERNEL\_CALL\_EXIT} trace event that included all information that our modified \texttt{KERNEL\_CALL\_ENTER} event included. Since there is no need for the \texttt{KERNEL\_CALL\_ENTER} events, they were disabled. Now tH is receiving a compact stream of \texttt{KERNEL\_CALL\_EXIT} events containing all the information it needs as shown in Figure \ref{fig:bits1} and described in section \ref{mod}.  This eliminated the need for most of the other trace data previously required and more than one third of the code in tH required to build its internal data structures. \texttt{\_NTO\_TRACE\_KERCALLENTER} was replaced with a \texttt{\_NTO\_TRACE\_KERCALLEXIT}, \texttt{\_KER\_CONNECT\_ATTACH}, \texttt{\_KER\_CONNECT\_DETACH}  and \texttt{\_NTO\_TRACE\_THRUNNING} were no longer required. tH is now much leaner (40\% of the code was eliminated), less computationally intensive (an average of 1.2\% versus 20\%), and more powerful as our results will show.

\subsection{The 32-bit versus 16-bit Message Header Dilemma}\label{32-2}
Figure \ref{fig:bits1} shows the three 32-bit values sent in the trace data when the \texttt{MsgSend} kernel call is invoked by a process. The last value is the first 32-bits of the message payload. Our intention is to differentiate between them for profiling. All messages to the QNX kernel (\textit{procnto}) use a 16-bit message type in their header, and the rest of the message contains the data, including the second 16-bits of the message head. Our decision to include the additional 16-bits of message data was made based on the assumption that usually, specialized safety-critical software does not have too much variance in their messages. Too much variance means that we will not be able to easily profile the system. Using our test harness, we ran some field evaluations using the 32-bit value and determined that the average number of different 32-bit message types is around ten. Table \ref{tab:32} shows the number of different 32-bit message headers for some of the system processes we profiled. Some of these processes are quite complex such as Parallax, a process responsible for decision making in autonomous vehicles, devb-umass \cite{devb}, QNX's mass storage device driver, drm-intel, QNX's direct rendering manager for Intel Graphics \cite{drm}, Screen \cite{screen2}, QNX's graphics compositing windowing subsystem, and fs-nfs3 \cite{nfs3}, QNX's network file system driver. There doesn't seem to be a high variability in the different types of messages, therefore, using the entire 32-bits might be suitable. However, one test made us re-evaluate this decision: the QNX port of Google's Blink browser engine (BlinQ) \cite{chromium2}.

 \begin{table}[ht]
  \begin{center}
  	\begin{tabular}{|l|l|l|}
  		\hline
  		Process Name & \# Distinct Kernel calls & Total kernel calls\\
  		\hline
 launcher & 10 & 7630\\
 qterminal & 20 & 12615\\
 devb-umass & 16 & 13520\\
 io-hid & 1 & 1707578\\              
 slog2info & 6 & 88418\\
 parallax & 9 & 117872451\\
 drm-intel & 8 & 35086592\\
 screen & 11 & 7313114\\
 analogclock & 5 & 442815\\
 fs-nfs3 & 17 & 9372245\\
 slogger2 & 3 & 433\\ 
  		\hline
  	 	\end{tabular}
    \caption{Total number of distinct kernel calls of 32bit message headers}
    \label{tab:32-2}
  \end{center} 
  \end{table}	

Even though our work here targets specialized safety-critical software with a very small and limited functionality, pushing the limits during this design phase helps us gain further insight into the data we are attempting to model.

Table \ref{tab:blink} shows the trace data collected from the web browser for approximately 10 hours. The browser was left running an active HTML5 openGL test. The first column shows the destination process, followed by the 16-bit message header and the 32-bit header. The last row shows their total count. In total, Blinq sent 61688997 messages to 11 different processes in the system. There were 52 distinct 16-bit message types and 85 distinct 32-bit message types. The maximum ratio between them is 1:9 (1 16-bit header has 9 different 16-bit pays loads). As an example, in the process with index 1, the process manager is sent one message of a single 16-bit header, 0x0100, however if we include the entire 32-bit header with the extra 16 bits of the data payload we'd find five extra distinct types: 0x0001, 0x0000, 0x0003, 0x0005, and 0x0007. 
    
Although the results in Table \ref{tab:32-2} are not conclusive, we still opted for reducing the message header to 16-bits for the first round of experiments to avoid non-deterministic behavior resulting from the greatly varying payloads. The entire 32-bit payload and perhaps even more will be part of our experiments in the future.

\subsubsection{A Summary of the Technical Contributions}
The following list summarizes the technical contributions of this work:
\begin{enumerate}
    \item Thread Homeostasis, tH, is an anomaly detector based on pH's system call profiling concept was created for the QNX operating system.
    \item tH profiles running system threads.
    \item In addition to system calls, tH builds models from sequences of message identifiers used in interprocess communication.
    \item A novel approach to uniquely identify an IPC message was created.
    \item The QNX microkernel was modified to customize the trace data generated. This allowed tH to run full-time with low CPU usage overhead while both collecting and parsing a continuous stream of compact trace data in real-time with low latency.
\end{enumerate}

\subsection{Usage}
\subsubsection{Starting up the anomaly detector}
Ideally, tH should be started on system boot right after the kernel and before any other programs have started. This enables tH to capture all system calls made by any process right from its inception. However, tH can be started at any point in time, either before or after the process(es) being monitored have started.

\subsubsection{Configuration file}
A configuration file is required as a command-line argument for tH. JSON was chosen to represent the configuration file since it is trivial to extend and add support for future configuration options.

The list below shows a sample configuration file. The configuration file structure and description are as follows:
\begin{itemize}
\item{\textbf{buf\_size}}: Allows configuring the kernel tracing buffer size. Not Implemented.
\item{\textbf{win\_size}}: The size of the sliding window. The current default is 8 and the maximum is 32. This option has been added so that we could run future experiments with different window sizes.
\item{\textbf{mon\_list}}: A comma-separated list of processes to monitor. If this list is empty, all processes are monitored by default, otherwise, the monitoring is only limited to this list. A few sub-options identify a process:
\begin{itemize}
    \item{The sub-field \textbf{id} is the process identifier}
    \item{\textbf{type} specifies whether \textbf{id} contains a process name or a process ID (currently only process name is supported).}
\end{itemize}
\item{\textbf{exc\_list}}: This is a comma-separated list of processes to exclude from monitoring. Not Implemented.
\item{\textbf{prof\_path}}: The full path to profile data. Existing profiles will be read from there and new profiles will be created at this path.
\item{\textbf{notify}}: Will be used to specify a specific process to notify when anomalies occur. Not Implemented.
\item{\textbf{normal\_wait}}: This is the time (in seconds) that tH will keep the profiles frozen before it deems it normal.
\end{itemize}

\begin{lstlisting}[language=xml, caption={Configuration file example.}, captionpos=b,  basicstyle=\tiny]
{
    "buf_size": 64,
    "win_size": 8,
    "mon_list": [
        {
            "id": "proc/boot/io-bluetooth",
            "type": 2,
            "desc": "bluetooth driver",
            "win_size": 8,
            "notify": 1
        },
        {
            "id": "proc/boot/btman",
            "type": 2,
            "desc": "bluetooth manager",
            "win_size": 8,
            "notify": 1
        }
    ],
    "exc_list": [
    ],
    "prof_path": "/home/myqnx7/tH_rootdir",
    "notify": 1,
    "normal_wait": 180
}
\end{lstlisting}

\subsubsection{Status Information}
As mentioned earlier, QNX persistent publish and subscribe (PPS) is used as the main interface for tH. Figure \ref{fig:pps_dir} shows the PPS interface director listing. There are two different types of information that tH presents, a general overview of the system using the \textbf{status} object and specific thread information using the \textbf{per-process} object. The \textbf{status} object contains a summary and general status information and thread specific information that is represented by objects with matching process IDs, concatenated with the thread ID names. The listing below shows a part of the output from the status object as read from command line:

\begin{figure}[ht]
	\centering
	\includegraphics[width=\textwidth]{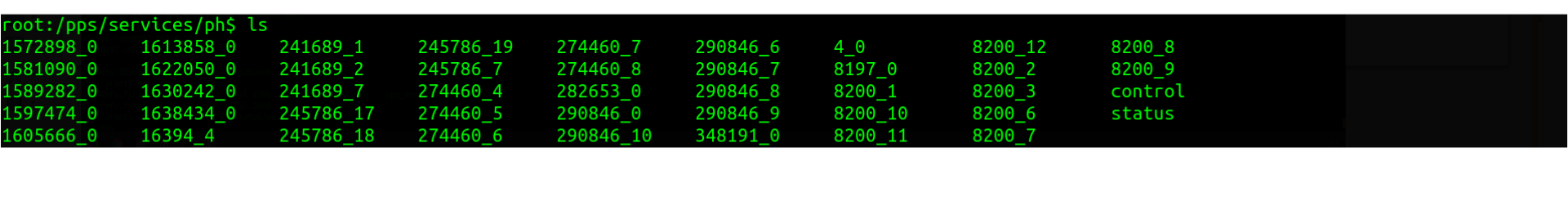}
	\caption{PPS interface directory list.}
	\label{fig:pps_dir}
\end{figure}

\begin{lstlisting}[language=xml, caption={PPS process information example.}, captionpos=b,  basicstyle=\tiny]
@status
anom_count:n:0
pid_241689::DETECTING:NORMAL:0
pid_241689_tid_1::DETECTING:NORMAL:0
pid_241689_tid_2::DETECTING:NORMAL:0
pid_241689_tid_7::DETECTING:NORMAL:0
pid_274460::LEARNING:THAWED
pid_274460_tid_4::DETECTING:NORMAL:0
pid_274460_tid_5::DETECTING:NORMAL:0
pid_274460_tid_6::LEARNING:THAWED
pid_274460_tid_7::DETECTING:NORMAL:0
pid_274460_tid_8::LEARNING:FROZEN
state::running
\end{lstlisting}

The entries under the status object are as follows:
\begin{itemize}
\item{\textbf{anom\_count}}: The total number of anomalies found across all threads and process so far.
\item{\textbf{pid\_\textit{number}}}: The status of process with ID \textit{number}. A process can be in several modes: normal, learning and frozen or learning and thawed. If the process is normal, this means that all its threads have normalized. If all the threads have frozen, the process's state is LEARNING:FROZEN. Otherwise, the process remains in the LEARNING:THAWED status.
\item{\textbf{pid\_\textit{number}\_\textit{tid}}}: Similar to the process ID information above but specific to thread ID.
\item{\textbf{state}}: tH's running status, running or none if tH is disabled.
\end{itemize}

\begin{lstlisting}[language=xml, caption={PPS process information example.}, captionpos=b,  basicstyle=\tiny]
anomalies:n:0
frozen:n:1
last_mod_count:n:1230
normal_count:n:1300
path::./test_client
sequences:n:24
state::NORMAL
time_to_normal:n:1000
train_count:n:1300
test_count:n:1000
tid:1
\end{lstlisting}

Under each process ID the following entries and their description are found:
\begin{itemize}
\item{\textbf{anomalies}}: The number of anomalies seen during testing.
\item{\textbf{frozen}}: 1 if the thread is currently frozen, 0 otherwise.
\item{\textbf{last\_mod\_count}}: The number of system calls that have occurred since last modification.
\item{\textbf{normal\_count}}: The number of normal calls seen while in training phase before the profile is locked.
\item{\textbf{path}}: The location of the process binary on the filesystem.
\item{\textbf{sequences}}: The number of sequences that have inserted (not the number of look-ahead pairs).
\item{\textbf{state}}: The state of the thread. Could be one of: normal, frozen or thawed.
\item{\textbf{time\_to\_normal}}: The amount of time taken for this thread to fully train, normalize and lock the profile.
\item{\textbf{train\_count}}: Total number of system calls seen during training.
\end{itemize}

\section{Field Evaluation}\label{evaluation_chapter}

After having verified the expected behaviour of tH using the test harness, we now evaluate it on real in-field software. The evaluation is split into two parts: the learning phase and the detection phase. Testing whether tH will be able to determine a behavioral pattern, freeze a profile and then normalize, is an important deciding factor on whether tH is practical for in-field use. If processes do not have deterministic behaviour, tH will not be able to normalize or will require a very long time until it thinks it has observed all possible combinations of messages and kernel calls. The argument here is that embedded software has enough deterministic behaviour that can be profiled by our detector. More importantly, a lack of deterministic behavior would mean that tH would detect anomalies in perfectly normal data. The purpose of this experiment is to evaluate the false-positive rate generated by tH while running perfectly normal software. 

Even though we run a few experiments to evaluate tH's ability to detect true positives, these experiments are in no way extensive or in-depth. The creation of an unbiased faulty dataset for evaluating an anomaly detector correctly is no trivial task. Due to timing constraints, we purposefully left this type of evaluation to a future work where a proper fault-injection framework will be used to inject faults into the running software for an attempt to introduce behavioral anomalies observable by tH.

\subsection{The Evaluation Data}
\label{evaluation}
Ideally, tH would be evaluated on a system with some level of safety-critical functionality, after all, most of the systems that use QNX are safety-critical systems. Originally, we intended to test tH on one of QNX's autonomous vehicles \cite{qnx_auto}: create a software load with our kernel change, build a system identical to the one on the vehicles including tH, upload it to the vehicle then train and test tH while driving around the city. However, we were bound by ethical and legal constraints due to the high risk level associated with such an experiment. Thankfully, we found the next best alternative, simulating driving around the city using real-life, high resolution sensor data.

\begin{figure}
	\centering
	\includegraphics[scale=0.55]{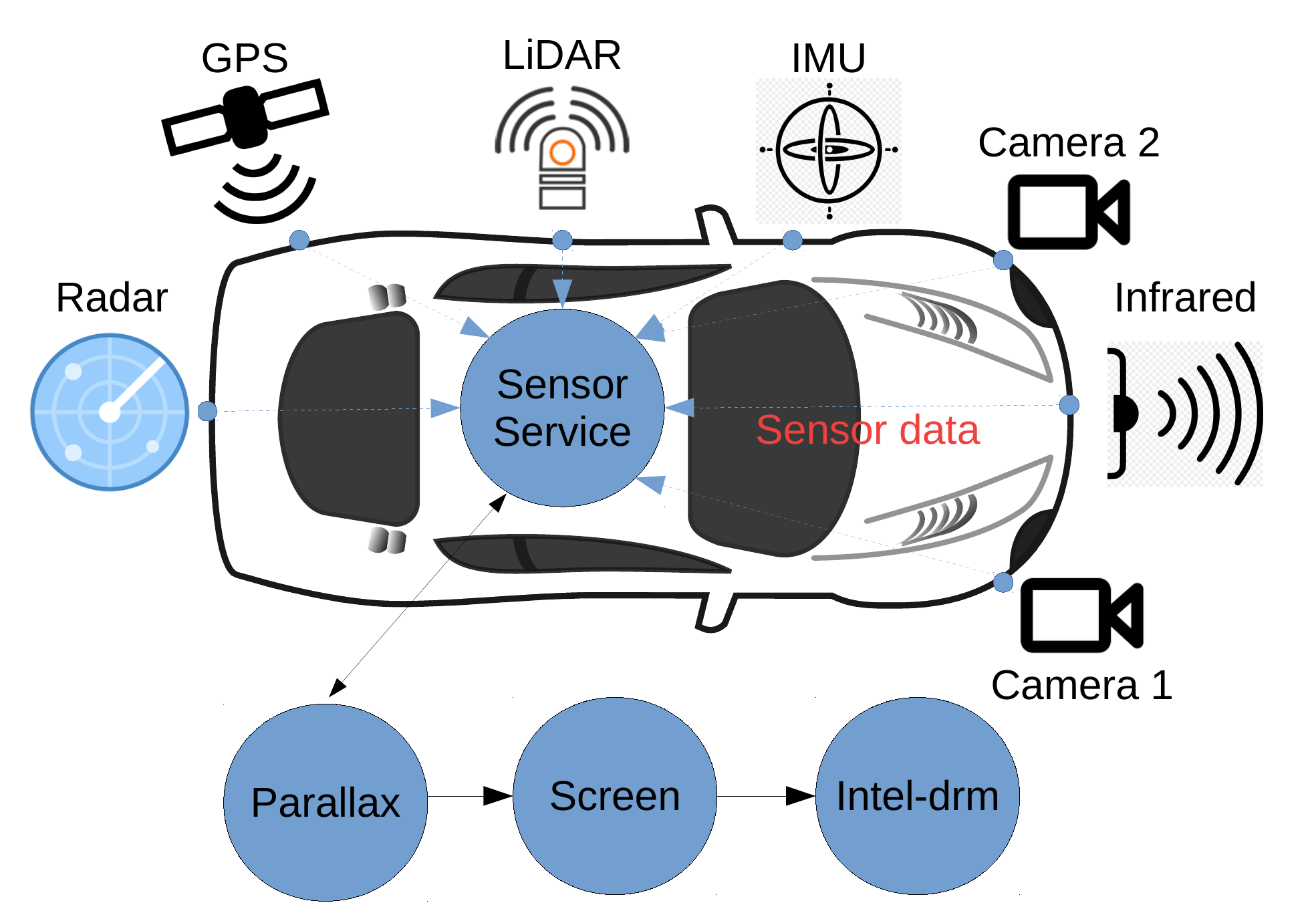}
	\caption{The various sensors and processes used to record and replay sensor data.}
	\label{fig:proximity}
\end{figure}

While driving the QNX demo autonomous vehicle \cite{qnx_auto} around various cities, the Blackberry QNX team recorded some of the data received from various sensors used by the vehicle. The sensors generating the high-resolution data included a GPS, an Inertial Measurement Unit (IMU), an Infrared (IR) sensor, a LiDAR, a Radar, and left and right cameras. The data was recorded for later offline analysis and simulations. Each recording session usually lasts for a few seconds, since the sensors generate a very high volume of data. This data can be replayed offline to the exact same software stack that is responsible for part of the autonomous driving, street object detection. From the software's perspective, the car believes it is driving over that same section of the map with the exact same sensor inputs even though it is sitting idle in the lab.

QNX's Advanced Driver Assistance Systems (ADAS) \cite{adas} represent a core software component for autonomous (and non-autonomous) vehicles running QNX. The Sensor service software is part of this package \cite{sensor2}. Sensor is used to capture data from various inputs, including multiple cameras around the vehicle, radar, LiDAR, IMU and GPS sensors and present it to other system components. In order to provide the ability to simulate real-world data of a self-driving car, the Sensor software was designed to sensor inputs from high-definition recordings instead of reading the live feed from a hardware sensor.

\begin{figure}[ht]
	\centering
	\includegraphics[scale=0.8]{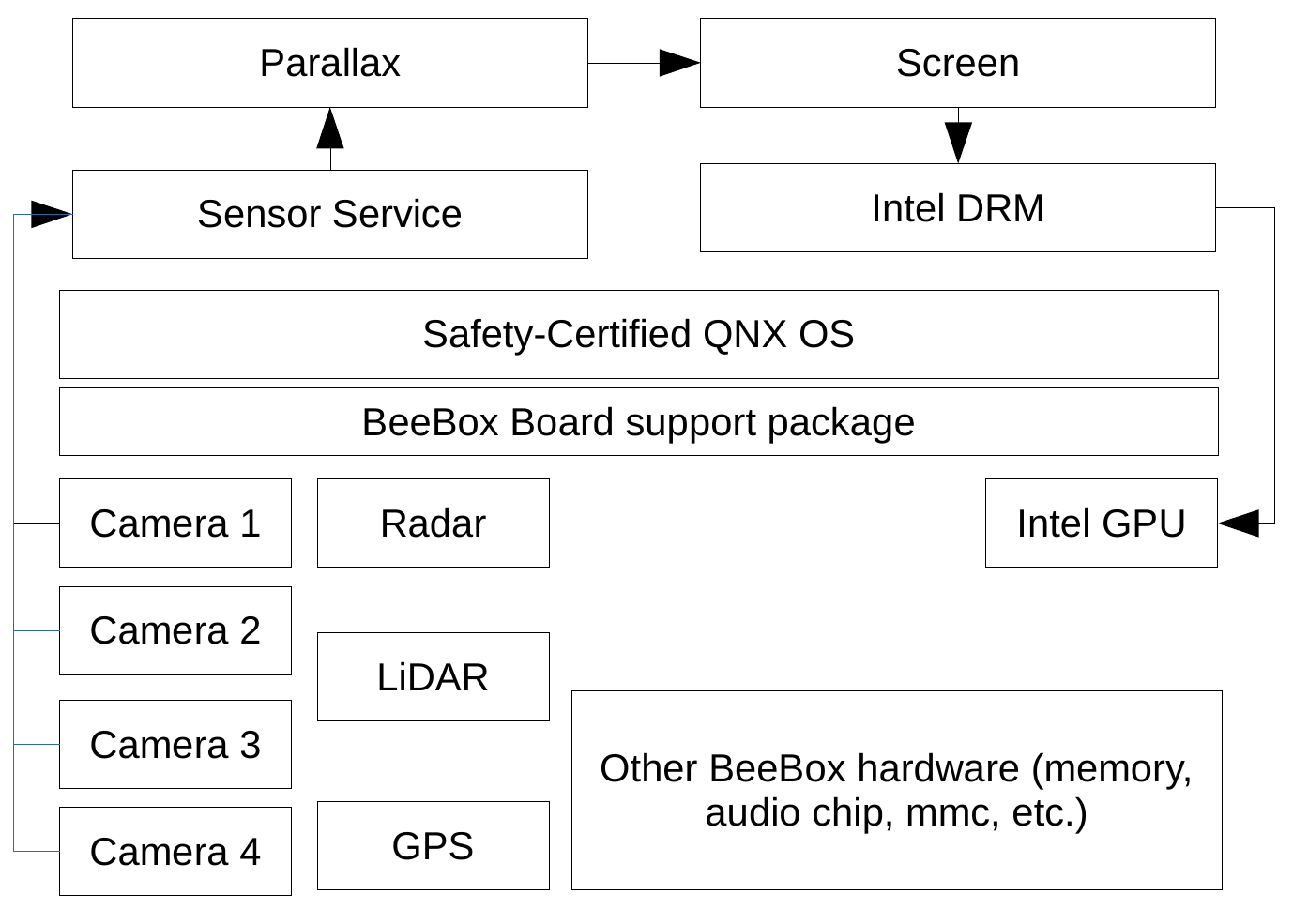}
	\caption{Parallax/Sensor System overview.}
	\label{fig:sensor_stack}
\end{figure}

\begin{figure}[ht]
	\centering
	\includegraphics[width=\textwidth]{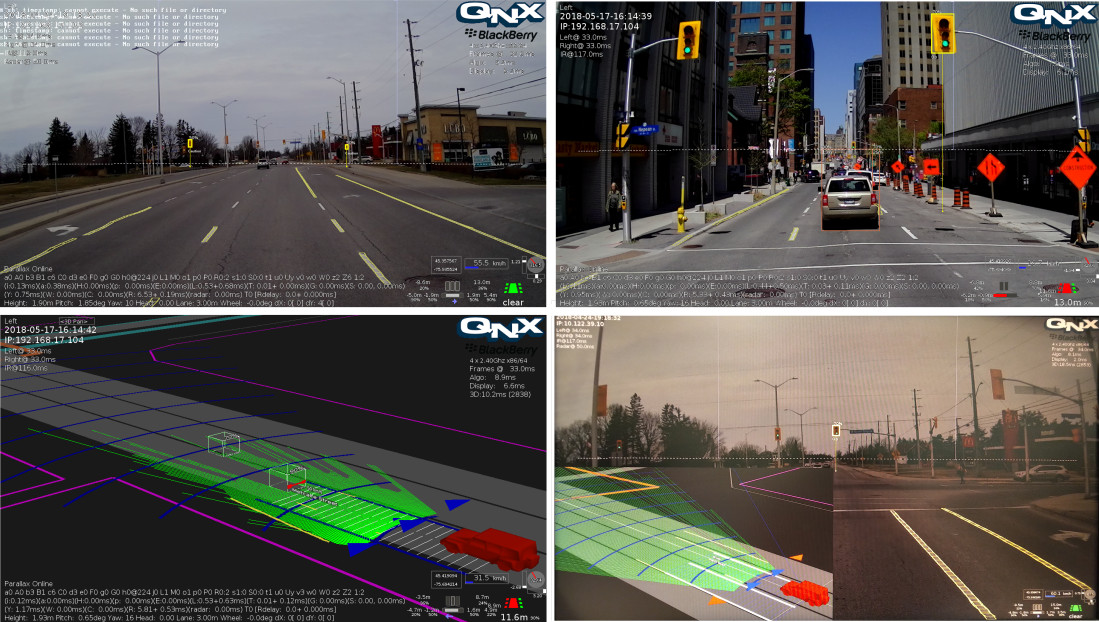}
	\caption{Parallax screenshots.}
	\label{fig:parallax_sc}
\end{figure}

Sitting on top of Sensor is another layer of software that would be making decisions based on the captured data streams from the multitude of sensors around the car. Fortunately, QNX has its own demo implementation of such decision making software, Parallax. Parallax is an internally developed, \footnote{Currently for internal use and demo purposes.}, highly complex piece of software used for analyzing images and various sensor data in order to detect objects, patterns, and pedestrians. QNX ADAS customers would usually have their own algorithms that enable autonomous vehicles to make the right decision. Figures \ref{fig:sensor_stack} and \ref{fig:proximity} show an overview of how all the components connect together. The Sensor services collects data from all the input sources. Parallax asks Sensor for the data, makes its calculations, then outputs the data to the screen using the Screen process \cite{screen2}. Since we run our tests on an Intel-based system, the screen uses Intel Direct Rendering Manager (DRM) server to utilize the GPU \cite{drm}. The sensor data is stored on a USB key and hence the use of the USB disk driver devb-umass \cite{devb}. All of this runs on the safety-certified QNX operating system and the board support package layer (system bootup layer). Figure \ref{fig:parallax_sc} shows screen shots of various Parallax. 

The software stack is quite sophisticated and CPU intensive. At the time of this experiment, Parallax contained 42,443 lines of code in addition to 10 shared libraries it links to. Parallax consumes an average of 64\% of the CPU. The Sensor service has 2,107 lines of code and on average uses 13\% of the CPU. Screen consumes 2\% of the CPU and has 57,683 lines of code in addition to the libraries it links to. Intel-drm has 10,464 lines of code and consumes 11\% of the CPU.

The recorded data is divided into multiple datasets. Each dataset consists of 7 files, one for each sensor recording. A manifest file is added to each dataset containing the physical location on the USB key of each of the files was well as the type of sensor the data represents (radar, Lidar, GPS, IR, IMU or light/right Camera). This manifest file is passed to the Sensor process as part of its initialization, enabling it to read the corresponding dataset files. The length of each dataset ranges from 5 to 15 seconds of recordings, with an average size of 3 gigabytes per dataset. Because the length of the recordings is very short due to the high volume of data generated, the recordings for each dataset is replayed in a continuous loop. The car ``thinks" it is driving over the same area over and over again. In total, we had 16 datasets.

\begin{table}[ht]
  \begin{center}
  	\begin{tabular}{|l|p{12cm}|}
  		\hline
  		\textbf{Process Name} & \textbf{Process Description} \\
  		\hline
  		devb-umass & The driver for a USB mass-storage interface \cite{devb}. All of the rerecorded sensor data were stored on a drive and hence the inclusion of this driver.\\
  		io-hid & Manager for human interface devices such as a touch screen or a mouse \cite{hid}\\
  	    slog2info & A QNX client to display messages from the system log \cite{slog2info2}.\\
  	    io-pkt & The network driver \cite{iopkt2}\\
  		\hline
  	\end{tabular}
    \caption{Other QNX system binaries involved in the evaluation and their description} 
    \label{tab:process description}
  \end{center} 
  \end{table}
  
\subsubsection{The Run}
The field tests were run on an Intel x86\_64 Beebox \cite{beebox2}. The Beebox used has a quad-core 2.08GHz Intel® Celeron® 64-bit processor with 2 MB L2 of Cache, 2GB of ram and a 32GB eMMC. The system has an Intel® HD Graphics for Intel® Celeron® Processor. In addition to the services mentioned above, Table \ref{tab:process description} shows all the other binaries that are not a core part of the test but where utilized during the test runs. 

After the system has booted up, a USB key containing all the sensor data that Parallax and Sensor will be streaming was mounted. The Screen and Intel-drm services were then started, followed by Sensor. To start streaming data and simulate driving, Parallax was started. Finally, tH was started over a remote SSH session.

\subsection{The Learning Phase}
During the learning phase, we attempt to teach tH the behavior of the software stack. Initially, we selected three datasets for training. The datasets were over 7.5 gigabytes in size for a total 28 seconds of driving. 

For the \textbf{first dataset}, the simulation was run in a continuous loop until the critical processes under test had normalized. tH was run with a time-to-normal value of 3,600 seconds, that is, profiles would remain frozen for 1 hour before they normalize if they haven't seen new sequences within that time period. The choice of this particular time-to-normal value was not arbitrary. In the previous version of tH, where we profiled processes as opposed to threads, we ran our experiments with different time-to-normal values (180s, 600s, 1200s, 3600s,10800s) over seven datasets for each value. Table \ref{tab:fp_data2} along with the graphs in Appendix \ref{fig:plot2} show the false positives generated during testing. We concluded that the time-to-normal values did not have any significant effect on the false-positive rate for our testing datasets. Even though the tests ran for a long time, they were, in reality, a few seconds of real intensive sensor data replays run in a continuous loop. After normalization was complete for the first dataset, the training profiles were saved to disk. 

 \begin{table}
  \begin{center}
  	\begin{tabular}{|l|l|l|l|l|l|}
  		\hline
  		\textbf{Process Name} & 180s & 600s & 1200s & 3600s & 10800s\\
  		\hline
\textbf{Parallax} & 5.97E-06\% & 81E-06\% & 2.18E-06\%	& 5.32E-06\% & 4.38E-06\%\\
Sensor & 7.87E-05\% &	3.54E-05\% & 	1.49E-04\% & 8.32E-05\% & 3.25E-05\%\\
Intel-drm & 2.47E-06\% & 2.75E-06\%	 & 2.99E-06\% & 4.93E-06\% & 1.93E-06\%\\
Screen & 1.68E-05\% & 7.66E-06\% & 1.98E-05\% & 1.94E-05\% & 2.15E-05\%\\
devb-umass & 1.55E-03\% & 6.71E-03\% & 7.23E-03\% & 7.91E-03\% & 7.54E-03\%\\
io-pkt-v6-hc & 3.48E-01\% & 2.11E+00\% & 1.83E+00\% & 0.00E+00\% & 0.00E+00\%\\
  		\hline
  	\end{tabular}
    \caption{\textbf{Average} false positives for different time-to-normal profiles for profile-based models}
    \label{tab:fp_data2}
  \end{center} 
  \end{table}

In order to augment \textbf{datasets 2 and 4} after the profiles have been normalized and locked, we ran tH in learning mode. As described in Section \ref{dualmodes}, this allows tH to add more data to the already locked profiles. After no new behavior has been detected for over an hour, tH was manually terminated and allowed to save its augmented profiles to disk. A tool was developed to parse through the binary profiles and produce human-readable results.

It is worth noting that many of the processes in the system did not normalize or were not profiled at all. This was due to the fact that they were inactive during our test, such as the PCI-server or the pipe process. Others were profiled but tH was terminated once our main processes under test normalized and did not get a chance to see enough kernel calls during their lifetime, such as the SSH daemon. Even though we SSH'd into the machine to run tH, there was hardly any traffic on the network to exercise the daemon and build a profile.

As we incrementally added more traing datasets, we could see more threads normalizing as threads that were inactive before have now become active. 

\subsubsection{Using Multiple Datasets For Training}\label{first}
After profiling the system behavior using the three datasets, we decided to evaluate our training methodology before putting tH through the real testing rounds and producing the final results. We ran testing on two datasets, using the augmented training profiles. Both tests ran for four hours each, 960 times longer than each loop iteration. Tables \ref{tab:dataset4} and \ref{tab:dataset5} show the results. For the first test set, Parallax, Sensor, Screen and Intel-drm, slog2info, io-hid and io-pkt showed no anomalies for the whole run. The Anomaly/Train Count column shows the total number of calls made during this period; all processes' threads have been highly active during the testing phase. However, the Sensor process has one out of six threads showing anomalies with a total of three anomalies in 9379052 calls. The USB disk driver, devb-umass, had three anomalous threads out of 11 active threads, with a rate of four anomalies in 56315359 calls. This comes to a total of 4/40 anomalous thread. The second dataset shows similar results: Parallax has two out of six anomalous threads at a rate of 0.00000166\%, Sensor has two anomalous threads out of five, at a rate of 0.000035306\% and the USB disk driver has 3/11 anomalous threads with a rate of 0.005429176\%. This was quite a high percentage. We came to the quick realization that training using three datasets was not enough: tH was not experiencing enough software behaviour and was generating a higher rate of false-positives that we have initially anticipated. 

\begin{table}
  \begin{center}
  	\begin{tabular}{|l|l|l|}
  		\hline
  		Process Name-Thread ID & Anomaly/Train Count (\%) & Total False Positives\\
  		\hline
  	    Parallax-0 & 0/7342025 (0\%)&  \\
  	    Parallax-6 & 0/416263 (0\%)&  \\
  	    Parallax-7 & 0/416222 (0\%)&  \\
  	    Parallax-8 & 0/416210 (0\%)&  \\
  	    Parallax-9 & 0/24646353 (0\%)&  \\
  	    Parallax-10 & 0/53123282 (0\%)& \\
  	    Parallax-11 & 0/0 (NA)& \\
  	    Parallax-12 & 0/0 (NA)& 0/86360355 \\
  		\hline
  	    Sensor-4       & 0/1681517 (0\%)&   \\
  	    Sensor-5       & 0/1685225 (0\%)&   \\
  	    Sensor-6       & 3/973325 (0\%)&   \\
  	    Sensor-7       & 0/2241408 (0\%)&   \\
  	    Sensor-8       & 0/2797577 (0\%)&   \\
  	    Sensor-9       & 0/2797577 (0\%)&   \\
  	    Sensor-10       & 0/0 (NA)&   \\
  	    Sensor-11       & 0/0 (NA)& 3/9379052 \\
  		\hline
  	    Screen-1       & 0/773724 (0\%)&   \\
  	    Screen-2       & 0/25059466 (0\%)&   \\
  	    Screen-7       & 0/276072 (0\%)& 0/26109262 \\
  		\hline
  	    Intel-drm-7    & 0/117363 (0\%)&   \\
  	    Intel-drm-14    & 0/25645754 (0\%)&   \\
  	    Intel-drm-15    & 0/0 (NA)&   \\
  	    Intel-drm-17    & 0/0 (NA)&   \\
  	    Intel-drm-18    & 0/28809443 (0\%)&   \\
  	    Intel-drm-19    & 0/22448047 (0\%)&   \\
  	    Intel-drm-20    & 0/26402958 (0\%)& 0/103423565 \\
  		\hline
  		devb-umass-1   & 0/55776515 (0\%)&   \\
  		devb-umass-2   & 1/795 (0.125\%)&   \\
  		devb-umass-3   & 0/533095 (0\%)&   \\
  		devb-umass-6   & 0/769 (0\%)&   \\
  		devb-umass-7   & 3/732 (0.409\%)&   \\
  		devb-umass-8   & 0/756 (0\%)&   \\
  		devb-umass-9   & 0/748 (0\%)&   \\
  		devb-umass-10   & 1/702 (0.142\%)&  \\
  		devb-umass-11   & 0/685 (0\%)&  \\
  		devb-umass-12   & 0/552 (0\%)&  \\
  		devb-umass-13   & 0/10 (0\%)& 4/56315359 \\
  	    \hline
  	    io-pkt-v6-hc & 0/39 (0\%) & 0  \\
  	    slog2info    & 0/28745 (0\%) & 0  \\
  	    io-hid       & 0/552127 (0\%) & 0  \\
  		\hline
  	\end{tabular}
    \caption{False positive results for testing dataset 4 - 2.9 gigabytes of sensor recordings}
    \label{tab:dataset4}
  \end{center} 
  \end{table}

\begin{table}
  \begin{center}
  	\begin{tabular}{|l|l|l|}
  		\hline
  		Process Name-Thread ID & Anomaly/Train Count (\%) & Total False Positives\\
  		\hline
  	    Parallax-0 & 1/6188982 (0.0000001\%)&  \\
  	    Parallax-6 & 0/350115 (0\%)&  \\
  	    Parallax-7 & 0/350117 (0\%)&  \\
  	    Parallax-8 & 0/350115 (0\%)&  \\
  	    Parallax-9 & 0/14424971 (0\%)&  \\
  	    Parallax-10 & 1/38584735 (0.00000002\%)& \\
  	    Parallax-11 & 0/0 (NA)& \\
  	    Parallax-12 & 0/0 (NA)& 1/60249035 \\
  		\hline
  	    Sensor-4       & 0/1681517 (0\%)&   \\
  	    Sensor-5       & 2/1505501 (0.000001\%)&   \\
  	    Sensor-6       & 3/898650 (0.00003\%)&   \\
  	    Sensor-7       & 0/1972340 (0\%)&   \\
  	    Sensor-8       & 0/2439137 (0\%)&   \\
  	    Sensor-9       & 0/0(NA)&   \\
  	    Sensor-10       & 0/0 (NA)&   \\
  	    Sensor-11       & 0/0 (NA)& 3/8497145 \\
  		\hline
  	    Screen-1       & 0/653087(0\%)&   \\
  	    Screen-2       & 0/21155044 (0\%)&   \\
  	    Screen-7       & 0/275981(0\%)& 0/22084112 \\
  		\hline
  	    Intel-drm-7    & 0/141064 (0\%)&   \\
  	    Intel-drm-14    & 0/0 (NA)&   \\
  	    Intel-drm-15    & 0/0 (NA) &   \\
  	    Intel-drm-17    & 0/28305912 (0\%)&   \\
  	    Intel-drm-18    & 0/14033855 (0\%)&   \\
  	    Intel-drm-19    & 0/22365322 (0\%)&   \\
  	    Intel-drm-20    & 0/24906751 (0\%)& 0/89752904 \\
  		\hline
  		devb-umass-1   & 0/36017(0\%)&   \\
  		devb-umass-2   & 1/690 (0.144\%)&   \\
  		devb-umass-3   & 0/13496 (0\%)&   \\
  		devb-umass-6   & 0/695 (0\%)&   \\
  		devb-umass-7   & 1/669 (0.149\%)&   \\
  		devb-umass-8   & 0/749 (0\%)&   \\
  		devb-umass-9   & 0/680 (0\%)&   \\
  		devb-umass-10   & 1/714 (0.14\%)&  \\
  		devb-umass-11   & 0/741 (0\%)&  \\
  		devb-umass-12   & 0/695 (0\%)&  \\
  		devb-umass-13   & 0/111 (0\%)& 3/55257 \\
  	    \hline
  	    io-pkt-v6-hc & 0/38 (0\%)& 0  \\
  	    slog2info    & 0/28745 (0\%)& 0  \\
  	    io-hid       & 0/551968 (0\%)& 0  \\
  		\hline
  	\end{tabular}
    \caption{False positive results for testing dataset 5 - 0.5 gigabyte of sensor recordings}
    \label{tab:dataset5}
  \end{center} 
  \end{table}
  
\subsubsection{Thread Pools}
After carefully investigating the source code for the anomalous processes, we concluded that their use of thread pools is the root cause of the anomalies we observed. Consider a thread pool consisting of three threads, de-queing a queue of tasks numbered 1 through 10. At the first run, thread 1 works at 100\% of the  time and performs all the work. If during training, threads 2 and 3 never performed any work or performed less work that usual, tH will build a profile for them that does not accurately reflect their possible behavior. To account for this variation, we need to stress the system and exercise all the threads so that we can learn all their possible behaviors. Given that the datasets we have for training have a short duration (15 seconds each on a loop), this wasn't made possible even when training over multiple datasets. The use of thread pools or thread pool-like behaving threads run in a non-deterministic behavior as a result of either thread schedule and/or the variation in the coming workload for each thread.
Since the short duration of the datasets was a limiting factor to experiencing the software behavior in full during training on a small number of datasets, an obvious solution was to modify the training methodology and add more datasets to the training. 

\subsubsection{Stress Testing the Disk Driver} \label{stress}
The USB disk driver, devb-umass showed the greatest number of anomalies during the initial testing phase. Almost every single test run showed a few anomalies in a devb-umass thread. It is highly dependent on multiple threads sharing the same work queues (thread pools). One way to exercise all the legal behavior of all its threads is to stress test the driver. Fortunately, QNX has a series of tools for such a task. The stress tests were run and the driver's profiles saved. From that point on, devb-umass showed \textbf{0 anomalies} in every single preliminary test run \footnote{Preliminary test runs were executed for one hour.}.

\subsubsection{The 50/50 Split}
After augmenting the training with one more dataset, we re-ran the test on the same dataset we showed in Table \ref{tab:dataset5}. The new testing results with the augmented training shows an improvement. The previously anomalous threads in the disk driver, as well as the Sensor and Parallax processes, have completely disappeared. However, new anomalies have been introduced in another Sensor thread and an Intel-drm thread, which was previously inactive, and became more active based on the work load and thread scheduling priorities. We concluded that indeed, adding more training data does improve the quality of the behavioral models and reduce the false positives detected.

As a result we split our datasets into two sets at random, a set for training and another for testing. Training was performed on datasets 1 through 7 and testing was performed on datasets 8 through 16. tH was run with all the training datasets and allowed to augment its locked profiles with new behavior. The learning was terminated manually after 12 hours of seeing no new behavior. The total size of the datasets was 4.8 gigabytes.

\subsubsection{Testing on the Training datasets - A Quick Sanity Check} \label{second}
After augmenting the training with the extra datasets we ran the testing on all datasets we had used for training for over two hours each. Initially, after learning on a dataset, running a test on the same training dataset would still yield, on average, one or two anomalous threads with an average of three anomalies each. We could see clearly that some threads remain inactive (0 anomalies/0 total message count) during training, then becoming active while testing the exact same dataset. As explained previously in section \ref{first}, by design, this is due to the thread pool-like behaviour of some threads. Different threads were being non-deterministically scheduled to handle incoming tasks and tH was not getting the chance to observe their behavior during training. Adding more training datasets, allowed thread pool participating threads to experience more work and thus allow tH to learn a more complete picture of their possible behavior. As the training datasets were incrementally added, we have seen the thread anomalies drop down to \textbf{0} anomalous threads for all seven training datasets over a two-hour testing period. This is not to claim that we have learnt the behavior of the thread pools in full, but rather that we have learnt enough, given our limited training datasets' sizes. We are now ready to start testing.

\subsubsection{tH Usage Statistics}
To collect CPU usage statistics,  we ran top and sampled tH's CPU usage. tH uses on average, 1.12\% of the CPU and 4.5 MBs of memory at runtime (stack and heap space). The save disk profiles had a total size of 17.7 KB. This is quite a small footprint for such a complex system.

\subsubsection{Process Normalization Results}\label{normalization}
Table \ref{tab:normalization} shows a summary of the number of threads that normalized per-process along with their average and maximum normalization times. There was an error in calculating the normalization times in two of the intel-drm's threads and their data was excluded. On average, all active threads in the main processes (Parallax, Sensor, Screen, Intel-drm) normalized in 1.8 hours. The thread with the highest normalization time took 13 hours, an outlier compared to the rest of the threads. Initially, some threads did not normalize during training on the first and main training dataset but normalized later while augmenting the training with the extra datasets as described in Section \ref{first}. In addition, some threads were not active during training, for example, Parallax's main thread or idle threads that are part of a thread pool and did not get a chance to execute. As a result, there is a lack of full thread normalization count (the Normal/Total Threads column). The per-thread process normalization times are shown in the Appendix \ref{tab:normalization_details}. The total number of threads in the processes under test is 81 (active and non-active) out of which 45 where active. 100\% of all active threads in the system normalized (45 threads). The data shows that Intel-drm sent 70,152 messages per second, the rate of messages sent during the training period, followed by Screen, Parallax, devb-umass and then Sensor in descending order of message rates.

\begin{table}
  \begin{center}
  	\begin{tabular}{|l|l|l|l|l|l|}
  		\hline
  		Process         & Normal (active)           & Avg. Time     & Max Time    & Avg. Msg    & Msg\\
  		Name &              /Total Threads          & to Normal     & to Normal   & Count       & /Second\\
  		\hline
  	    Parallax     & 8/11 & 7972  & 19190 & 357577012 &  18633 \\ 
  	    Sensor       & 7/10 & 9744 &  46482 & 37964068  & 3896 \\
  	    Screen       & 4/14 & 3724.5 & 4091 & 262041049 &  70356 \\
  	    Intel-drm    & 10/21 & 3474 & 5353 &  243709074 &  70152 \\
  		devb-umass   & 13/13 & 7587 & 23007 & 115235735 &  1189 \\
  	    \hline
  	    io-pkt-v6-hc & 1/5 & 70936 & 70936 &  19133 &  0.29 \\
  	    slog2info    & 1/1 & 3708 & 3708 &    762018 &  205 \\
  	    io-hid       & 1/6 & 3602 & 3602 &    14688381 & 4077 \\
  		\hline
  	\end{tabular}
    \caption{Normalization statistics}
    \label{tab:normalization}
  \end{center} 
  \end{table}

Table \ref{tab:normalization_perprofile} shows the normalization data per profile. As expected, the time taken to normalize with the per-profile method was much lower. The profile sees the data from all the different threads and combines them into one profile. The data rate for the entire process is much higher than its per-thread counterpart and thus tH freezes and normalizes much faster. As well, the compact model has more permutations from all the thread system calls and is less likely to detect that a sequence is anomalous (as we'll prove later). This decreases the likelihood of a process to thaw after it has been frozen.

It is important to realize that there is a single in-kernel trace data buffer, therefore the fact that we're monitoring the entire system greatly affects the time taken to normalize. What is important and should remain constant, is the number of events taken to normalize. We have implemented an additional feature that allows tH to instruct the kernel to monitor the Parallax process only. This way, there would be no other trace data generated except for Parallax's. Running tH with this configuration showed normalization times that are slightly less and on average have the same number of events generated before a process normalizes.

\begin{table}
  \begin{center}
  	\begin{tabular}{|l|l|}
  		\hline
  		Process & Avg. Normalization\\
  		Name    & in Seconds\\
  		\hline
  	    Parallax        & 31040\\
  	    Sensor          & 3603 \\
  	    Screen          & 3611\\
  	    Intel-drm       & 10922\\
  		devb-umass      & 55119\\
  	    io-pkt-v6-hc    & 27061\\
  	    slog2info       & 3619\\
  		io-hid          & 3602\\
  		\hline
  	\end{tabular}
    \caption{Normalization statistics for the per-process profiling method}
    \label{tab:normalization_perprofile}
  \end{center} 
  \end{table}

\subsection{The Testing Phase} \label{testing}

tH's ability to profile software behavior is highly dependent on whether it will think of normal data as anomalous or benign. The testing sets are high-resolution recordings of several sensor input data of the autonomous vehicle driving at different locations. The total size of the test data was 4.8 gigabytes for a maximum of 15 seconds of driving (a minimum of 5). All the tests ran in a continuous loop for exactly 12 hours. A script was developed to start and terminate the tests. Ideally, tH would have been trained well enough and would not find any mismatches in the system call sequences. Running the test for 12 hours each was not an arbitrary decision. We initially started running the tests for over 24 hours and noticed that in all the runs, we were not recording any new anomalies past the 3 hour run-time. In addition, it was not feasible to run all these tests for a longer period of time due to time constraints. Running the tests for longer shows less false positive rates as the number of messages keep increasing and the anomalies generated remain constant.

The tests were run remotely via a secure remote terminal (SSH) into the embedded system over the intranet. Hence the inclusion of the io-pkt, the network drive, in our results.

Table \ref{tab:alldatasets} shows the results per testing dataset. The first column in the table is a combination of the process name and thread ID, followed by the number of anomalies reported out of the total calls seen during testing. The last column shows the average false-positives for the entire process. Table \ref{tab:summary} shows the summary of the results. 

The total number of tested running threads is 45. All the threads in the core processes (Parallax, Sensor, and Screen), showed \textbf{zero} anomalies in all the test runs. The GPU driver, drm-Intel showed anomalies in two of the six datasets: both anomalies appeared in thread 21 (1/45 threads), 1/186,878,492 (0.000000535\%) and 2/195,985,082 (0.00000102 \%) anomalies. The USB flash driver showed anomalies in three of the six datasets: in two datasets, thread 4 (1/45 threads), 3/2315 (0.129589633\%) and 2/66i (3.03030303\%) anomalies and in the other dataset, threads 4 and 21 (2/45 threads) showed 1/71 (1.408450704\%) and 1/1444 (0.069252078\%) anomalies respectively. The other support processes such as io-okt, slog2info and io-hid showed no anomalies in all six datasets.
\newline
\par

\setlength\LTleft{-1.0in}
\setlength\LTright{-1in plus 1 fill}
  	\begin{longtable}{|l|l|l|l|l|l|l|}
  	
  		\hline
  		Process -  & Set & Set  & Set  & Set & Set & Set  \\
  		Thread ID & 1 & 2 & 3 & 4 &  5 &  6  \\
  		\hline
  	    Parallax-0 & 0/33812744     & 0/23949869      & 0/14236830    & 0/33812322    & 0/33281734    & 0/35395620 \\
  	    Parallax-6 & 0/0            & 0/0             & 0/0           & 0/1058102     & 0/1000216     & 0/1076343 \\
  	    Parallax-7 & 0/962748       & 0/1090750       & 0/1057923     & 0/967162      & 0/881667      & 0/940404 \\
  	    Parallax-8 & 0/854818       & 0/967274        & 0/898050      & 0/937632      & 0/848803      & 0/909797 \\
  	    Parallax-9 & 0/831194       & 0/931742        & 0/839093      & 0/22114515    & 0/26658432    & 0/31374106 \\
  	    Parallax-10 & 0/37921481    & 0/31912732      & 0/18644677    & 0/98737722    & 0/92637253    & 0/103117400 \\
  	    Parallax-11 & 0/86724565    & 0/97495376      & 0/74821427    & 0/0           & 0/0           & 0/0 \\
  	    Parallax-12 & 0/0           & 0/0             & 0/0           & 0/0           & 0/0           & 0/0 \\
  	    \textbf{Total \%}       & \textbf{0}\%    &\textbf{0}\%  &\textbf{0}\%   &\textbf{0}\%  &\textbf{0}\%   &\textbf{0}\%   \\
  		\hline
  	    Sensor-4   & 0/4752335 &    0/5037517 & 0/4865351 & 0/4812122               & 0/4762835     & 0/5024051\\      
  	    Sensor-5   & 0/4782273 &    0/5037517 & 0/4854295 & 0/4812122               & 0/4796520     & 0/5035120\\    
  	    Sensor-6   & 0/2930950 &    0/2929943 & 0/2932986 & 0/17843295              & 0/17612668    & 0/18533258\\    
  	    Sensor-7   & 0/17976442 &   0/6753514 & 0/1727998 & 0/15520688              & 0/15247503    & 0/16113234\\  
  	    Sensor-8   & 0/15554401 &   0/8433901 & 0/1728123 & 0/2312811               & 0/2326762     & 0/2370077\\  
  	    Sensor-9   & 0/2317209 &    0/3402894 & 0/3402435 & 0/0                     & 0/0           & 0/0\\
  	    Sensor-10  & 0/0 &          0/0       & 0/0       & 0/0                     & 0/0           & 0/0\\  
  	     \textbf{Total \%}      & \textbf{0}\%    &\textbf{0}\%  &\textbf{0}\%   &\textbf{0}\%  &\textbf{0}\%   &\textbf{0}\%   \\
  		\hline
  	    Screen-1   & 0/1162670585   & 0/950560598   & 0/525434677   & 0/1406358189  & 0/1258094712  & 0/914563944\\  
  	    Screen-2   & 0/57460340     & 0/65387103    & 0/57329720    & 0/63350821    & 0/59402457    & 0/64231233\\  
  	    Screen-6   & 0/834574       & 0/835745      & 0/835865      & 0/837104      & 0/836297      & 0/833634\\  
  	    Screen-7   & 0/0            & 0/0           & 0/0           & 0/0           & 0/0           & 0/0\\  
  	    \textbf{Total \%}       & \textbf{0}\%    &\textbf{0}\%  &\textbf{0}\%   &\textbf{0}\%  &\textbf{0}\%   &\textbf{0}\%   \\
  		\hline
  	    Intel-drm-6     & 0/489132      & 0/497661      & 0/479823  & 0/535294      & 0/513495 & 0/460392\\
  	    Intel-drm-7     & 0/0           & 0/0           & 0/0       & 0/0           & 0/0 & 0/0\\
  	    Intel-drm-14    & 0/0           & 0/0           & 0/0       & 0/0           & 0/0 & 0/0\\
  	    Intel-drm-15    & 0/0           & 0/0           & 0/13066559 & 0/0          & 0/0 & 0/0\\
  	    Intel-drm-16    & 0/33930946    & 0/35469330    & 0/29320667 & 0/34288888   & 0/31343971 & 0/37341598\\
  	    Intel-drm-17    & 0/31299497    & 0/36844188    & 0/31406731 & 0/34851092   & 0/33176156 & 0/38257197\\
  	    Intel-drm-18    & 0/32759761    & 0/34380182    & 0/32782650 & 0/35443673   & 0/31070878 & 0/30335954\\
  	    Intel-drm-19    & 0/31920048    & 0/35005735    & 0/30326565 & 0/33859799   & 0/32979838 & 0/35310895\\
  	    Intel-drm-20    & 0/29229022    & 0/35312620    & 0/30117836 & 0/33361612   & 0/32393358 & 0/33681805\\
  	    Intel-drm-21    & 1/27250086    & 0/33010963    & 2/28484251 & 0/33716209   & 0/32107750 & 0/31759378\\
  	    \textbf{Total \%}       & \textbf{0}\%    &\textbf{0}\%  &\textbf{0}\%   &\textbf{0}\%  &\textbf{0}\%   &\textbf{0}\%   \\
  		\hline
  		devb-umass-1    & 0/129514955   & 0/162852297   & 0/197252847   & 0/127576290      & 0/139912922    & 0/146890056\\
  		devb-umass-2    & 0/2320        & 0/2280        & 0/2174        & 0/2292           & 0/2246         & 0/4971\\
  		devb-umass-3    & 0/1179673     & 0/1518491     & 0/1097283     & 0/1109493        & 0/1618036      & 0/1300704\\
  		devb-umass-4    & 0/72          & 0/46          & 3/2315        & 0/50             & 2/66           & 1/71\\
  		devb-umass-6    & 0/2225        & 0/2312        & 0/2217        & 0/2366           & 0/2343         & 0/2333\\
  		devb-umass-7    & 0/2288        & 0/2297        & 0/2189        & 0/2304           & 0/2250         & 0/2361\\
  		devb-umass-8    & 0/2153        & 0/2303        & 0/2136        & 0/2268           & 0/2290         & 0/2259\\
  		devb-umass-9    & 0/2189        & 0/2087        & 0/2043        & 0/2178           & 0/2215         & 0/2221\\
  		devb-umass-10    & 0/2050       & 0/2193        & 0/1982        & 0/2144           & 0/2196         & 0/2134\\
  		devb-umass-11    & 0/1855       & 0/1989        & 0/1925        & 0/1902           & 0/1902         & 0/1918\\
  		devb-umass-12    & 0/1524       & 0/1274        & 0/1314        & 0/1420           & 0/1521         & 1/1444\\
  		devb-umass-13    & 0/475        & 0/180         & 0/590         & 0/345            & 0/203          & 0/149\\
  		devb-umass-14    & 0/51         & 0/9           & 0/31          & 0/52             & 0/21           & 0/9\\
  	    \textbf{Total \%}      & \textbf{0}\%    &\textbf{0}\%  &\textbf{0}\%   &\textbf{0}\%  &\textbf{0}\%   &\textbf{0}\%   \\
  	    \hline
  	    io-pkt-v6-hc  & 0/115           & 0/116         & 0/115 & 0/116             & 0/116 & 0/116\\
  	    slog2info     & 0/85998         & 0/86237       & 0/86048 & 0/86443         & 0/86255 & 0/86008\\
  	    io-hid        & 0/1669183       & 0/1671475     & 0/1671758 & 0/1674230     & 0/1672609 & 0/1667283\\
  		\hline
    \caption{False-positive results for all testing datasets}
    \label{tab:alldatasets}
  	\end{longtable}

\begin{table}[H]
  	\begin{tabular}{|l|c|c|c|c|c|c|}
  		\hline
        \multicolumn{7}{|c|}{Total False-Positives Per Process Per Dataset}\\
  		\hline
  		Process & 1 & 2 & 3 & 4 & 5 & 6 \\
  		\hline
  	    Parallax        & 0 & 0                 & 0             & 0 & 0             & 0 \\
  	    Sensor          & 0 & 0                 & 0             & 0 & 0             & 0 \\
  	    Screen          & 0 & 0                 & 0             & 0 & 0             & 0  \\
  	    Intel-drm       & 1/186,878,492 & 0                 & 2/195,985,082  & 0 & 0             & 0 \\
  		devb-umass      & 0  & 0                 & 3/198,369,046  & 0 & 2/141,548,211  & 2/148,210,630\\
  	    io-pkt-v6-hc    & 0 & 0                 & 0             & 0 & 0             & 0  \\
  	    slog2info       & 0 & 0                 & 0             & 0 & 0             & 0  \\
  		io-hid          & 0 & 0                 & 0             & 0 & 0             & 0  \\
  		\hline
  	\end{tabular}
    \caption{Summary of the testing results over all datasets after 12 hours of repeated testing}
    \label{tab:summary}
  \end{table}
  
\subsection{Brief Fault Detection Tests}  
As mentioned before, we have not conducted an intensive investigation into the true purpose of the anomaly detector: detecting real true-positive anomalies. We realize that this is a limitation to our work presented here. The creation of these tests requires a careful study so that the results are not biased in our favor and true real-world anomalies are tested. We leave this to future investigations. However, there are some brief tests we could conduct to validate in a way the effectiveness of our proposal. This is not conclusive in anyway, but nevertheless serves its purpose.

\subsubsection{Removing a Sensor}\label{remove}
For this test, we attempt to simulate disconnecting the right and left cameras by removing their file from the system. After starting one of the datasets (dataset 1), the right and left camera files were removed. As a result of interacting with the file system residing on the USB flash drive in a way that was not experienced during learning, \textbf{two} threads of the USB driver showed anomalous behaviors: thread 11 generated \textbf{seven} anomalies and thread 7 generated two anomalies. Immediately following the file removal, Sensor thread 5 generated \textbf{10} anomalies as a result of not being able to read the sensor input data (the camera files).

Another test we conducted was removing sensors from the configuration before we started the test run. All the sensors were removed except for the first camera. After 5 minutes, the test was terminated. Parallax's thread 6 was showing a constant rate of anomalies to come to a total of 777254/19842367 (3.917143554\%). This is an undoubtedly detectable rate.

\subsubsection{Physically Interacting with the System}\label{interact}
During training tH, we did not interact with the car system at all, we let it ``drive" on its own without any interference. Thus tH is not expecting this sort of behavior. In this test we physically interacted with the system in order to see if tH will notice any difference in the way the system behaves.

\subsubsection{The Keyboard}
Parallax supports the ability to enable and disable different views on the screen (3D object rendering view, LiDAR view, etc.). As expected, tH did not like our interaction and deemed the system behaviour abnormal. After one key press, Parallax's thread 1 showed \textbf{18} anomalies. By the time we got to the letter q, thread 1 showed \textbf{519} anomalies and slog2info's thread 1 had \textbf{81} anomalies and Screen's thread 2 had \textbf{14} anomalies. Intel-drm's threads 16 and 17 showed \textbf{three} and \textbf{seven} anomalies respectively.

\subsubsection{The Mouse}
Table \ref{tab:mouse} shows the results after connecting and moving a mouse. A constant flurry of anomalies was generated and we terminated the test after a few seconds.

\begin{table}[H]
\begin{center}
  	\begin{tabular}{|l|c|c|}
  		\hline
  		Process & Anomalies & Thread ID(s) \\
  		\hline
  	    Parallax        & 974 & 0 \\
  	    Sensor          & 0 & NA\\
  	    Screen          & 1529 & 6  \\
  	    Intel-drm       & 6 & 16 \\
  		devb-umass      & 9 & 10\\
  	    io-pkt-v6-hc    & 0 & NA \\
  	    slog2info       & 0 & NA\\
  		io-hid          & 24 & 1\\
  		\hline
  	\end{tabular}
    \caption{Summary of the anomalies generated after connecting and moving a mouse for a few seconds}
    \label{tab:mouse}
\end{center}
  \end{table}
  
The anomalies generated for both the keyboard and mouse tests were a constant rate of anomalies. For this reason, it did not make sense to show the anomaly as a percentage of the total calls.

\subsubsection{Changing the restart order}\label{restart}
In our experience, embedded systems have a pre-defined static list of process boot order. This list does not usually change. Part of the way tH works is by using process IDs to distinguish between the different messages a process sends (see \ref{implementation}). If a process does not have the same ID, for example via changing the restart order of processes on a system, or replacing a binary with different one with the same name, tH would be able to identify this quickly. In this test, we changed the boot order by starting the ssh daemon before all the other processes under test. This resulted in Parallax, Sensor, Screen, drm-Intel, devb-umass, io-pkt, slog2info and io-hid having a different process ID. This caused tH to generate a continuous storm of anomalies; every call from everything was anomalous, and we had to terminate the test after a few seconds.

\subsubsection{Overheating}\label{overheating}
One of our test runs generated an exceptionally high amount of anomalies. Much investigation was performed to determine the root cause. It turns out that we erroneously disabled the CPU fan in the BIOS settings and neglected to turn it back on. This fortunate coincidence led to a discovery that thread behavior changes as the hardware overheats. For this test, we disabled the CPU fan and wrapped the BeeBox in a bit of clothing to cause overheating. We then ran one of the testing datasets with a previously zero false positive rate across all threads. The test was terminated after 38 minutes as an excessive amount of heat was generated and we did not want to cause permanent damage to the hardware. The results are shown in Table \ref{overheating}. Parallax thread 9 shows a 0.09\% anomaly rate while Screen's thread 2 had 0.02\%. Intel-drm threads 16 to 21 show an average anomaly rate of 0.01\% for 38 minutes of runtime. In addition, drm-Intel creates an extra thread (thread 22) to deal with the slowing down GPU, which is regarded as an anomaly. The USB disk driver had two anomalous threads (3 and 1) with 19 (0.12\%) and 11 (0.0001\%) anomalies respectively.

\begin{table}[!h]
\begin{center}
  	\begin{tabular}{|l|l|}
  		\hline
  		Process-Thread ID & Anomalies \\
  		\hline
  	    Parallax-9        & 782/996802 (0.078450886\%)\\
  	    Sensor-*          & 0 (0\%)\\
  	    Screen-2          & 335/1910893 (0.017531071\%)\\
  	    Intel-drm-16      & 103/998486  (0.010315618\%)\\
  	    Intel-drm-17      & 65/951152   (0.006833818\%)\\
  	    Intel-drm-18      & 167/1341177 (0.012451749\%)\\
  	    Intel-drm-19      & 119/1190857 (0.009992804\%) \\
  	    Intel-drm-20      & 59/1243944  (0.004742979\%) \\
  	    Intel-drm-21      & 60/912983   (0.006571864\%)\\
  		devb-umass-1      & 10/9328783 (0.000107195\%)\\
  		devb-umass-3      & 29/25087 (0.11559772\%)\\
  	    io-pkt-v6-hc-*    & 0 (0\%) \\
  	    slog2info-*       & 0 (0\%)\\
  		io-hid-*          & 0 (0\%)\\
  		\hline
  	\end{tabular}
    \caption{Summary of the overheating testing results after 38 minutes}
    \label{tab:overheating}
\end{center}
  \end{table}

\section{Discussion}\label{discussion}
In this chapter we discuss the test results produced in Chapter \ref{evaluation_chapter}. For the evaluation of our implementation, we mainly focus on the false positives generated when testing using clean data that is known to be normal and contains no anomalies. The true positive anomaly tests we conducted were brief and simple and we plan to expand on them as part of future work.

\subsection{Results Analysis}

\subsubsection{Determinism}

As the results in Section \ref{testing} show, Parallax, Sensor, Screen and the auxiliary processes (io-pkt, slog2info and io-hid) show no anomalies. This goes to show that all of these processes have deterministic behavior and that tH was able to reliably learn what that is. We have conclusive evidence that shows that all these processes are highly active in terms of CPU usage, as shown in Section \ref{evaluation} and the high number of issued system calls and messages sent as shown in Table \ref{tab:alldatasets}. 

Normalization results in Section \ref{normalization} show an average normalization time of 1.8 hours for Parallax, Sensor, Screen and Intel-drm. This is another indicator of behavioral determinism in the threads under test. In comparison, when we attempted to profile a QNX port of a web browser (based on Google Blink), the average normalization time for the normalized threads was orders of magnitude larger. Clearly, a web browser is a software not intended to perform a critical function and we did not expect it to have the same level of behavioral determinism.

To gain further insight, we collected extra runtime data. If the distinct messages communicated between processes had high variance, there was no hope that we could have profiled its behavior and detected anomalies. Consider a thread that sends thousands of different types of messages at random times. Profiling its behavior using short sequences of message headers would not yield great results; it would take a very long time to build a profile of all possible sequences. Such a profile would then be useless while testing for anomalies, since all possible sequences are permissible. Table \ref{tab:connectivity} shows the number of distinct kernel calls each process makes, followed by the distinct message header types sent to processes other than the process manager (non-kernel calls). The final column is the number of system processes each process communicates with, apart from the process manager. 

The average unique kernel call types is 2 and the average unique non-kernel call message type is 1.78. This confirms that using the 16-bit message header for profiling the message sequence could potentially be a valid choice since the variance in unique message types is very low. Table \ref{tab:perprocess} has details of the collected data including the list of the different messages sent, the process indexes, the channel ID, and the node ID they are sent to, along with the count for each of them.  

The data also shows that processes are highly connected. On average, the number of system processes each process communicated with is 2.2 (maximum of 7), this is not surprising in a non-monolithic system with a microkernel architecture. All drivers and processes are highly communicating system processes that rely on sending messages to perform their core functions. A perturbation in the communication between them could be a powerful indicator of anomalies.

\begin{table}[ht]
  \begin{center}
  	\begin{tabular}{|l|l|l|l|}
  		\hline
  		Process & Distinct & Distinct & Communicating\\
  		Name & Kernel Calls (\%) & Non-Kernel Calls (\%) & Processes\\
  		\hline
  	    Parallax & 2 (28.5\%) & 5 (71.5\%) & 4\\
  	    Sensor & 3 (42.8\%) & 4 (57.2\%) & 2\\
  	    Screen & 1 (25\%) & 3 (75\%) & 3\\
  	    Intel-drm & 7 (100\%) & 0 (0\%) & 0\\
  		devb-umass & 3 (75\%) & 1 (25\%) & 7\\
  	    io-pkt-v6-hc & 1 (100\%) & 1 (100\%) & 2\\
  	    devc-pty & 1 (100\%) & 0 (0\%)  & 0 \\
  	    slog2info & 0 (0\%) & 1 (100\%) & 1\\
  		io-hid & 0 (0\%)  & 1 (100\%) & 1\\
  		\hline
  	\end{tabular}
    \caption{Process Connectivity Data}
    \label{tab:connectivity}
  \end{center} 
  \end{table}

\subsubsection{The False Positives}
As discussed in section \ref{testing}, the GPU and USB drivers show one anomalous thread on average in a few datasets. After further investigation, we concluded the three main reasons for such anomalies are: thread pools, overheating, and using a low quality USB flash drive.

\subsubsection{Thread Pools}\label{pools}
Most of the processes under test create two different types of threads. The first type is single-tasked threads, or threads that only have one type of job to do. These threads always show deterministic behavior and have 0\% false-positives. The second type is multi-tasked threads; much like thread pools, they process a queue of incoming requests. These requests can vary in nature and cause the threads to behave differently according to the task being handled. A simple display of the running threads from the command prompt shows thread names that indicate they are part of a thread pool to process requests. Listing \ref{lst:pidin} shows a snippet of such output. Given that the work load per queued task varies and that the thread scheduling is up to the operating system, the behavior of a thread that is part of a thread pool becomes less deterministic. The thread's workload or the types of tasks it performs will vary every time the software is run. This can be easily observed from the total count of messages each process emits. Table \ref{tab:alldatasets} shows the different total message counts seen by every thread and how they radically differ. Parallax thread 6 was completely dormant in the first three datasets and then became active while running the other three datasets. Thread 9 in the Sensor process was very active in the first three datasets and had no activity during testing of the last three datasets. 

Due to the thread work load varying on every run, we hypothesized that this was the main reason as to why the threads in drm-Intel and devb-umass show anomalies. Devb-umass thread 21 clearly named \texttt{regmgr\_thread\_pool}, confirming our theory that it is indeed part of a thread pool. This is not a problem, as tH can be easily modified to handle threads pools in two different ways:
\begin{enumerate}
    \item \textit{By training over more radically different datasets}. As we have shown in Section \ref{second}, adding more training sets to the system enhances the profiling capability. Ideally, over-stressing the system in such a way that all threads that are part of a thread pool are over loaded and experience all practically possible types of requests in all possible orders. In practice, it is not unlikely to see safety-critical systems exhaustively tested to their limits in such a way that all software behavior is exhibited. As a matter of fact, we were able to obtain some of the stress tests that QNX performs on USB disk driver (devb-umass) and use them for training as described in Section \ref{stress}. As a result, we have seen less anomalies reported by any of the USB disk driver's threads. The training datasets we have obtained, even though they amount to a total of 7.5 gigabytes, are still limited. Each test is roughly 15 seconds in length, hardly giving tH the chance to experience the full system behavior. A future endeavour would be to deploy tH on one of QNX's self-driving vehicles and training would occur continuously while driving around, for extended periods of time. We believe this would yield a much better training experience and a more realistic training environment. The training datasets we used are as close to reality as we could possibly get but are not as close as we would have hoped.
    \item \textit{By naming process threads accordingly and having tH use thread names as identifiers of thread types}. Listing \ref{lst:pidin} shows the output of a shell command display the thread names of all system processes. As can be seen from the thread names, most of the processes under test have a thread pool or thread pool like implementation: Sensor's \texttt{SensorRegMgr} threads, Screen's \texttt{screen-msg} threads and drm-intel's \texttt{resmgr\_thread\_pool} threads. A code inspection of Parallax also shows that the unnamed threads are part of a thread pool. For a future implementation, we propose that all single-tasked threads in the process are uniquely named. Threads belonging to a thread pool or ones that are intended to be used in parallel with other identical threads performing the same tasks should either be named similarly or left unnamed. This would allow tH to recognize these threads and build a \textbf{single profile} for all the threads belonging to a pool and a  \textbf{profile per thread} for all other threads in the process. This way, the behavior of these threads is regarded as one; similar to the original pH's single per-process profile but for a specific thread subset. Such a change would be trivial to implement in tH and would reduce the need to try and learn all possible permutations of behavior that a thread can experience. Figure \ref{fig:thread_pool} depicts a representation of this concept.
    
\begin{figure}
	\centering
	\includegraphics[width=\textwidth]{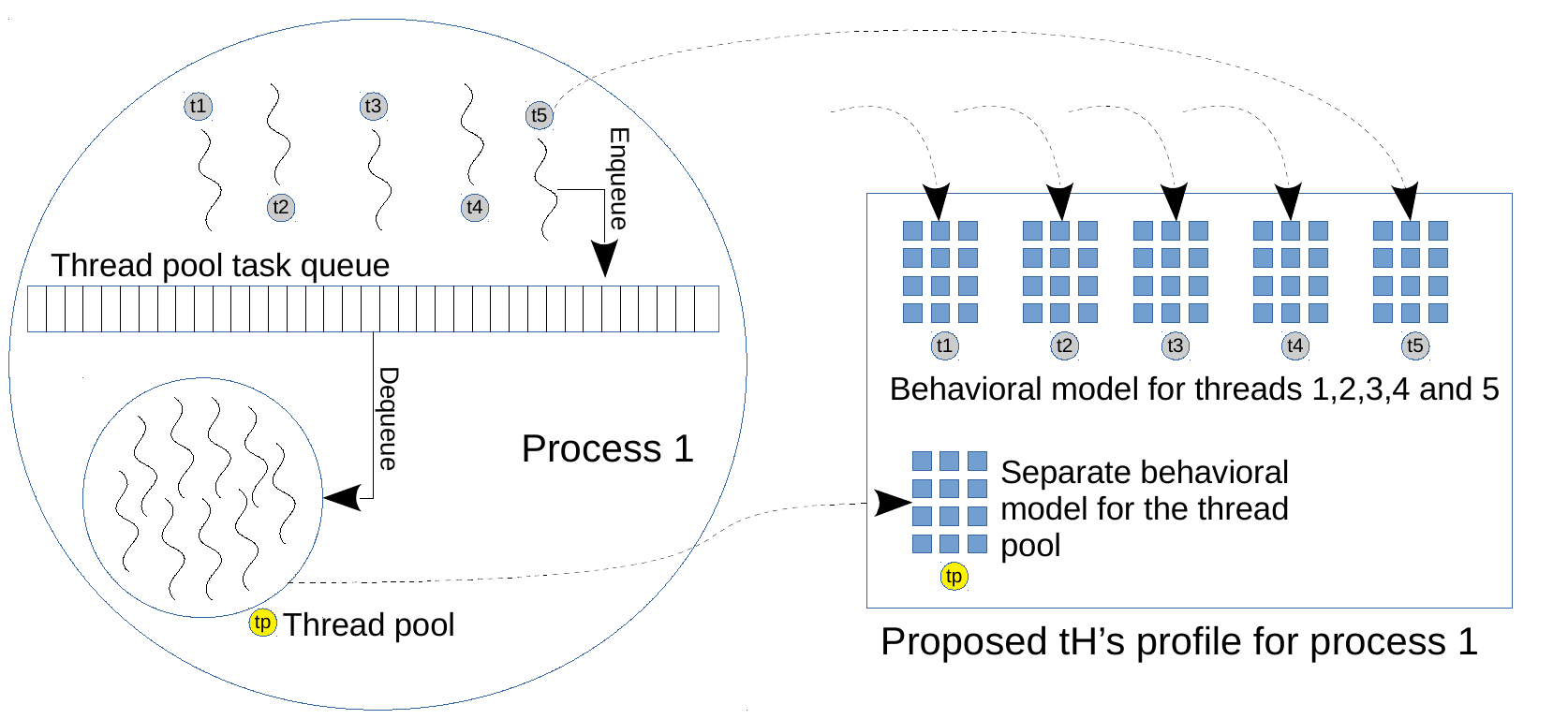}
	\caption{Proposed per-thread pool profile.}
	\label{fig:thread_pool}
\end{figure}

\begin{lstlisting}[caption={Output of process thread names.},label={lst:pidin},language=Java, captionpos=b,  basicstyle=\tiny]
# pidin threads
root:/$ pidin threads
     pid tid name               thread name
    8200   1 oc/boot/devb-umass xpt_signal_handler            
    8200   2 oc/boot/devb-umass USBdi_event_handler           
    8200   3 oc/boot/devb-umass fsys_resmgr                   
    8200   4 oc/boot/devb-umass umass_driver_thread           
    8200   5 oc/boot/devb-umass async_io                      
    8200   6 oc/boot/devb-umass fsnotify_thread               
    8200   7 oc/boot/devb-umass fsys_resmgr                   
    8200   8 oc/boot/devb-umass fsys_resmgr                   
    8200   9 oc/boot/devb-umass fsys_resmgr                   
    8200  10 oc/boot/devb-umass fsys_resmgr                   
    8200  12 oc/boot/devb-umass fsys_resmgr                   
    8200  13 oc/boot/devb-umass fsys_resmgr                   
  241689   1 sbin/screen        screen-monitor                
  241689   2 sbin/screen        drm-event-handler             
  241689   3 sbin/screen        screen-dpy-3                  
  241689   4 sbin/screen        screen-msg                    
  241689   5 sbin/screen        screen-msg                    
  241689   6 sbin/screen        screen-msg                    
  241689   7 sbin/screen        screen-hid                    
  241689   8 sbin/screen        screen-msg                    
  241689   9 sbin/screen        screen-msg                    
  241689  10 sbin/screen        screen-msg                    
  241689  12 sbin/screen        screen-msg                    
  241689  13 sbin/screen        screen-msg                    
  241689  14 sbin/screen        screen-msg                    
  241689  15 sbin/screen        screen-msg                    
  245786   1 sbin/drm-intel     drm-monitor                   
  245786   2 sbin/drm-intel     kernel timer                  
  245786   3 sbin/drm-intel     system wq                     
  245786   4 sbin/drm-intel     system unbound wq             
  245786   5 sbin/drm-intel     system long wq                
  245786   6 sbin/drm-intel     tasklet wq                    
  245786   7 sbin/drm-intel     i915                          
  245786   8 sbin/drm-intel     i915-dp                       
  245786   9 sbin/drm-intel     i915-isr                      
  245786  10 sbin/drm-intel     i915_modeset                  
  245786  11 sbin/drm-intel     i915-userptr-acquire          
  245786  12 sbin/drm-intel     i915/signal:0                 
  245786  13 sbin/drm-intel     i915/signal:1                 
  245786  14 sbin/drm-intel     i915/signal:2                 
  245786  15 sbin/drm-intel     i915/signal:4                 
  245786  16 sbin/drm-intel     resmgr_thread_pool            
  245786  17 sbin/drm-intel     resmgr_thread_pool            
  245786  19 sbin/drm-intel     resmgr_thread_pool            
  245786  20 sbin/drm-intel     resmgr_thread_pool            
  245786  21 sbin/drm-intel     resmgr_thread_pool            
  245786  22 sbin/drm-intel     resmgr_thread_pool            
  274460   1 sensor             SensorService                 
  274460   2 sensor             SensorResMgr                  
  274460   3 sensor             SensorResMgr                  
  274460   4 sensor             SensorResMgr                  
  274460   5 sensor             5                             
  274460   6 sensor             6                             
  274460   7 sensor             7                             
  274460   8 sensor             8                             
  274460   9 sensor             9                             
  274460  10 sensor             SensorResMgr                  
  286749   1 Parallax           Main                          
  286749   2 Parallax           2                             
  286749   3 Parallax           3                             
  286749   4 Parallax           4                             
  286749   5 Parallax           5                             
  286749   6 Parallax           6                             
  286749   7 Parallax           Camera                        
  286749   8 Parallax           Analysis                      
  286749   9 Parallax           Hud                           
  286749  10 Parallax           10                            
  286749  11 Parallax           Algo                          
  \end{lstlisting}

\end{enumerate}

\subsubsection{Overheating and High Sensitivity}\label{overheat_discussion}
As the overheating test in section \ref{overheating} show; tH is sensitive to fluctuations in the heat produced by the hardware, in particular when the hardware (GPU, USB flash drive) overheats. Given the high throughput of data continuously being processed for 12 hours, it is not surprising for the hardware to overheat. As well, the QNX threads, and in particular the ones belonging to the flash drive and the GPU, react appropriately when they experience slowdowns as a result of their respective hardware behaving differently. One could argue that tH is overly-sensitive, predicting a problematically high false positive rate in the field. To this we argue, i) embedded safety-critical systems must be tested to their limit, including any behavior resulting from acceptable levels of overheating. Such behavior would thus be learnt during the training phase. Abnormal behavior resulting from the acceptable overheating thresholds being crossed is perfectly acceptable. We claim that sensitivity to behavioral disturbances as a result of overheating is a desired feature when it comes to safety-critical systems, ii) our plan to create thread-pools as mentioned in Section \ref{fig:thread_pool} might reduce this sensitivity since all threads would be contributing to a single thread profile. This is a reduction in sensitivity that we are not sure is desired. This needs to be investigated and tested further after the single profile for all thread pools are implemented. Only then could an appropriate decision be made. One potential solution is to provide the single thread-pool profile as a configurable option. That way, drivers that deal with hardware could have this feature disabled to remain highly sensitive to overheating or other unpredictable hardware fluctuations and disabled for others where appropriate.

\subsubsection{Low Quality USB Flash Drive}
Similar to the behavior change seen as a result of overheating, the USB flash driver we used for our experiments was a low quality USB 2.0 flash driver. As the USB driver experiences slowdowns in the transfer rates of the flash driver, they reorganize themselves and behave differently in order to deal with the slowdowns they're experiencing. Migrating the data to the onboard NVMe drive and rerunning one of the datasets that showed anomalies in the USB flash drive eliminated the anomalies. For future work, all of our testing data will be moved to the highly performing NVMe drive.

\subsection{True Positive Tests}
Even though the true-positive tests we ran were non-exhaustive and incomplete, the results of the few tests we ran were positive. All the tests show a perturbation in the learnt normal behavior when an unknown and previously unseen event occurs. A particularly interesting result is the activation of anomalies in io-hid, the human-interface device driver responsible for handling the keyboard and mouse, when the mouse and keyboard are plugged and unplugged then used. The anomalies could be seen over almost the \textbf{entire} network including slog2info, the error and status logging service. This supports our hypothesis that viewing the system as a network of threads has more power in detecting anomalies over isolating specific components. The network anomaly has a much lower threshold and provides a higher confidence in true anomalies being classified. The Sensor and the network drivers generate no anomalies when moving the mouse, however. Even though this is not conclusive, it provides some evidence that the system is not overly sensitive. This is because Sensor and the network driver have nothing to do with the mouse movement and thus were not disturbed. The overheating test results in Section \ref{overheating} show an unprecedented amount of anomalies. Without doubt, the behavior of the threads does change as the hardware overheats as discussed in Section \ref{overheat_discussion}.

\subsection{Limitations}

\subsubsection{Limitations of Our Testing Approach}
One of the limitations we had during the learning phase was the short duration of the learning datasets. Even though 15 seconds of driving produced 3.5 gigabytes of data, we still believe that this was not enough to exercise all the processes' threads in order to test a real-life driving scenario. As shown from the testing results, this might have been enough to produce zero false positives for the testing datasets for Parallax, Sensor and Screen, but this does not provide solid proof that the same behavior will be produced when driving for longer periods of time while detecting anomalies. As mentioned in Section \ref{second}, multiple datasets were added to achieve these false positive results. Given the high volume of data generated by the various sensors and the high definition cameras, it was infeasible to record an extended length of driving period. For future work, we intend to install tH on a system that reads the sensor data in parallel to the original data collection methods on an actual self-driving car and building the training profiles as the car navigates the streets.

\subsubsection{The Dynamic Thread Creation}
One of the issues that we will face in the future, are processes that dynamically create threads at runtime. As a matter of fact, we have experienced threads being created as a result of the overheating test in Section \ref{overheating}; drm-intel has created a new thread, thread 22 to handle the slowing down GPU. Since we build profiles for every thread in the system, we heavily rely on thread IDs to identify the correct thread learning structure. During learning, threads created dynamically at runtime pose no issues; the appropriate structure will be created and assigned to their thread ID. However, a dynamically created thread at testing time with an ID that has never been observed during testing is regarded as an anomaly. We do not believe this behavior is incorrect. The ability to create threads at runtime should be allowed, however, we argue that the system should be tested to its full potential, with the maximum number of threads created. Typically, more threads are dynamically added to a process to support a heavier workload and potentially increase parallelism. A safety-critical system should be subjected to extreme workloads in order to test all conditions and our training dataset was deficient in this regard. Given this exhaustive testing, dynamic thread creation would pose no problem to tH. Further to this, the use of per-thread look-ahead pairs table versus one table per process, allows us to easily differentiate between threads that we have profiles for, i.e. they normalized during training, and ones that are dynamically created at runtime and have no training profiles. This prevents newly created threads from generating false-positives, and allows us to isolate their emergent behavior from other well-behaved threads with normalized profiles. 

\subsubsection{Detecting Internal Process Corruption}
One of the limitations of our detection method is its inability to detect internal process defects if they don't result in a behavioral change. Monitoring a process at the system call and message sending level does not detect or prevent corruptions within the process itself if this does not cause any behavioral change from an outside observer's perspective, be it in the kernel or another service. An example would be a buffer overflow attempting to change the result of a calculation within the process' own address-space. However, if a vulnerability, or an error in the process were to cause any kind of system-wide malicious activity, such as spawning a root shell, it would certainly need to interface with the kernel by invoking system calls. Arguably, the former can potentially have dire effects on a safety-critical system. One could also argue that an internal process defect that does not result in a behavioral change, especially in a microservices-based architecture cannot possibly have much consequences. Clearly, this requires further investigation. What if the defect was internal to a driver that is the final layer in the message-chain and would then communicate incorrectly with a device? Can an internal fault happen without any early warning signs and behavioral changes in any other thread in the system leading to the fault? These are questions that we will be investigating in the future.

\subsection{Conclusion and Future Work}
In this thesis, we presented Thread Homeostasis (tH), a technique that profiles software to establish a baseline of normal behaviour in order to identify behaviour that falls outside the norm. tH works by efficiently modeling the behavior of process threads through the way they interact with the underlying operating system and other services. Such a profiling technique can be a unique identifier of how the system naturally behaves during its operation and can be used to identify when it misbehaves. Learning how to \textit{not} behave in the field could prove to be a powerful early warning system of an occurrence of a fault or an unintended consequence. Our work is motivated by two reasons. Firstly, engineering practices are often not adequate for developing safety critical software. This can, and has resulted, in faults and errors going unnoticed in safety critical software. Secondly, there is a gap in anomaly detection research for fault and error detection.

The autonomous vehicle demo software we tested had a high level of complexity and yet our results showed that threads exhibit a fair level of behavioral determinism suitable for profiling by tH. The results also show that profiling such behavior can be done in a matter of hours by tH. Most of the threads under test showed zero false positives indicating that tH was able to recognize that the threads behave similarly to its learned expectation. tH's minimal CPU usage and memory footprint along with these results serve as concrete evidence that tH has great potential as an online realtime anomaly detector for faults in safety-critical systems.

Our novel and efficient approach uses inter-process message headers and system calls for building behavioral models for every running thread on the system. This extends anomaly detection from being limited to one or a few processes running on the system to monitoring all threads running in a realtime microkernel message-passing operating system. The brief true positive detection tests we conducted show the power of viewing the system as a set of interconnected threads. Leveraging the nature of a message passing microkernel, it is much easier to see anomalies propagate and perturb many of the system processes and threads. This provides a more confident indicator of anomalies and reduces the threshold of the error tolerance of misclassifying a true anomaly. The per-thread profiles provide a more sensitive behavioral model than the per-process profiles. Since all the threads in a process conjointly update one profile, the model becomes highly dependent on the operating system scheduling algorithm and workloads. The per-thread profiles provide freedom from such possible deviations in behavior.

The exhaustive level of testing and verification that a safety-critical system should be subjected to, serves as an ultimate training ground for tH. tH could prove to be a useful tool for detecting the incompleteness of the software's testing and verification. Anomalies during this phase of the tests, before the product goes operational, are a very strong indicator that there exists system behaviors that have never been experienced before during testing and warrants an investigation.

This work has served its purpose of providing an efficient mechanism to extract the required information out of the QNX kernel and building compact profiles for modeling system threads' behavior. This sets the stage and lays the ground work for conducting further in-depth analysis and research into the suitability and applicability of this method for fault tolerance in safety-critical systems.

\subsubsection{Future Work}

\subsubsection{More Field Evaluations}
We do not consider the results in this thesis to be conclusive, however we consider them as evidence that tH has great potential. In order to provide concrete evidence, more real-life in-field tests are required. We intend to have tH run on QNX's self-driving autonomous vehicle in order to learn and detect anomalies while driving, thus experiencing more varied behaviors for an extended period of time.

\subsubsection{Eliminating Process IDs From the Training Profiles}
Currently, we use the process index/ID when building the behavioral profiles (see chapter \ref{implementation}). This prevents the process restart order from changing as shown in section \ref{restart}. While we argued that this is a desired behavior, having the configurable option to be able to turn this feature on or off might be desired. In future tests, we intend to evaluate tH on more dynamic systems, such as desktop environments running the QNX operating system. We have designed and implemented a process index translation unit that uses process names to translate the process ID in the learnt profile. As part of a future effort we plan to integrate this in tH. Even though the intention of this anomaly detector is for tightly controlled, deterministic safety-critical systems that do not expect a process restart order to change, it is nevertheless a good idea to have this module implemented for greater flexibility and testing purposes.

\subsubsection{Testing with Different Sequence Window Sizes}
Currently, the window that slides over the sequence of calls during both training and testing is of size 8 (see Chapter \ref{implementation}). The size of this window has a profound effect on the training models. Imagine a hypothetical window size equal to the total number of calls emitted by a thread. This would cause all possible permutations of all the events to be included in the model and considered to be normal behavior, thus over generalizing. In this case, no anomalies would be detected, be it true or false positives. On the contrary, a window size of just one would be too restrictive. A thread must emit this exact system call sequence, otherwise, its behavior would be anomalous. For future work, we plan to experiment with different sliding window sizes and conduct analysis on the results. The configurable window size currently implemented in tH allows us to do this with ease.

\subsubsection{Correlating Anomalies to Source Code}
A current limitation of our system is the difficulty of tracking back to the root cause of the anomalies observed. Indeed, if an anomaly occurred, we can determine the destination and receiving threads and the type of message sent. However, we cannot determine the location in the source code that caused such a message or system calls. We plan to investigate the possibility of correlating anomalies with exact source code locations. 

\subsubsection{Different Message Header Size}
The current tH implementation only uses the 16-bit message header sent between processes (see Section \ref{32-2}). We plan to investigate the effects of including more data from the message has on profiling. Our expectation is that there might be too much variance in the message data, which might render profiling difficult, if not impossible.

\subsubsection{Having a Profile Per Named Thread}
As discussed in Section \ref{fig:thread_pool}, besides having a profile per thread, we plan on creating a profile for every thread pool in the process. This aims at reducing the effects of the non-deterministic behavior shown by thread pools that share the same tasks. This frees the profiles from dependence on environmental factors introduced by varying workloads and thread scheduling differences.

\subsubsection{Fault Injection}
As part of evaluating tH's efficiency, we plan on investigating whether tH will be able to detect injected faults in critical software. Fault-injection testing is a research area in its own right. Setting up and conducting the proper fault-injection tests is a topic that we intend to investigate in the future.

\bibliographystyle{plain}
\bibliography{simple}

\appendix

\section{Normalization data per-thread}
  \begin{center}
\setlength\LTleft{-0.0in}
  	\begin{longtable}{|l|l|l|l|}
  		\hline
  		Process Name-Thread ID & Time Taken to Normalize (seconds) & Message Count\\
  		\hline
  	    Parallax-0 & 19190 & 204402279 \\
  	    Parallax-6 & 3603 &  828673725\\
  	    Parallax-7 & 3603 & 8659174  \\
  	    Parallax-8 & 3603 & 8659174  \\
  	    Parallax-9 & 17404 & 8659174  \\
  	    Parallax-10 & 4830 &  828673725  \\
  	    Parallax-11 & 4395 & 144215119  \\ 
  	    Parallax-12 & 7148 & 828673725  \\
  		\hline
  	    Sensor-4       & 3613 &  44098948   \\
  	    Sensor-5       & 3613 &   44072755  \\
  	    Sensor-6       & 46482 &  25819587    \\
  	    Sensor-7       & 3613 &  59425036   \\
  	    Sensor-8       & 3613 & 74189300   \\
  	    Sensor-9       & 3646 &  13631151\\
  	    Sensor-10      & 3631 &   4511711\\
  		\hline
  	    Screen-1       & 3602 & 422036776\\
  	    Screen-2       & 4091 & 618783165\\
  	    Screen-6       & 3602 & 2238679 \\
  	    Screen-7       & 3603 & 5105574 \\
  		\hline
  	    Intel-drm-6    & 3603 & 1266248\\
  	    Intel-drm-7    & 3604 & 2132900\\
  	    Intel-drm-14    & 4214 & 433653431\\
  	    Intel-drm-15    & 218108 & 49858124\\ 
  	    Intel-drm-16    & 4625 &   88396741\\
  	    Intel-drm-17    & 5353 &  374491804\\
  	    Intel-drm-18    & 109963 & 307403101\\ 
  	    Intel-drm-19    & 5195 & 565311523\\
  	    Intel-drm-20    & 4267 & 533148727\\
  	    Intel-drm-21    & 3876 & 81428142\\
  		\hline
  		devb-umass-1   & 3612 & 1483818683\\
  		devb-umass-2   & 4971 & 20339\\
  		devb-umass-3   & 3869 & 14093250\\
  		devb-umass-4   & 23007 & 1289\\
  		devb-umass-6   & 6164 & 20089\\
  		devb-umass-7   & 5906 & 19779\\
  		devb-umass-8   & 6169 & 19782\\
  		devb-umass-9   & 6278 &  19133\\
  		devb-umass-10   & 5139 & 18627\\
  		devb-umass-11   & 5879 & 17239\\
  		devb-umass-12   & 5973 & 14659\\
  		devb-umass-13   & 21658 & 1551\\
  		devb-umass-14   & 588180 & 132\\ 
  	    \hline
  	    io-pkt-v6-hc & 70936 & 19133\\
  	    slog2info    & 3708 & 762018  \\
  	    io-hid       & 3602 & 14688381 \\
  		\hline
    \caption{Process Normalization Statistics for the Process's Active Threads}
    \label{tab:normalization_details}
  	\end{longtable}
  \end{center}

\section{Unused Kernel Trace Events by tH}
    \centering
\setlength\LTleft{-0.0in}
    \begin{longtable}{|l|p{8cm}|}
    \hline
         \textbf{Trace Event} & \textbf{Purpose}\\
         \hline
        \_NTO\_TRACE\_KERCALLENTER  & Kernel call enter trace data includes the kernel call numbers that a thread running on a core has generated, the core number and in case of a MessageSend() kernel call, the data would include the connection ID (coid). Using internal lists we built we could infer the channel, node and process ID.\\
         \hline
        \_\_KER\_CONNECT\_ATTACH & ConnectAttach() trace data, used to keep an internal list of connection IDs (coids). This allows us to know the channel, node and process ID that is attached to this coid and build our internal lists accordingly. This is used when \_NTO\_TRACE\_KERCALLENTER is received.Note that in case we receive a connection ID that is not in the list we would issue a kernel call to get the information we need (because we might have started tH after starting the process and we might have missed the ConnectAttach() kernel call.\\
         \hline
        \_\_KER\_CONNECT\_DETACH & ConnectDetach() trace data, used to deinitialize the internal structures initialized upon the receipt of \_\_KER\_CONNECT\_ATTACH.\\
         \hline
         \_NTO\_TRACE\_THRUNNING & Received when a thread becomes running on a core. Used when \_NTO\_TRACE\_KERCALLENTER is received to infer which process is running of the core.\\
         \hline
         \_NTO\_TRACE\_THDESTROY & Used to track thread destruction. Since we have a sequence list per-thread, it is important to destroy and free the per-thread sequence in case a new thread was created with the same thread ID.\\
         \hline
        \_NTO\_TRACE\_PROCCREATE & Tracks newly created processes in the system. This is important so that we can allocate and initialize the appropriate per-process structures.\\
         \hline
        \_NTO\_TRACE\_PROCCREATE\_NAME & Supplies names of the newly created processes in the system. This is important since the configuration file for tH can specify specific binaries to track by name. This is also important to track the exec()s in the system.\\
         \hline
         \_NTO\_TRACE\_PROCDESTROY & Tracks process destruction events. This is important to deinitialize and free our internal per-process structures.\\
         \hline
    \caption{Trace events required prior to the addition of kernel exit trace event}
    \label{tab:trace_events}
    \end{longtable}

\section{Trace data for Google's Blink Port on QNX}\label{appendixblinq}

  \begin{center}
\setlength\LTleft{-0.0in}
  	\begin{longtable}{ |l|l|p{4cm}|p{1.6cm}|}
  		\hline
  		Rx Proc. Indx & 16-bit Msg Head & 32-bit Msg Head & Count\\
  		\hline
\multirow{14}{*}{1 (procnto)} & 0x0102 & 0x100100-00100102 & 1\\
& 0x0116 & 0x100100-00040116 & 232634\\

& \multirow{6}{*}{0x0100} & 0x100100-\textbf{0002}0100 & 147428\\
& & 0x100100-\textbf{0001}0100 & 1325 \\
& & 0x100100-\textbf{0000}0100 & 1223\\
& & 0x100100-\textbf{0003}0100 & 94255\\
& & 0x100100-\textbf{0005}0100 & 9\\
& & 0x100100-\textbf{0007}0100 & 48\\

& \multirow{2}{*}{0x0106} & 0x100100-00100106 & 588 \\
& & 0x100100-00100106 & 47069\\

& 0x0104 & 0x100100-00080104 & 47223\\

& \multirow{3}{*}{0x0041} & 0x100100-00010041 & 1039\\
& & 0x10010000070041 & 22145\\
& & 0x100100-00000041 & 175843\\
& 0x0115 & 0x100100-003c0115 & 184807\\
& 0x0110 & 0x100100-00180110 & 94140\\
& 0x0074 & 0x100100-00000074 & 97\\
& 0x0002 & 0x100100-00100002 & 1\\
& 0x0000 & 0x100100-00000000 & 6281788\\
& 0x0040 & 0x100100-00000040 & 175773\\
& 0x001a & 0x100100-0001001a & 6\\
\hline

\multirow{2}{*}{2} & 0x0102 & 0x200100-\textbf{0010}0102 & 8\\ 
 & 0x0100 & 0x200100-00020100 & 1\\
\hline

\multirow{2}{*}{3} & 0x0101 & 0x300100-\textbf{0010}0101 & 8531344\\
 & 0x0102 & 0x300100-\textbf{0010}0102 & 8531507\\
 & 0x0105 & 0x300100-\textbf{0040}0105 & 19837741\\
\hline

\multirow{12}{*}{8} & \multirow{2}{*}{0x0102} & 0x800300-\textbf{0018}0102 & 11283\\
& &0x800300-\textbf{0010}0102 & 313\\
 & \multirow{2}{*}{0x0101} & 0x800300-\textbf{0018}0101 & 2178\\
 & & 0x800300-\textbf{0010}0101 & 332\\
 & 0x0116 & 0x800300-\textbf{0004}0116 & 2712\\
 & \multirow{3}{*}{0x0100} & 0x800300-\textbf{0000}0100 & 97\\
 & &0x800300-\textbf{0001}0100 & 110\\
 & &0x800300-\textbf{0002}0100 & 2667\\
 & 0x0106 & 0x800300-00100106 & 11113\\
 & 0x010f & 0x800300-000c010f & 149\\
 & \multirow{5}{*}{0x0100} & 0x800300-00010100 & 968\\
 & & 0x800300-00030100 & 74\\
 & & 0x800300-00000100 & 879\\
 & & 0x800300-00040100 & 18\\
 & & 0x800300-00050100 & 9\\
 
 & 0x0104 & 0x800300-00080104 & 623\\
 & 0x0110 & 0x800300-00180110 & 12171\\
 & 0x0119 & 0x800300-00080119 & 88\\
 & 0x010b & 0x800300-000c010b & 15\\
 & 0x0108 & 0x800300-00080108 & 6\\
\hline

9 & 0x0101 & 0x900100-\textbf{0010}0101 & 53429\\
\hline

\multirow{10}{*}{15} & \multirow{2}{*}{0x0113} & 0xf00200-\textbf{001c}0113 & 476\\
& & 0xf00200-\textbf{0018}0113 & 290\\
& \multirow{2}{*}{0x0102} & 0xf00200-\textbf{0014}0102 & 134\\
& & 0xf00200-\textbf{0010}0102 & 1719597\\
& \multirow{2}{*}{0x0101} & 0xf00200-\textbf{0014}0101 & 6432\\
& & 0xf00200-\textbf{0010}0101 & 1745946\\
& 0x0105 & 0xf00200-\textbf{0040}0105 & 1281981\\
& 0x0116 & 0xf00200-\textbf{0004}0116 & 414\\
& \multirow{2}{*}{0x106} & 0xf00200-00100106 & 5154194\\
& & 0xf00200-00100106 & 859466\\
& 0x0100 & 0xf00200-00020100 & 382\\
& 0x0115 & 0xf00200-003c0115 & 10\\
& 0x010d & 0xf00200-0044010d & 10\\
\hline

\multirow{5}{*}{20} & 0x0106 & 0x1400100-00100106 & 18\\
& 0x0102 & 0x1400100-\textbf{0010}0102 & 3421622\\
& 0x101 & 0x1400100-\textbf{0010}0101 & 22\\
& 0x0116 & 0x1400100-\textbf{0004}0116 & 13\\ 
& \multirow{3}{*}{0x0100} & 0x1400100-00020100 & 3\\
& & 0x1400100-00010100 & 11\\
& & 0x1400100-00000100 & 11\\
\hline

21 & 0x0100 & 0x1500100-\textbf{0002}0100 & 134\\
\hline

\multirow{9}{*}{27} & \multirow{9}{*}{0x0113} & 0x1b00400-\textbf{0060}0113 & 131\\
& & 0x1b00400-\textbf{0010}0113 & 12652\\
& & 0x1b00400-\textbf{004c}0113 & 379416\\
& & 0x1b00400-\textbf{0028}0113 & 458\\
& & 0x1b00400-\textbf{0138}0113 & 13\\
& & 0x1b00400-\textbf{00a8}0113 & 13\\
& & 0x1b00400-\textbf{0020}0113 & 39\\
& & 0x1b0040-\textbf{000a}00113 & 48\\
& & 0x1b00400-\textbf{0048}0113 & 13\\
\hline

28 & 0x0106 & 0x1c0050-000100106 & 25907766\\
\hline

\multirow{4}{*}{29} & 0x0102 & 0x1d00100-\textbf{0010}0102 & 1\\ 
& 0x0116 & 0x1d00100-\underline{0004}0116 & 2\\
& 0x100 &  0x1d00100 - 00020100 & 1\\
& 0x0100 & 0x1d00100 - 00000100 & 1\\
\hline
    \caption{Trace data for Google's Blink Port on QNX after approx. 10 hours}
    \label{tab:blink}
  	\end{longtable}
  \end{center}

\section{Normalization Details Per-Process}
\begin{center}

\setlength\LTleft{-0.0in}
\setlength\LTright{-1in plus 1 fill}
    \begin{longtable}{|l|l|l|l|l|}
    \hline
\multicolumn{5}{|c|}{\textbf{./Parallax}}\\
\hline
\multicolumn{2}{|l|}{Total calls} & \multicolumn{3}{|l|}{420329147}\\
\multicolumn{2}{|l|}{Total calls to normalize} & \multicolumn{3}{|l|}{243554806 (57.94\%)}\\
\multicolumn{2}{|l|}{Num Non-MsgSend calls} &  \multicolumn{3}{|l|}{0 (0.000000\%)} \\
\multicolumn{2}{|l|}{Total MsgSends to procnto} & \multicolumn{3}{|l|}{239620764 (57.007885\%)}\\
\multicolumn{2}{|l|}{Total MsgSends to non-procnto} & \multicolumn{3}{|l|}{180708383 (42.992111\%)}\\
\multicolumn{2}{|l|}{Num processes msg receiving} &  \multicolumn{3}{|l|}{4}\\
\multicolumn{2}{|l|}{Num distinct MsgSend types to procnto} &  \multicolumn{3}{|l|}{2 (28.571430\%)} \\
\multicolumn{2}{|l|}{Num distinct MsgSends types to non-procnto} &  \multicolumn{3}{|l|}{5 (71.428574\%)} \\
\hline
MSG & PID & CHID & ND & COUNT\\
\hline
0x106 & 29 & 2 & 0 & 53092090 (12.63\%)\\
0x100 & 1 & 1 & 0 & 227240164 (54.06\%)\\
0x100 & 8 & 3 & 0 & 75693679 (18.00\%)\\
0x106 & 27 & 5 & 0 & 41818368 (9.94\%)\\
0x113 & 26 & 4 & 0 & 10087307 (2.39\%)\\
0x0 & 1 & 1 & 0 & 12380600 (2.94\%)\\
0x113 & 26 & 4 & 0 & 16939 (00.0040\%)\\

\hline
\multicolumn{5}{|c|}{\textbf{./proc/boot/devb-umass}}\\
\hline
\multicolumn{2}{|l|}{Num Non-MsgSend calls} &  \multicolumn{3}{|l|}{0 (0.000000\%)} \\
\multicolumn{2}{|l|}{Total MsgSends to procnto} & \multicolumn{3}{|l|}{31708 (0.011081\%)}\\
\multicolumn{2}{|l|}{Total MsgSends to non-procnto} & \multicolumn{3}{|l|}{286113706 (99.988922\%)}\\
\multicolumn{2}{|l|}{Num processes msg receiving} &  \multicolumn{3}{|l|}{7}\\
\multicolumn{2}{|l|}{Num distinct MsgSend types to procnto} &  \multicolumn{3}{|l|}{3 (75.000000\%)} \\
\multicolumn{2}{|l|}{Num distinct MsgSends types to non-procnto} &  \multicolumn{3}{|l|}{1 (25.000000\%)} \\
\hline
MSG & PID & CHID & ND & COUNT\\
\hline
0x113 & 7 & 1 & 0 & 286113706 (99.98\%)\\
0x41 & 1 & 1 & 0 & 15865 (0.0055\%)\\
0x10e & 1 & 1 & 0 & 12 (0.000000042\%)\\
0x40 & 1 & 1 & 0 & 15831 (0.0055\%)\\

\hline
\multicolumn{5}{|c|}{\textbf{./proc/boot/io-hid}}\\
\hline
\multicolumn{2}{|l|}{Total calls} & \multicolumn{3}{|l|}{3091954}\\
\multicolumn{2}{|l|}{Total calls to normalize} & \multicolumn{3}{|l|}{7339 (0.23\%)}\\
\multicolumn{2}{|l|}{Num Non-MsgSend calls} &  \multicolumn{3}{|l|}{0 (0.000000\%)} \\
\multicolumn{2}{|l|}{Total MsgSends to procnto} & \multicolumn{3}{|l|}{0 (0.000000\%)}\\
\multicolumn{2}{|l|}{Total MsgSends to non-procnto} & \multicolumn{3}{|l|}{3091954 (100.000000\%)}\\
\multicolumn{2}{|l|}{Num processes msg receiving} &  \multicolumn{3}{|l|}{1}\\
\multicolumn{2}{|l|}{Num distinct MsgSend types to procnto} &  \multicolumn{3}{|l|}{0 (0.000000\%)} \\
\multicolumn{2}{|l|}{Num distinct MsgSends types to non-procnto} &  \multicolumn{3}{|l|}{1 (100.000000\%)} \\
\hline
MSG & PID & CHID & ND & COUNT\\
\hline
0x113 & 7 & 1 & 0 & 3091954 (100.00\%)\\

\hline
\multicolumn{5}{|c|}{\textbf{./proc/boot/slog2info}}\\
\hline
\multicolumn{2}{|l|}{Total calls} & \multicolumn{3}{|l|}{158556}\\
\multicolumn{2}{|l|}{Total calls to normalize} & \multicolumn{3}{|l|}{398 (0.24\%)}\\
\multicolumn{2}{|l|}{Num Non-MsgSend calls} &  \multicolumn{3}{|l|}{0 (0.000000\%)} \\
\multicolumn{2}{|l|}{Total MsgSends to procnto} & \multicolumn{3}{|l|}{0 (0.000000\%)}\\
\multicolumn{2}{|l|}{Total MsgSends to non-procnto} & \multicolumn{3}{|l|}{158556 (100.000000\%)}\\
\multicolumn{2}{|l|}{Num processes msg receiving} &  \multicolumn{3}{|l|}{1}\\
\multicolumn{2}{|l|}{Num distinct MsgSend types to procnto} &  \multicolumn{3}{|l|}{0 (0.000000\%)} \\
\multicolumn{2}{|l|}{Num distinct MsgSends types to non-procnto} &  \multicolumn{3}{|l|}{1 (100.000000\%)} \\
\hline
MSG & PID & CHID & ND & COUNT\\
\hline
0x101 & 2 & 1 & 0 & 158556 (100.00\%)\\

\hline
\multicolumn{5}{|c|}{\textbf{./sbin/drm-intel}}\\
\hline
\multicolumn{2}{|l|}{Total calls} & \multicolumn{3}{|l|}{579689879}\\
\multicolumn{2}{|l|}{Total calls to normalize} & \multicolumn{3}{|l|}{4035021 (0.69\%)}\\
\multicolumn{2}{|l|}{Num Non-MsgSend calls} &  \multicolumn{3}{|l|}{0 (0.000000\%)} \\
\multicolumn{2}{|l|}{Total MsgSends to procnto} & \multicolumn{3}{|l|}{579689879 (100.000000\%)}\\
\multicolumn{2}{|l|}{Total MsgSends to non-procnto} & \multicolumn{3}{|l|}{0 (0.000000\%)}\\
\multicolumn{2}{|l|}{Num processes msg receiving} &  \multicolumn{3}{|l|}{0}\\
\multicolumn{2}{|l|}{Num distinct MsgSend types to procnto} &  \multicolumn{3}{|l|}{7 (100.000000\%)} \\
\multicolumn{2}{|l|}{Num distinct MsgSends types to non-procnto} &  \multicolumn{3}{|l|}{0 (0.000000\%)} \\
\hline
MSG & PID & CHID & ND & COUNT\\
\hline
0x106 & 1 & 1 & 0 & 58160861 (10.03\%)\\
0x116 & 1 & 1 & 0 & 58160417 (10.03\%)\\
0x47 & 1 & 1 & 0 & 114405790 (19.73\%)\\
0x100 & 1 & 1 & 0 & 116321534 (20.06\%)\\
0x40 & 1 & 1 & 0 & 58160567 (10.03\%)\\
0x43 & 1 & 1 & 0 & 116320663 (20.06\%)\\
0x41 & 1 & 1 & 0 & 58160047 (10.03\%)\\

\hline
\multicolumn{5}{|c|}{\textbf{./sbin/screen}}\\
\hline
\multicolumn{2}{|l|}{Total calls} & \multicolumn{3}{|l|}{150581874}\\
\multicolumn{2}{|l|}{Total calls to normalize} & \multicolumn{3}{|l|}{579764 (0.38\%)}\\
\multicolumn{2}{|l|}{Num Non-MsgSend calls} &  \multicolumn{3}{|l|}{0 (0.000000\%)} \\
\multicolumn{2}{|l|}{Total MsgSends to procnto} & \multicolumn{3}{|l|}{12842020 (8.528265\%)}\\
\multicolumn{2}{|l|}{Total MsgSends to non-procnto} & \multicolumn{3}{|l|}{137739854 (91.471741\%)}\\
\multicolumn{2}{|l|}{Num processes msg receiving} &  \multicolumn{3}{|l|}{3}\\
\multicolumn{2}{|l|}{Num distinct MsgSend types to procnto} &  \multicolumn{3}{|l|}{1 (25.000000\%)} \\
\multicolumn{2}{|l|}{Num distinct MsgSends types to non-procnto} &  \multicolumn{3}{|l|}{3 (75.000000\%)} \\
\hline
MSG & PID & CHID & ND & COUNT\\
\hline
0x106 & 27 & 5 & 0 & 131913569 (87.60\%)\\
0x101 & 27 & 5 & 0 & 4280285 (2.84\%)\\
0x113 & 10 & 3 & 0 & 1546000 (1.02\%)\\
0x0 & 1 & 1 & 0 & 12842020 (8.52\%)\\

\hline
\multicolumn{5}{|c|}{\textbf{./sensor}}\\
\hline
\multicolumn{2}{|l|}{Total calls} & \multicolumn{3}{|l|}{74087213}\\
\multicolumn{2}{|l|}{Total calls to normalize} & \multicolumn{3}{|l|}{182439 (0.24\%)}\\
\multicolumn{2}{|l|}{Num Non-MsgSend calls} &  \multicolumn{3}{|l|}{0 (0.000000\%)} \\
\multicolumn{2}{|l|}{Total MsgSends to procnto} & \multicolumn{3}{|l|}{360645 (0.486784\%)}\\
\multicolumn{2}{|l|}{Total MsgSends to non-procnto} & \multicolumn{3}{|l|}{73726568 (99.513206\%)}\\
\multicolumn{2}{|l|}{Num processes msg receiving} &  \multicolumn{3}{|l|}{2}\\
\multicolumn{2}{|l|}{Num distinct MsgSend types to procnto} &  \multicolumn{3}{|l|}{3 (42.857143\%)} \\
\multicolumn{2}{|l|}{Num distinct MsgSends types to non-procnto} &  \multicolumn{3}{|l|}{4 (57.142860\%)} \\
\hline
MSG & PID & CHID & ND & COUNT\\
\hline
0x101 & 8 & 3 & 0 & 73546247 (99.26\%)\\
0x116 & 8 & 3 & 0 & 72128 (0.09\%)\\
0x100 & 1 & 1 & 0 & 72130 (0.09\%)\\
0x100 & 8 & 3 & 0 & 72128 (0.09\%)\\
0x109 & 8 & 3 & 0 & 36065 (0.04\%)\\
0x41 & 1 & 1 & 0 & 144257 (0.19\%)\\
0x40 & 1 & 1 & 0 & 144258 (0.19\%)\\

\hline
\multicolumn{5}{|c|}{\textbf{./sbin/io-pkt-v6-hc}}\\
\hline
\multicolumn{2}{|l|}{Total calls} & \multicolumn{3}{|l|}{216}\\
\multicolumn{2}{|l|}{Total calls to normalize} & \multicolumn{3}{|l|}{170 (78.7\%)}\\
\multicolumn{2}{|l|}{Num Non-MsgSend calls} &  \multicolumn{3}{|l|}{0 (0.000000\%)} \\
\multicolumn{2}{|l|}{Total MsgSends to procnto} & \multicolumn{3}{|l|}{2 (0.925926\%)}\\
\multicolumn{2}{|l|}{Total MsgSends to non-procnto} & \multicolumn{3}{|l|}{214 (99.074074\%)}\\
\multicolumn{2}{|l|}{Num processes msg receiving} &  \multicolumn{3}{|l|}{2}\\
\multicolumn{2}{|l|}{Num distinct MsgSend types to procnto} &  \multicolumn{3}{|l|}{1 (50.000000\%)} \\
\multicolumn{2}{|l|}{Num distinct MsgSends types to non-procnto} &  \multicolumn{3}{|l|}{1 (50.000000\%)} \\
\hline
MSG & PID & CHID & ND & COUNT\\
\hline
0x101 & 9 & 1 & 0 & 214 (99.07\%)\\
0x40 & 1 & 1 & 0 & 2 (0.92\%)\\

\hline
\multicolumn{5}{|c|}{\textbf{./proc/boot/devc-pty}}\\
\hline
\multicolumn{2}{|l|}{Total calls} & \multicolumn{3}{|l|}{43}\\
\multicolumn{2}{|l|}{Total calls to normalize} & \multicolumn{3}{|l|}{39 (90.69\%)}\\
\multicolumn{2}{|l|}{Num Non-MsgSend calls} &  \multicolumn{3}{|l|}{0 (0.000000\%)} \\
\multicolumn{2}{|l|}{Total MsgSends to procnto} & \multicolumn{3}{|l|}{43 (100.000000\%)}\\
\multicolumn{2}{|l|}{Total MsgSends to non-procnto} & \multicolumn{3}{|l|}{0 (0.000000\%)}\\
\multicolumn{2}{|l|}{Num processes msg receiving} &  \multicolumn{3}{|l|}{0}\\
\multicolumn{2}{|l|}{Num distinct MsgSend types to procnto} &  \multicolumn{3}{|l|}{1 (100.000000\%)} \\
\multicolumn{2}{|l|}{Num distinct MsgSends types to non-procnto} &  \multicolumn{3}{|l|}{0 (0.000000\%)} \\
\hline
MSG & PID & CHID & ND & COUNT\\
\hline
0x17 & 1 & 1 & 0 & 43 (100.00\%)\\

\hline
\multicolumn{5}{|c|}{\textbf{./usr/sbin/sshd} (Did not normalize)}\\
\hline
\multicolumn{2}{|l|}{Total calls} & \multicolumn{3}{|l|}{1067}\\
\multicolumn{2}{|l|}{Total calls to normalize} & \multicolumn{3}{|l|}{ (\%)}\\
\multicolumn{2}{|l|}{Num Non-MsgSend calls} &  \multicolumn{3}{|l|}{0 (0.000000\%)} \\
\multicolumn{2}{|l|}{Total MsgSends to procnto} & \multicolumn{3}{|l|}{143 (13.402061\%)}\\
\multicolumn{2}{|l|}{Total MsgSends to non-procnto} & \multicolumn{3}{|l|}{924 (86.597939\%)}\\
\multicolumn{2}{|l|}{Num processes msg receiving} &  \multicolumn{3}{|l|}{}\\
\multicolumn{2}{|l|}{Num distinct MsgSend types to procnto} &  \multicolumn{3}{|l|}{14 (29.787233\%)} \\
\multicolumn{2}{|l|}{Num distinct MsgSends types to non-procnto} &  \multicolumn{3}{|l|}{33 (70.212761\%)} \\
\hline
MSG & PID & CHID & ND & COUNT\\
\hline
0x102 & 15 & 2 & 0 & 134 (12.55\%)\\
0x105 & 3 & 1 & 0 & 161 (15.08\%)\\
0x105 & 15 & 2 & 0 & 178 (16.68\%)\\
0x105 & 4 & 1 & 0 & 159 (14.90\%)\\
0x101 & 4 & 1 & 0 & 96 (8.99\%)\\
0x10d & 15 & 2 & 0 & 2 (0.18\%)\\
0x116 & 15 & 2 & 0 & 30 (2.81\%)\\
0x116 & 3 & 1 & 0 & 24 (2.24\%)\\
0x116 & 1 & 1 & 0 & 16 (1.49\%)\\
0x100 & 1 & 1 & 0 & 24 (2.24\%)\\
0x100 & 8 & 3 & 0 & 9 (0.84\%)\\
0x109 & 8 & 3 & 0 & 4 (0.37\%)\\
0x101 & 8 & 3 & 0 & 6 (0.56\%)\\
0x116 & 8 & 3 & 0 & 4 (0.37\%)\\
0x72 & 1 & 1 & 0 & 2 (0.18\%)\\
0x101 & 15 & 2 & 0 & 32 (2.99\%)\\
0x102 & 3 & 1 & 0 & 4 (0.37\%)\\
0x102 & 1 & 1 & 0 & 2 (0.18\%)\\
0x115 & 1 & 1 & 0 & 6 (0.56\%)\\
0x115 & 15 & 2 & 0 & 4 (0.37\%)\\
0x115 & 3 & 1 & 0 & 10 (0.93\%)\\
0x115 & 4 & 1 & 0 & 4 (0.37\%)\\
0x116 & 4 & 1 & 0 & 6 (0.56\%)\\
0x100 & 4 & 1 & 0 & 1 (0.09\%)\\
0x10d & 4 & 1 & 0 & 1 (0.09\%)\\
0x106 & 15 & 2 & 0 & 3 (0.28\%)\\
0x11 & 1 & 1 & 0 & 1 (0.09\%)\\
0x102 & 4 & 1 & 0 & 1 (0.09\%)\\
0x115 & 15 & 2 & 0 & 8 (0.74\%)\\
0x106 & 15 & 2 & 0 & 6 (0.56\%)\\
0x100 & 1 & 1 & 0 & 25 (2.34\%)\\
0x100 & 3 & 1 & 0 & 2 (0.18\%)\\
0x10d & 3 & 1 & 0 & 2 (0.18\%)\\
0x100 & 15 & 2 & 0 & 2 (0.18\%)\\
0x10d & 15 & 2 & 0 & 2 (0.18\%)\\
0x12 & 1 & 1 & 0 & 2 (0.18\%)\\
0x17 & 1 & 1 & 0 & 5 (0.46\%)\\
0x0 & 1 & 1 & 0 & 24 (2.24\%)\\
0x115 & 3 & 1 & 0 & 5 (0.46\%)\\
0x41 & 1 & 1 & 0 & 6 (0.56\%)\\
0x106 & 8 & 3 & 0 & 9 (0.84\%)\\
0x100 & 8 & 3 & 0 & 11 (1.03\%)\\
0x71 & 1 & 1 & 0 & 4 (0.37\%)\\
0x13 & 1 & 1 & 0 & 22 (2.06\%)\\
0x40 & 1 & 1 & 0 & 4 (0.37\%)\\
0x115 & 4 & 1 & 0 & 3 (0.28\%)\\
0x106 & 4 & 1 & 0 & 1 (0.09\%)\\

\hline
    \caption{Normalization details per-process} 
    \label{tab:perprocess}
    \end{longtable}
\end{center}

\section{Average False-Positive Graphs}\label{graph}
\label{fig:plot2}

Below are graphs of the average false-positive percentages over all datasets for each process.

\begin{figure}
\includegraphics[keepaspectratio=true,scale=0.65]{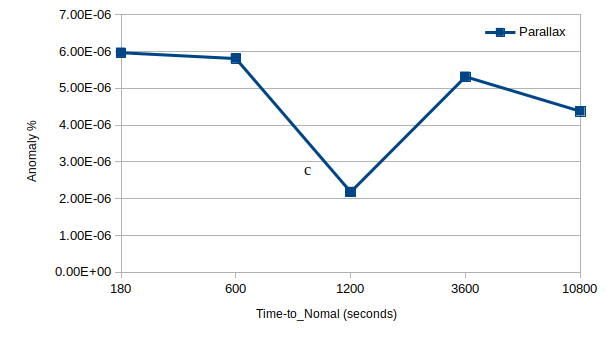}
\end{figure}

\begin{figure}
\includegraphics[keepaspectratio=true,scale=0.65]{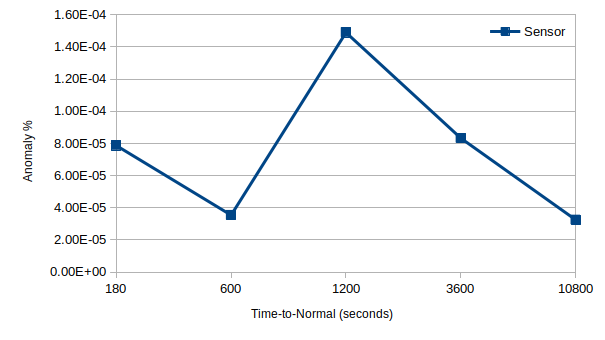}
\end{figure}

\begin{figure}
\includegraphics[keepaspectratio=true,scale=0.65]{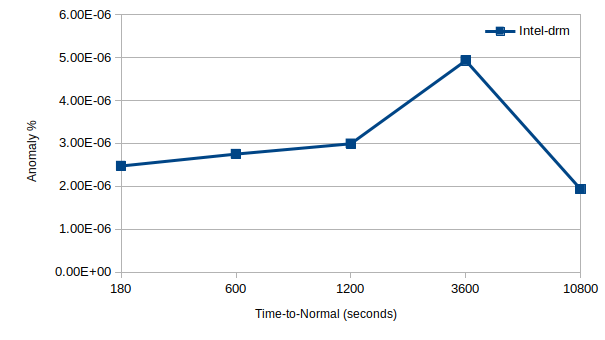}
\end{figure}

\begin{figure}
\includegraphics[keepaspectratio=true,scale=0.65]{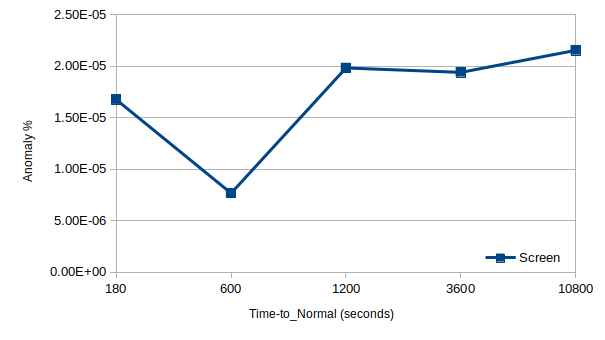}
\end{figure}

\begin{figure}
\includegraphics[keepaspectratio=true,scale=0.65]{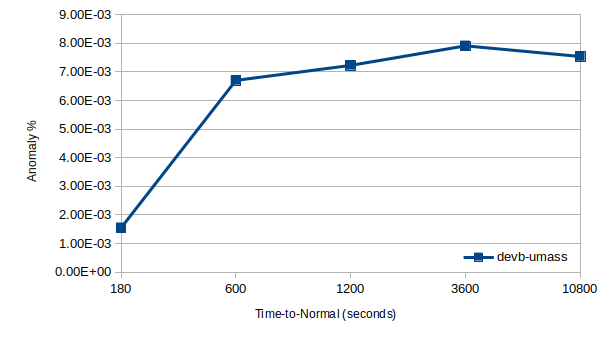}
\end{figure}

\begin{figure}
\includegraphics[keepaspectratio=true,scale=0.65]{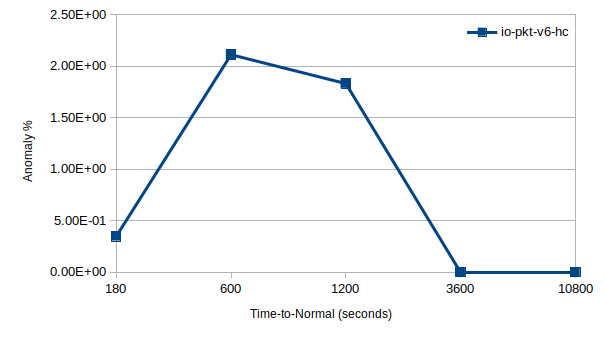}
\end{figure}

\end{document}